# Perturbation centrality and Turbine: a novel centrality measure obtained using a versatile network dynamics tool

**Kristóf Z. Szalay and Peter Csermely**[*]

Department of Medical Chemistry, Semmelweis University, P O Box 260., H-1444 Budapest 8, Hungary

**ABSTRACT**

Analysis of network dynamics became a focal point to understand and predict changes of complex systems. Here we introduce Turbine, a generic framework enabling fast simulation of any algorithmically definable dynamics on very large networks. Using a perturbation transmission model inspired by communicating vessels, we define a novel centrality measure: perturbation centrality. Hubs and inter-modular nodes proved to be highly efficient in perturbation propagation. High perturbation centrality nodes of the Met-tRNA synthetase protein structure network were identified as amino acids involved in intra-protein communication by earlier studies. Changes in perturbation centralities of yeast interactome nodes upon various stresses well recapitulated the functional changes of stressed yeast cells. The novelty and usefulness of perturbation centrality was validated in several other model, biological and social networks. The Turbine software and the perturbation centrality measure may provide a large variety of novel options to assess signaling, drug action, environmental and social interventions.

**INTRODUCTION**

In the last decade the network approach became a widely used methodology to study complex systems. As an example, protein structure networks, where network nodes represent amino acids, and edges symbolize their physical distance, are increasingly used to describe conformational changes, formation of protein complexes, drug binding and enzyme action [1–3]. Recently several programs have been introduced for the construction of protein structure networks from 3D structural data and for their analysis [4,5]. Protein-protein interaction networks (interactomes) provide a great help to understand the molecular mechanism of cellular functions, the development of diseases and drug design [6]. In protein-protein interaction networks such as BioGRID [7], STRING [8], and DIP [9], nodes are proteins, and edges denote their physical interactions.

Network dynamics is necessary to understand the changes of complex systems, and therefore became a hot topic of network studies [10,11]. A number of programs have been developed for the calculation of certain aspects of network dynamics, such as network simulation tools based on Boolean dynamics [12–16], the random-walk based ITM-Probe [17], the law of mass action-based PerturbationAnalyzer [18], or the biological system modeling tool, BIOCHAM [19]. However, to our knowledge, no stand-alone program exists, which can easily integrate any dynamical models together with any types of starting perturbations, and can also provide the complete time-domain simulation results, not only the summative end-result. Recently the complex network dynamics simulation tool, Conedy was introduced [20]. Conedy is a Python extension enabling researchers already using computational dynamics to add networks to their repertoire. However, a complete toolkit is still missing having built-in algorithms, analysis tools and visualization, to enable life and network scientists to add network dynamics to their studies.

---

[*]Corresponding author: Csermely, P. (csermely.peter@med.semmelweis-univ.hu)

Our in-house developed program, Turbine, is able to accommodate a large variety of network dynamic simulations. Any real-world network can be imported to the program and perturbations can be introduced at any nodes or node-combinations at the beginning, at any time during the simulation, individually, repeatedly, or continuously. This allows the analysis of the effect of different environmental changes on network dynamics. Computational optimizations allow the simulation of large networks (in the range of millions of nodes and edges) on a standard desktop-grade computer. The initial phase of Turbine development was mentioned in an earlier conference lecture summary [21]. Here we introduce the fully developed program (freely downloadable from here: http://turbine.linkgroup.hu), and show that its results on the importance of hubs and inter-modular nodes in the propagation of perturbations reflect well both our intuitive expectations and former knowledge. We defined a new measure of dynamic network centrality, and termed it as perturbation centrality. Perturbation centrality correctly identified substrate binding sites and amino acids participating in allosteric signaling in protein structure network networks. Changes of perturbation centrality were in agreement with the known functional changes of the yeast protein-protein interaction network after stress. The novelty and usefulness of perturbation centrality was validated in several other model, biological and social networks. Turbine is an integrative and versatile tool to simulate network dynamics and to predict the effects of environmental changes, signaling stimuli, drugs or social interventions.

## RESULTS

In the simulations using our network dynamics program called Turbine, we used a model termed "communicating vessels". The basic idea behind this model was that intensive physical variables (e.g. temperature) tend to perform an equalization-like dynamics behaving like communicating vessels (see the detailed description in **Methods**). The communicating vessels model gives an exponential decay of perturbations (see **Supplementary Results** of **Text S1**), which is in agreement with several earlier findings [22,23].

**Modules limit perturbation propagation**

The intuitive impression that modules limit the information spread in complex networks was described in multiple papers, and was shown in many simulations [24–27]. The equations of the communicating vessels model (where every affected node dissipates an equal amount of energy in every time-step of the simulation) suggest that the more nodes are affected by a given perturbation, the faster the perturbation becomes dissipated. Taken together these two considerations, our expectation was that a network with rather distinct modules (termed as pronounced modules) will propagate and dissipate perturbations slower than a network with tightly interconnected modules (termed as fuzzy modules). To test whether the communicating vessels model of Turbine shows this property, we used the benchmark graph generating tool of Lancichinetti and Fortunato [28] to generate unweighted and undirected scale-free, modular benchmark networks (hereafter called as benchmark graphs) with different ratios of inter-modular edges (**Table S7** of **Text S1**). The benchmark graph with pronounced or fuzzy modules had 5% or 40% of inter-modular edges, respectively. We have used a new measure termed "fill time" for comparison. The fill time of node $i$ is the time needed to raise the energy level of 80% of the nodes above 1 unit while continuously adding energy to node $i$. **Figure 1A** shows fill times calculated on 7 random generations of these benchmark graphs using the same perturbations starting from each node in separate simulations. Fill times of all nodes and all 7 benchmark graphs with different random seeds were averaged. As expected, sparse inter-modular edges of the pronounced modules delayed the propagation of perturbation, resulting in a 4.8 times larger fill time as compared to those observed in benchmark graphs with fuzzy modules (**Figure 1A**). **Supplementary Results** in **Text S1** show that fill time is highly correlated with closeness centrality, making fill time useful as a verification of the model rather than a novel centrality measure.

Next we assessed the propagation of single perturbations using the same model, but adding a dissipation term to the communicating vessels dynamics. **Figures 1B through 1E** are illustrations of the propagation of dissipated perturbations using image snapshots of the Turbine viewer program after 50 time-steps of the simulation. The starting module of the benchmark graphs with pronounced modules trapped the initial perturbation, if the size of the perturbation was sufficiently high (1,000,000 units, **Figure 1D**). This 'module encapsulation' effect was entirely missing from the benchmark graphs with fuzzy modules (**Figures 1C and 1E**), and was also not observable, when the starting perturbation was low intensity (10,000 units). Thus, roughly the same number of nodes was affected by the perturbation in benchmark graphs having either pronounced (**Figure 1B**) or fuzzy modules (**Figure 1C**) if the initial perturbation was low-intensity (10,000 units). On the contrary, a high-intensity starting perturbation (1,000,000 units) affected much less nodes in benchmark graphs having pronounced modules (**Figure 1D**) as compared to those having fuzzy modules (**Figure 1E**). After applying a high intensity perturbation to benchmark graphs with fuzzy modules almost all nodes became perturbed after the 50 time-steps shown (**Figure 1E**).

The right two sets of bars of **Figure 1A** show the differences in perturbation dissipation in a quantitative manner. Here a measure termed "silencing time" was calculated as the first time step when all nodes had an energy value less than 1. The same perturbation was started from each node of the 7 random representations of benchmark graphs in separate simulations, and their silencing times were averaged for all nodes and for all the 7 graphs. Bars with capital letters refer to the benchmark graphs shown on **Figure 1B through 1E.** Benchmark graphs with pronounced modules dissipated low intensity perturbations only slightly slower than benchmark graphs with fuzzy modules (cf. bars "B" and "C" on **Figure 1A**). This is in agreement with the approximately same number of nodes affected after 50 time-steps in the same pair of simulations (**cf. Figures 1B and 1C**). On the contrary, high intensity perturbations were dissipated dramatically (2.6 times) slower by benchmark graphs with pronounced modules as opposed to those with fuzzy modules (cf. bars "D" and "E" on **Figure 1A**). These results clearly indicated that pronounced modular structures trap perturbations in agreement with earlier results [27]. The difference between the behavior of low-, and high-intensity perturbations arises from the fact that perturbations are decaying exponentially with the distance from their origin (for a proof see the analysis of **Text S1**). Thus, when the perturbation is of relatively low-intensity (compared to the size of the module and the dissipation rate) it is dissipated without reaching the boundaries of its module of origin. In the case of high-intensity perturbations, a significant amount of energy transverses the boundary of its module of origin, which makes the modular-encapsulation effect observable. Module encapsulation of perturbations was also tested using the random-walk based ITM-Probe method [17], with very similar results (see **Supplementary Results**, **Table S7** of **Text S1**, and **Figures S1 through S3** of **Text S1**).

**Definition of perturbation centrality as the reciprocal of silencing time**

Prompted by the utility of silencing time to characterize the propagation and dissipation of perturbations (**Figure 1**), and utilizing our former observation that the reciprocal of fill time was correlated with closeness centrality (**Table S1** of **Text S1**), we defined a centrality-type measure for dissipated perturbations, and termed it as perturbation centrality. Our aim was to conceive a measure that takes into account both local (weighted degree) and more global (modular position) perturbation properties. It was also important that the measure should be easy to understand and calculate, and that its properties should be almost independent of the size of the network. As a result of our initial studies (**Table S2** of **Text S1**) we have found that setting the initial perturbation value to $10*n$ units ($n$ being the number of nodes in the network) achieves all of these goals. Thus, perturbation centrality of node $i$ was defined as the reciprocal of silencing time retrieved by using a Dirac delta type starting perturbation of $10*n$ units, where $n$ is the number of nodes in the network, using a dissipation value of 1. Silencing time of node $i$ was the first time when all of the nodes had an energy value less than a pre-set minimal threshold after an initial perturbation started from node $i$. We selected this threshold as "1", the minimum sensible value with the dissipation also being set to 1 (note that the dissipation value is the minimal threshold, since after reaching this value all energy of the network will be dissipated in the next step).

Following the above definition, we have tried to find the location of our newly conceived perturbation centrality measure on the "centrality landscape" by testing its correlation with established centrality measures in selected networks. We have tested perturbation centrality against closeness centrality, which is the average geodesic distance from the given node to all other nodes; betweenness centrality, the number of shortest paths between any two nodes passing through the tested node; community centrality [29], a measure high in the cores of network communities and PageRank, an iterative measure coined by Brin and Page [30], where nodes "vote" on each other *via* their edges in proportion with their degree. We have also tested the correlations between perturbation centrality and degree (as well as weighted degree), since these measures can also be interpreted as local centrality measures. These correlations between perturbation centrality and other centrality-type measures of different real-world networks are shown in **Table 1**. There was a considerable correlation between perturbation and closeness centralities. However, the strength of this correlation was noticeably less than that between the reciprocal of fill time and closeness centrality (average correlations were 67% and 89.5%, respectively, with explained change, $R^2$ values of 45% and 80%, respectively). A similarly high correlation was observable between perturbation and community centralities, as well as between perturbation centrality and weighted degree suggesting that nodes in key community locations and/or hubs may be among the best dissipators of perturbations. Correlations between perturbation centrality and either PageRank or betweenness centrality were smaller (but noticeable) than that between perturbation and closeness centralities.

On one hand, data of **Table 1** indicate that perturbation centrality is a more local centrality measure than closeness centrality. On the other hand perturbation centrality is a more global centrality measure than weighted degree or PageRank. Thus perturbation centrality is a novel, mesoscopic-type centrality measure characterizing the information transfer capability of a given node (or edge: see **Supplementary Results** and **Figure S10** of **Text S1**) in a network. A visual representation of the relationship between perturbation centrality and the other centrality measures shows a unique position of perturbation centrality further supporting the novelty of perturbation centrality (**Figure S4** of **Text S1**).

**Hubs and inter-modular nodes have a central role in perturbation dissipation**

Next we investigated which types of nodes are best for dissipating perturbations. On one hand, we observed large differences in the average perturbation efficiency of modular networks with differing ratios of inter-modular nodes (**Figure 1**). On the other hand, hubs have been proven to have a high information transmission efficiency [31,32]. Based on these considerations we defined 4 node categories: 1.) intra-modular non-hubs; 2.) intra-modular hubs; 3.) inter-modular non-hubs and 4.) inter-modular hubs. We defined hubs as nodes with degrees in the top 10%, and inter-modular nodes as nodes with at least 40% inter-modular edges. Intra-modular non-hubs correspond to roles R1 and R2 (ultra-peripheral and peripheral nodes) in the previous representation of Guimerà *et al.* [33]; intra-modular hubs correspond to role R5 (provincial hubs), while inter-modular non-hubs and inter-modular hubs correspond to roles R3/R4 (non-hub connectors and kinless nodes) and R6/R7 (connector and kinless hubs), respectively. **Figure 2** shows that, in agreement with our expectations, inter-modular non-hubs had a larger perturbation centrality than intra-modular non-hubs in benchmark networks with pronounced modules. Importantly, in networks with pronounced modules inter-modular non-hubs had larger perturbation centrality than intra-modular hubs. On the contrary, in benchmark networks with fuzzy modules hubs in any modular position had a larger perturbation centrality than non-hubs. These observations are again in agreement with expectations and earlier findings [27]. The explanation of these findings is that in fuzzy modules the modular structure did not restrict the propagation of perturbations, so it is not surprising that intra-modular hubs dissipated perturbations faster than inter-modular non-hubs.

Importantly, there was a large (87%) difference between the perturbation centrality of intra-modular hubs versus the effect of inter-modular non-hubs in a wide variety of real-world networks (**Supplementary Results** and **Table S8** of **Text S1**), suggesting that from a perturbation perspective real-world networks resemble the benchmark graphs with fuzzy modules more, than the benchmark graphs with pronounced modules. (Note that the same observation was obtained, when we compared the low-intensity and high-intensity silencing times – see **Table S2** of **Text S1**).

**Perturbation centrality uniquely identifies all key regions of Met-tRNA synthase**

Prompted by the general applicability of the perturbation centrality measure to characterize real-world networks, next we compared the perturbation centralities with structural and functional properties of nodes in two types of biological networks in detail. The comparison of substrate-free with substrate-bound forms of proteins gives an important system to study the changes in perturbation differences in the respective protein structure networks.

**Figure 3** shows residues with top 20% increase of different centralities upon substrate binding of Met-tRNA synthase. Red and yellow symbols of **Figure 3A** represent residues with highly increased *perturbation* centrality. Residues marked with yellow symbols are within a distance of 4.5Å from the tRNA$^{Met}$. Note that perturbation centrality increase highlights the active site and both binding sites of the tRNA. On the contrary, residues with highly increased *closeness* centrality (**Figure 3B**) are smeared around the active site, and residues with the highest increase of *betweenness* centrality (**Figure 3C**) are scattered all around the protein. The fact that perturbation centrality was increased most at the two substrate binding sites of tRNA$^{Met}$ upon binding, indicates that substrate-induced changes in protein structure help a better spread of perturbations caused by substrate binding. This self-amplification may be an important contributor to the propagation of binding-induced conformational changes and allosteric mechanisms.

The tight secondary structures of α-helices had larger perturbation efficiency than the more flexible loops. Importantly, perturbation centrality proved to be better at distinguishing secondary structures, signaling amino acids, as well as amino acids whose importance was proven experimentally than betweenness or closeness centralities (**Figures S6** through **S8** and **Table S9** of **Text S1**).

**Various stress types induce different perturbation dissipating regions of the yeast interactome**

As a continuation of the characterization of substrate-induced changes in protein structure networks, we assessed perturbation centralities in a well-characterized change of the interactome. In our earlier studies stress-induced changes in mRNA expression resulted in a marked re-configuration of yeast interactome modules [34,35]. Here we calculated perturbation centralities for all nodes in the Database of Interacting Proteins yeast interactome (release 2005), using stress-induced mRNA changes [36,37] to calculate the edge-weights of the stressed yeast interactome as described before [35]. The observation of Mihalik and Csermely [35] that communities of the interactome become more separated under heat shock is expected to induce a lower average perturbation centrality (due to the module encapsulation effect shown before hindering the propagation of perturbations). Indeed, a major change was observed: the average perturbation centrality of the heat-shocked interactome was $6.07*10^{-4}$, 40.5% lower than the average perturbation centrality of the unstressed interactome ($1.02*10^{-3}$, α=0.05, p=$2.2*10^{-16}$, Wilcoxon rank-sum test).

We also observed a marked difference of the proteins with highest perturbation centrality in heat-shock compared to the other stress types. Only 42 of the 100 unstressed top perturbation centrality nodes appeared in the top 100 nodes of heat-shocked cells. However, 65 of the unstressed top nodes appeared in the oxidatively-stressed interactome, and 68 in the osmotically-stressed interactome. At the same time, 77 of the top 100 perturbation centrality nodes were the same in the oxidatively- and the osmotically-stressed interactome, while the number of matching nodes was only 30 and 34, when we compared the oxidatively- and the osmotically-stressed interactome against the heat-shocked interactome, respectively. These results are visualized in the Venn-diagram of **Figure 4**. These data prompted us to perform a detailed investigation of the functions of the top perturbation dissipator proteins in unstressed and stressed yeast interactomes.

**Functional assessment of key perturbation dissipator proteins in unstressed and stressed yeast interactomes**

For the assessment of the function of proteins having the highest perturbation centrality in the yeast interactome before and after various types of stresses, we used the g:Profiler tool [38], which performs a statistical enrichment analysis to find over-representation of Gene Ontology terms, biological pathways, or regulatory DNA elements in a set of genes or proteins. Only those terms were taken as significant, where the p-value was less than 0.05 after applying Bonferroni correction. Using this method, the top 100 nodes having the largest perturbation centrality in the unstressed interactome had 11 enriched terms, which was extended differently in heat shock and oxidative/osmotic stress (**Table S3** of **Text S1**).

The enrichment analysis on the nodes having the top 100 largest *increase* in perturbation centrality in heat-shock, oxidative and osmotic stress resulted in 28, 22 and 34 enriched terms, respectively. *Carbohydrate metabolism*, *trehalose metabolic process* and *glycogen metabolic process* were upregulated in all types of stresses, which is in agreement of previous findings [39]. Importantly, the term *response to stimulus* appeared in all three types of stresses, and *response to stress* appeared in heat-shock and osmotic stress (**Table S4** of **Text S1**). Assessment of the function of proteins having the largest *decrease* in their perturbation centralities in various stress conditions indicated the down-regulation of *ribosome synthesis* and *protein translation* after stress (**Table S5 Text S1**), which are also well-known changes in stress [40].

Our results on protein structure and protein-protein interaction networks highlight the usefulness of perturbation centrality to identify functionally important nodes in biological networks, and show that the preferred way of comparing perturbation centralities in two similar networks is to compare the largest *changes* rather than the largest absolute values.

## DISCUSSION

We introduced a new method for the analysis of network dynamics. This new software called "Turbine" (http://turbine.linkgroup.hu) substantially extends the preliminary version of the program published as a conference summary [21]. A dynamic model termed "communicating vessels" was created to assess the propagation of perturbations in networks. To characterize network properties, two new measures were defined. "Fill time" characterizes the propagation-efficiency of un-dissipated perturbations. "Perturbation centrality" of a node is defined as the reciprocal of the time characterizing the dissipation of a perturbation starting from the given node in the network. Both the reciprocal of fill time and perturbation centrality were shown to be centrality-type measures. Fill time correlated well with closeness centrality. On the contrary, perturbation centrality could not be substituted with any standard network centrality measure. Perturbation centrality correctly identified hubs and bridges (inter-modular nodes) as key determinants of the speed of perturbation dissipation. Nodes having a high importance in the information transmission in protein structure networks and in protein-protein interaction networks were also characterized by high perturbation centrality values.

Network dynamics has already been assessed using a number of computational tools. Boolean dynamics [12–16] is a binary model, where every node can assume either an active or an inactive state, making Boolean dynamics a generalization of cellular automata on complex networks. Despite its simplicity, Boolean dynamics has been very successful in modeling cellular signaling networks, and identifying underlying causes of pathogenic responses [13,41]. However, there are also a handful of programs for non-Boolean dynamics. ITM-Probe [17] is based on random-walks; PerturbationAnalyzer [18] is a Cytoscape plug-in using the law of mass action; BIOCHAM [19] and Conedy [20] are more complex network dynamic tools. Turbine combines several advantages of the former options with large versatility, richness of output data, efficient use of memory and fast running time enabling the analysis of large networks (**Table S6** of **Text S1**). Turbine is able to accommodate a large number of other dynamical models than the communicating vessel model used in this paper. Turbine can handle multiple, repeated or continuous perturbations introduced at the beginning of the simulation or at any later time-steps. Moreover, the network structure may also be changed during the simulations.

Despite its apparent simplicity, the communicating vessels model well recapitulated the expected dissipation pattern of hub and inter-modular node perturbations on modular, scale-free benchmark graphs (**Figures 1** and **2**). These results were in agreement with earlier data [24–27].

Encouraged by these findings, we defined a novel type of dynamic centrality measure, and termed it as perturbation centrality. Perturbation centrality of node $i$ is the reciprocal of the silencing time of node $i$ with a starting excitation of $10*n$, where $n$ is the number of nodes in the network, setting both the dissipation and the silencing threshold to 1. Furthermore, the silencing time of node $i$ is the time needed to dissipate the perturbation starting from node $i$ at every node below a low residual perturbation threshold. The perturbation centrality measure has a rather straightforward centrality-type meaning. Intuitively thinking, the more central node $i$ is, the more nodes are reached by the perturbation started at node $i$. Thus perturbation started from a more central node is dissipated faster – since every node dissipates an equal amount of energy in each step – and so has a smaller silencing time than a perturbation started from a less central node.

Silencing time is not a continuous measure thus the precision of perturbation centrality has a lower bound. However, the parameters of the simulations were set achieving a rather good compromise between calculation speed and the resolution of perturbation centrality. Typical perturbation centrality values ranged from 0.33 (highest) to 0.0001 (lowest) depending on the analyzed network. These values corresponded to silencing times 30 and 10000. Note that the lowest perturbation centrality one can get depends on the number of simulated time-steps, i.e. the lowest perturbation centrality in an experiment with 2000 time steps is 1/2000=0.0005. The $n*10$ starting energy (where $n$ is the number of nodes) and the 1 dissipation rate parameters of the perturbation centrality were chosen in order to achieve that nodes in most networks can be characterized by silencing time values between 10 and 10000 steps. This translates to a perturbation centrality value between 0.1 and 0.0001 Thus, these parameters made a good compromise between the total time of simulation and the resolution of the perturbation centrality measure.

All correlations between perturbation centrality and other centrality measures were weaker than the high correlation between the reciprocal of fill time and closeness centrality (cf. **Tables 1** and **Table S1** of **Text S1**; see **Figure S4** of **Text S1** for a graphical representation). Thus perturbation centrality may capture novel dynamics-related features of central nodes (or in a very similar fashion, edges: see **Supplementary Results** and **Figure S10** of **Text S1**). In several case studies on protein structure networks and yeast protein-protein interaction networks we showed that this indeed may be the case. The distribution of perturbation centrality values was different in different networks, which may give an additional layer of network characterization (**Figure S5** of **Text S1**).

Perturbation properties of protein structures revealed by the Turbine model were in agreement with intuitive insights. The tight secondary structures of α-helices had large perturbation efficiency, while the more flexible loops had a lower efficiency of propagating and dissipating perturbations (**Figure S6**). Perturbation centrality discriminated secondary structures slightly better than closeness centrality and much better than betweenness centrality (**Figures S7** and **S8** of **Text S1**). Moreover, perturbation centrality, uniquely of the three tested centralities, highlighted all important segments of Met-tRNA synthase. The substrate binding-induced local increase in perturbation centralities may indicate a self-amplifying cycle, where substrate-induced changes might help a better spread of perturbations caused by substrate binding. Amino acids involved in allosteric communication in Met-tRNA synthetase [42], as well as amino acids with experimentally verified importance in its function [42] had significantly higher than average perturbation centrality in the protein structure network of the enzyme (**Figure S6**; $p=6*10^{-8}$ and $9.5*10^{-8}$ for amino acids involved in communication; bound and free conformation, respectively; p=0.0018 and 0.0022 for amino acids with experimentally verified importance; bound and free conformation, respectively using Wilcoxon rank-sum test, α=0.0125 adjusted with Bonferroni correction). These findings are in agreement with a number of earlier studies suggesting that perturbation efficiency plays a key role in intra-protein allosteric signaling, as well as showing that both binding sites and inter-domain, bridging amino acids play an especially important role in this process [43–46].

Differences between perturbation centralities in the interactomes of unstressed and stressed yeast (**Tables S4 and S5** of **Text S1**) were in agreement with our earlier data on the modular rearrangements of the yeast interactome upon stress [34,35] and with experimental data showing the down-regulation of yeast ribosome biogenesis and mRNA translation [40] , as well as the up-regulation of carbohydrate metabolism [39] after stress.

Considering the above results, the Turbine network dynamics tool and the perturbation centrality measure may have a number of highly interesting future applications. Studies on perturbations of various real world networks were used to assess network robustness [47]. Perturbation analysis was used in the identification of drug target candidates, including multi-target drugs or allo-network drugs [11,48–51]. Sequential perturbations have been suggested as a key modality of anti-cancer therapies [47]. Input signals with different dynamical profiles cause several non-trivial phenomena in signaling networks, such as kinetic insulation [52]. All these possibilities may be assessed by Turbine in the future and can be extended to ecosystems, social networks (infection spread, viral marketing) and engineered networks (power grids, Internet, etc.).

In conclusion, here we introduced Turbine, a new method for the analysis of network dynamics, and used it to study the propagation of perturbations in modular benchmark graphs and several types of real-world networks. We applied a dynamic model based on the concept of communicating vessels, and defined a new measure of dynamic network centrality, called as perturbation centrality. Hubs, inter-modular bridges and signal transducing amino acids were identified as nodes of high perturbation centrality, and were in agreement with a large number of earlier data. Changes of perturbation centrality in stressed yeast interactomes well described known functional changes after stress. The Turbine method and perturbation centrality open a large variety of options for future studies on network robustness, signaling mechanisms, drug design, as well as management of ecosystems, social and engineered networks.

## METHODS

**Brief description of the Turbine program**

An in-house designed program package, Turbine (http://turbine.linkgroup.hu) was used for the simulation of network perturbations. Turbine is a MATLAB and R-compatible toolkit for the analysis of network dynamics (including perturbations). Turbine contains multiple sub-programs written in C++, and a viewer written in C# to enable visual interpretation of the results. The program is using its own binary data and network format for performance reasons, but converters from multiple file formats, such as the *Pajek* network format or the *MATLAB/Octave* data format are also part of the default toolkit.

Turbine is based on a generalization of the systems theory approach [53] to complex networks. In the program we assign a state variable to all parameters of a network, which are expected to change during the simulation. Every node or edge has a separate value of every defined state variable. The effects of the state variables on each other (the evolution of the system in time) are determined by the particular *network dynamics model* used. In systems theory, this model is a set of ordinary or partial differential equations describing the change of the state variables in time, taking into account the effects of other state variables on the current component. In Turbine, any algorithm can be used as a model, making the user capable of simulating virtually any dynamics in any network. In the model, the user has to define the values of the state variables for the next step based on the values in the current step, which translates to creating a C++ function named PerStep(), which should return the values of the state vector for the $(n+1)^{th}$ time step given all preceding values in the $1^{st}$ through $n^{th}$..time steps. Of course, a model file may opt not to use all this information, and indeed, the communicating vessels model only uses the state vector of the previous, $n^{th}$ step, as it will be described in the difference form of the model equations of the communicating vessels model described in the following section. Turbine models are stored as extendable and replaceable DLL files. Multiple default models – such as "ripple" for testing wave-like propagation, "gossip" for testing binary probabilistic information spreading, and "XY" modeling the evolution of the Prisoner's Dilemma game in a network – are supplied with the Turbine program package available at our website: http://turbine.linkgroup.hu. Selecting a model for a given network requires background knowledge on the nature of the dynamics of the complex system represented by the network. In the future, we plan to introduce more specialized dynamics such as "integrate-and-fire" for neural networks to make model selection simpler.

For running a simulation, the user has to define the 1.) time of the individual steps (called as step-time); 2.) the total analysis time, which together with the step-time determines the number of performed iterations, and 3.) the starting values of the state variables. It is also possible to introduce perturbations to the system during the simulation. A perturbation can be applied to every state variable, separately for each node or edge, and the value corresponding to the perturbation at the current time step is added to the current value of the node's or edge's instance of the state variable. This way, any combination of node and edge perturbations can be added to the system (e.g. single, multiple, sequential or continuous), at any time, which allows the user to test the response of the network to any environmental effects, such as different drugs or drug-combinations, or any intrinsically generated effects, such as gene expression noise. The detailed description of the Turbine program, its User Guide and freely downloadable versions of Turbine, as well as their source codes are available at our web-site: http://turbine.linkgroup.hu.

**The communicating vessels model**

In the simulations of this paper, a model with only one state variable (energy) was used, to assess the dissipation of information (e.g. physical perturbations) in complex networks. The basic idea behind the model was that intensive physical variables (e.g. temperature) tend to perform an equalization-like dynamics behaving like communicating vessels. In the communicating vessels model network nodes represent the vessels and edges represent their connecting pipes. The algorithm of the model is as follows: in each time-step, every node transfers a proportion of its available energy through every available edge, proportional to 1.) the duration of the time-step; 2.) the weight of the edge (corresponding to 'pipe diameter'); and 3.) the difference of the state variables on the two ends of the edge (corresponding to 'pipe pressure'). There was a 'vaporization' effect, meaning that a constant amount of the available energy of every node could be dissipated in every step, which is an important property of most dynamical systems including molecular networks. Based on the above considerations, the differential form of the equation of the communicating vessel model is the following:

$$\frac{dS}{dt} = -\sum_{i=0}^{l}\left(\frac{S-S_i}{2}w_i\right) - D_0$$

where $S$ is the value of the state variable (energy) of the starting node of the edge, $S_i$ is the state variable (energy) of the node on the other end of the current edge, $w_i$ is the weight of the current edge, $l$ is the number of edges (degree) of the current node, and $D_0$ is the dissipation coefficient, which is kept constant for all nodes.

The differential form is only equal to the discrete difference equations which the algorithm uses if the step-time is infinitesimal. However, analyzing differential equations are often much easier, and this form is more familiar for systems theorists. For the sake of completeness, we have also included the difference equation form calculated by the algorithm below:

$$S[t+1] = -\sum_{i=1}^{l}\left(\frac{S[t]-S_i[t]}{2}w_i\right) - D_0$$

Variable names match the ones of the differential form: $S$ is the value of the state variable (energy) of the starting node of the edge, $S_i$ is the state variable (energy) of the node on the other end of the current edge, $w_i$ is the weight of the current edge, $l$ is the number of edges (degree) of the current node, and $D_0$ is the dissipation coefficient, which is kept constant for all nodes. This difference form of the equation shows an important criterion when choosing edge weights: the (absolute value of the) sum of weighted outdegrees should not exceed the reciprocal of the step-time ($1/\Delta t$) for any node, otherwise more energy will be propagated outwards than the amount contained in the node which – besides violating the conservation of energy – can destabilize the whole simulation. We have created a plugin for Turbine (*normalize mflow*) which can normalize any network to meet this criteria, keeping *relative* edge weights intact. This criterion can be summarized in the following equation:

$$-1 \leq \Delta t \sum_{i=1}^{l} w_i \leq 1$$

The described equations (and the algorithm) of the communicating vessels model can be naturally extended to directed graphs. This modification can be attained by separating the energy transfer on the inbound and outbound edges, like the following:

$$\frac{dS}{dt} = -\sum_{j=1}^{o}\left(\frac{S}{2}w_j\right) + \sum_{j=1}^{i}\left(\frac{S_j}{2}w_j\right) - D_0$$

where $S$ is the value of the state variable (energy) of the current node, $o$ is out-degree of the node, $w_j$ is the weight of the current edge, $S_j$ is the state variable (energy) of the current neighbor, $i$ is the in-degree of the current node, and $D_0$ is the (constant) dissipation coefficient. This model also allows the assessment of information propagation, silencing times and perturbation centrality in directed real-world networks such as the Internet, citation networks, or biological signaling networks.

This model provides a good starting point for the simulation of most network dynamics, if more detailed information is not available about the mechanism of the actual dynamic process. The DLL file containing the communicating vessels model is included with all Turbine packages, and is available on our web-site: http://turbine.linkgroup.hu.

**Turbine simulations**

Scripts for running all simulations with the exact parameters and networks used are downloadable from our web-site: http://turbine.linkgroup.hu.

**Calculation of fill time, silencing time and perturbation centrality**

Two types of tests were conducted on the target networks using the communicating vessels model described above. In the calculation of fill time one node was excited with 10,000 units of energy *in each time step*, and a $D_0=0$ constant dissipation was set. The speed of the propagation of the perturbation starting at the given node was characterized by the fill time of the network, which was defined as the time during the simulation when more than 80% of the nodes in the network had an energy value larger than 1. The fill time measure was calculated for each node of the network.

In the calculation of silencing time and perturbation centrality one node was excited with a given amount of energy *at the start of the simulation*, which was 10,000, 40,000 or 1,000,000 units as stated in the individual simulations, and a $D_0=1$ constant dissipation was set. Silencing time was defined as the first time, when all of the nodes had an energy value less than a pre-set threshold, which was 1 in all of our experiments. The speed of the dissipation of the perturbation starting at the given node was characterized by the perturbation centrality, defined as the reciprocal of the silencing time of an $n*10$ sized Dirac-delta-type perturbation, where $n$ is the number of nodes in the network, with the dissipation and the threshold for the silencing time set to 1 (the reasons of this choice can be found in **Table S2** of **Text S1**). All measures were calculated for each node of the network.

Turbine plug-ins to calculate the silencing time, the fill time, a script to calculate the perturbation centrality measure, as well as the User Guide can be downloaded from the web-site: http://turbine.linkgroup.hu.

**Generating protein structure networks from structure information**

Protein structure networks from the substrate-free and substrate-bound form of methionyl-tRNA synthetase enzyme (Met tRNA synthase [42]), as well as the substrate-free (1PO5, [54]) and imidazole-bound (1SUO, [55]) form of rabbit cytochrome P450 2B4 were built with the RINerator software [5], using the PDB files received from the authors of [42] in the case of MetRS, and using the published 1PO5 and 1SUO structures from the Protein Data Bank. The absolute value of interaction strengths was used for network building, since the companion program of RINerator called Probe returns negative interaction strengths for repulsive interactions, but our perturbation propagation model is assumed to only depend on the strength of the interaction rather than its repulsive or attractive nature.

**Characterization of proteins important in perturbation propagation in resting and stressed yeast cells**

For the functional characterization of proteins having the highest perturbation centralities (or highest changes in perturbation centralities) in resting yeast cells or yeast cells after various types of stresses, term enrichment analysis was performed using the R plugin of the g:Profiler [38] web service. g:Profiler uses terms from Gene Ontology, KEGG, and several pathway databases. Significant enrichment was stated when the p-value of a term was strictly less than 0.05 after applying Bonferroni correction for multiple testing.

**Statistical methods**

Statistical analyses including the calculation of means, medians, standard deviations, Welch two-sample t-tests, Wilcoxon rank-sum tests and correlation analyses were done using the R package [56].

**SUPPORTING INFORMATION**

Text S1: This supporting information (Text S1) contains 10 Supplementary Figures, 9 Supplementary Tables, Supplementary Results, Supplementary Methods as well as 39 Supplementary References.

## ACKNOWLEDGMENTS

Authors thank Amit Ghosh (Lawrence Berkeley National Laboratory, Berkeley CA, USA) and Saraswathi Vishveshwara (Molecular Biophysics Unit, Indian Institute of Science, Bangalore, India) for the *E. coli* Met-tRNA-synthetase 3D structural data, Balázs Szappanos (Biological Research Centre, Szeged, Hungary) for the metabolic network data, Ágoston Mihalik (Computational Cognitive Neuroimaging Group, University of Birmingham, UK) for the stressed yeast interactome data and members of the LINK-Group (www.linkgroup.hu) for their discussions and help.

## AUTHOR CONTRIBUTIONS

Conceived and designed the simulations: KZS, PC. Performed the simulations: KZS. Analyzed the data: KZS, PC. Wrote the paper: KZS, PC.

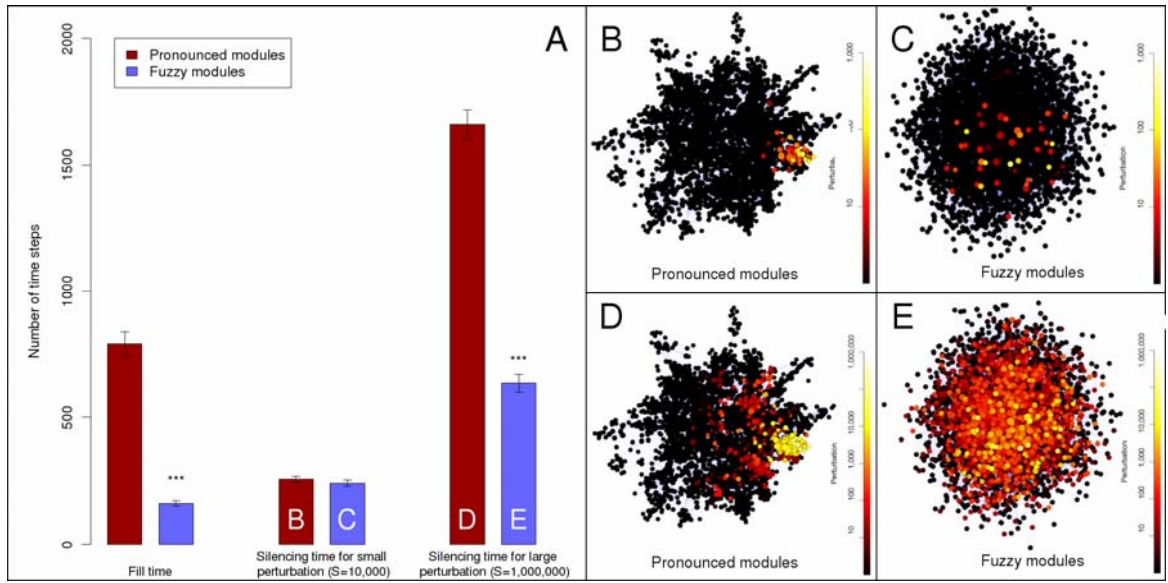

**Figure 1. Difference in perturbation propagation between benchmark graphs with pronounced and fuzzy modules.** Two times 7 randomly selected scale-free, modular benchmark graphs [28] were generated as described in **Supplementary Methods** and **Table S7** of **Text S1** with ratios of inter-modular edges of 0.05 (~300 of ~6,000 edges were inter-modular) and 0.4 (~2400 of ~6,000 edges were inter-modular), termed as "pronounced modules" and "fuzzy modules", respectively. Panel A: average fill times and silencing times, separately for the "fuzzy" and the "pronounced" group of networks. Fill times and silencing times were determined as described in **Methods**. Continuous perturbation intensity for fill time was 10,000 units, while initial perturbation intensities for silencing times were 10,000 or 1,000,000 units at low intensity or at high intensity perturbations, respectively. The three asterisk signs mark statistically significant differences with α=0.001. Dark red bars and light blue bars represent pronounced modules and fuzzy modules, respectively. Bar letter codes refer to Panels showing snapshots of perturbations with identical conditions. Panels B through E show image snapshots created by the Turbine viewer after 50 time-steps of the simulation, using a heat-based color map. (The order of colors marking the lowest to highest perturbation is: black → red → orange → yellow → white). Perturbation values were scaled logarithmically. Panels B and C show the effect of low intensity starting perturbations (S=10,000), while Panels D and E show the effect of high intensity starting perturbations (S=1,000,000). Panels B and D show benchmark graphs with pronounced modules, while Panels C and E show benchmark graphs with fuzzy modules.

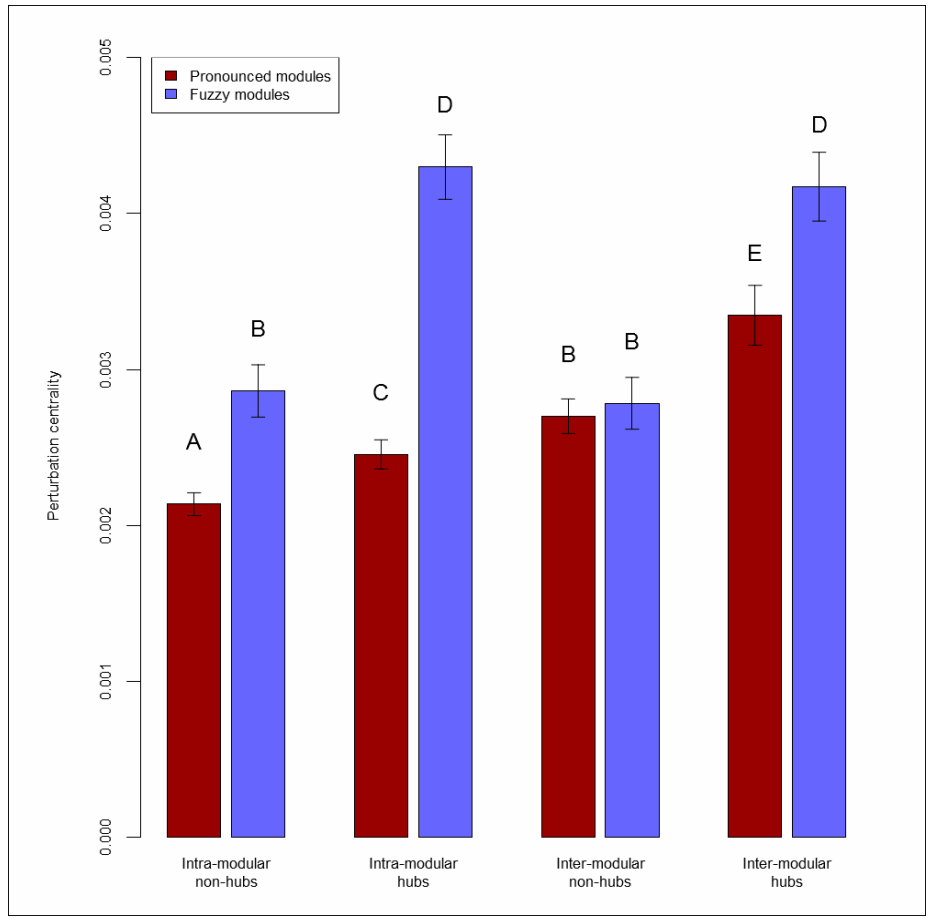

**Figure 2. Average perturbation centralities for different node types in benchmark graphs.** Scale-free, modular benchmark graphs [28] were created as described in **Supplementary Methods** and **Table S7** of **Text S1**. Average perturbation centralities were calculated as described in **Methods** using a starting perturbation of 40,000 units, since the benchmark networks contained 4,000 nodes. 4 node types were discriminated: intra-modular non-hubs, inter-modular non-hubs, intra-modular hubs and inter-modular-hubs, where hubs were nodes having a degree in the top 10%, and inter-modular nodes were nodes with more than 40% inter-modular edges. Different letters on top of the bars mark significantly different groups with α=0.01 (Wilcoxon rank-sum test). Dark red bars show results obtained using 7 randomly selected benchmark graphs with the ratio of inter-modular nodes set to 0.05, termed as pronounced modules, while light blue bars display data for 7 randomly selected benchmark graphs (with the same seed nodes as the ones used for pronounced modules) with ratio of inter-modular nodes set to 0.4, termed fuzzy modules.

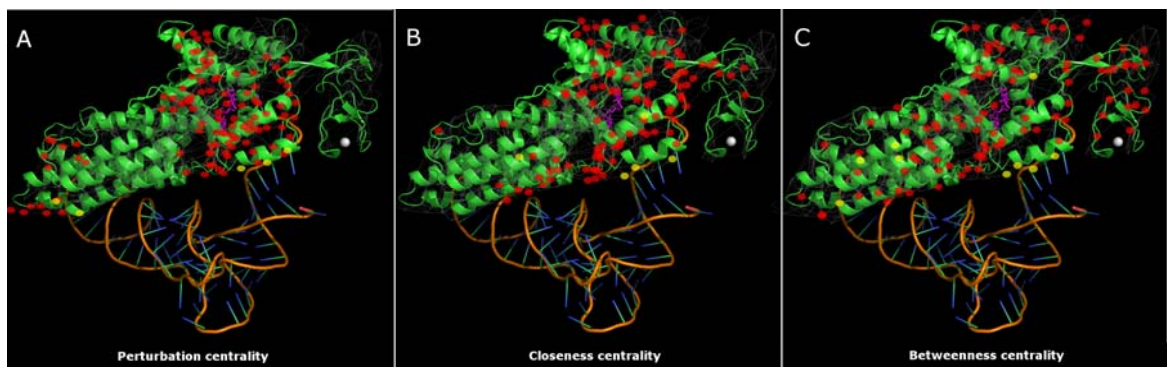

**Figure 3. Substrate binding-induced perturbation centrality changes mark important residues of *E. coli* Met-tRNA synthetase.** Protein structure networks of the substrate-free and substrate-bound forms of *E. coli* Met-tRNA synthetase protein were generated as described in the **Supplementary Methods** of **Text S1**. Perturbation centralities and the underlying protein structure network of Met-tRNA synthetase were calculated and visualized by the Turbine program as described in **Methods**, and were overlaid on the 3D image of the substrate-bound form of the protein (and its tRNA$^{Met}$ complex) generated with PyMOL [57] using ray-tracing. The bottoms of the images show the structure of tRNA$^{Met}$. The purple molecule in the middle of the protein structure is the substrate Met-AMP marking the active site of the enzyme, the white sphere on the right is the Zn$^{2+}$ ion. Red signs of Panels A, B and C mark amino acids having the highest *increase* of perturbation, closeness and betweenness centralities (top 20%) of the substrate-bound form compared to the substrate-free form, respectively. Yellow signs mark those contact amino acids, which are directly bound to the tRNA$^{Met}$, evidenced by an atomic distance of less than 4.5Å between any atom of the residue and the tRNA$^{Met}$, excluding hydrogens. To avoid overcrowding the image, only those contact amino acids are shown, which have a high increase of their centrality. A separate image showing all tRNA$^{Met}$-binding amino acids is shown in **Figure S9 of Text S1**. Note that red-labeled amino acids having the largest increase of perturbation centrality upon substrate binding (Panel A) are clustered around the active site and around both tRNA-binding sites, thus successfully discriminate all important parts of the protein. Amino acids showing the highest change in closeness centrality (Panel B) are smeared around the active site (which also occurs to be near the geometric center of the protein). Amino acids showing the highest change in betweenness centrality (Panel C) are scattered all around the protein.

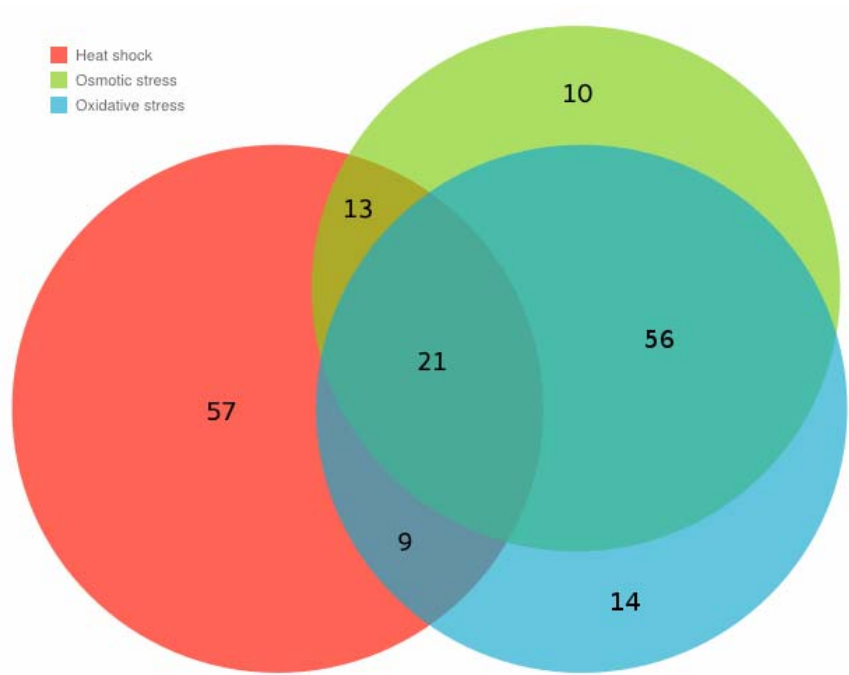

**Figure 4. Visualization of the difference among the three top 100 sets of proteins having the highest perturbation centrality in the DIP (2005) yeast interactome.**
Perturbation centralities were calculated for three stressed variations of the DIP (2005) yeast interactome according to **Methods**. The properties of the network as well as the method of generating its stressed versions are described in the **Supplementary Methods**. The sizes of the different areas of the diagram are roughly proportional to the number of proteins in the respective combination of the three sets. Numbers also show the number of proteins in different sets. This quantitative Venn diagram was generated using the Google Charts API.
(https://developers.google.com/chart/image/docs/gallery/venn_charts).
The red, green and blue circles show the sets of top 100 proteins having the highest perturbation centrality in the heat-shocked, osmotically- stressed and oxidatively-stressed networks, respectively. This figure illustrates the fact mentioned in the Section "Various stress types induce different perturbation dissipating regions of the yeast interactome" that the most important proteins in heat shock are substantially different from the most important proteins in the other two tested stress types (i.e. in osmotic and oxidative stresses).

**Table 1. Correlation between perturbation centrality and other centrality measures**

| Networks[a] | Closeness centrality | Betweenness centrality | Community centrality[b] | PageRank[c] | Degree | Weighted degree |
|---|---|---|---|---|---|---|
| Benchmark graphs with pronounced modules | **0.79** | 0.31 | 0.30 | *0.08* | *0.26* | *0.26* |
| Benchmark graphs with fuzzy modules | **0.79** | **0.76** | **0.83** | 0.67 | **0.83** | **0.83** |
| Substrate-free Met-tRNA synthetase protein structure network | **0.86** | 0.44 | 0.26 | 0.16 | 0.38 | 0.34 |
| Substrate-bound Met-tRNA synthetase protein structure network | **0.87** | 0.44 | 0.25 | 0.18 | 0.41 | 0.37 |
| Filtered Yeast Interactome | *0.09* | 0.33 | **0.80** | 0.47 | 0.67 | **0.85** |
| Database of Interacting Proteins yeast interactome (release 2005) | 0.62 | 0.56 | **0.84** | **0.73** | 0.66 | **0.72** |
| Database of Interacting Proteins yeast interactome (release 2010) | 0.67 | 0.41 | 0.63 | 0.47 | 0.52 | 0.65 |
| *E. coli* metabolic network | **0.72** | 0.31 | **0.97** | 0.67 | 0.59 | **0.99** |
| *B. aphidicola* metabolic network | **0.70** | 0.40 | **0.98** | **0.78** | **0.72** | **0.99** |
| School-friendship network | 0.68 | 0.43 | 0.69 | 0.58 | 0.68 | **0.71** |
| **Mean and standard error** | **0.67 (0.063)** | **0.44 (0.043)** | **0.65 (0.090)** | **0.48 (0.081)** | **0.57 (0.056)** | **0.69 (0.087)** |

Perturbation centrality was compared to other centrality measures calculated as described in **Supplementary Methods** of **Text S1**. Spearman correlations above r=0.7 are marked with bold letters, correlations below r=0.3 are marked with italics. Highest correlations were observed between perturbation centrality *versus* closeness centrality, community centrality [29] and weighted degree. This underlines the observations that besides geodesic distance (closeness centrality), modular position and degree also contribute to good perturbation properties. Note that measured correlations between perturbation and closeness centralities are much weaker than the correlations between the reciprocal of fill time and closeness centrality (mean is 0.895 in **Table S1** compared to 0.67 here, p=0.000487, Wilcoxon rank-sum test; correlations with closeness centrality failed the Shapiro normality test with p=0.0019)

[a]Network descriptions are given in **Supplementary Methods** and **Table S7** of **Text S1**.
[b]Community centrality was calculated using the LinkLand community detection method of the ModuLand family as described by Kovács *et al*. [29].
[c]PageRank values were calculated using the algorithm of the igraph library [58].

# Supplementary Information for

# Perturbation centrality and Turbine: a novel centrality measure obtained using a versatile network dynamics tool

Kristóf Z. Szalay and Peter Csermely[1]
*Department of Medical Chemistry, Semmelweis University, Budapest, Hungary*

**See the supporting website for further information: http://turbine.linkgroup.hu**

# Table of contents

---

[1] Corresponding author: Csermely, P. (csermely.peter@med.semmelweis-univ.hu)

# Supplementary Figures

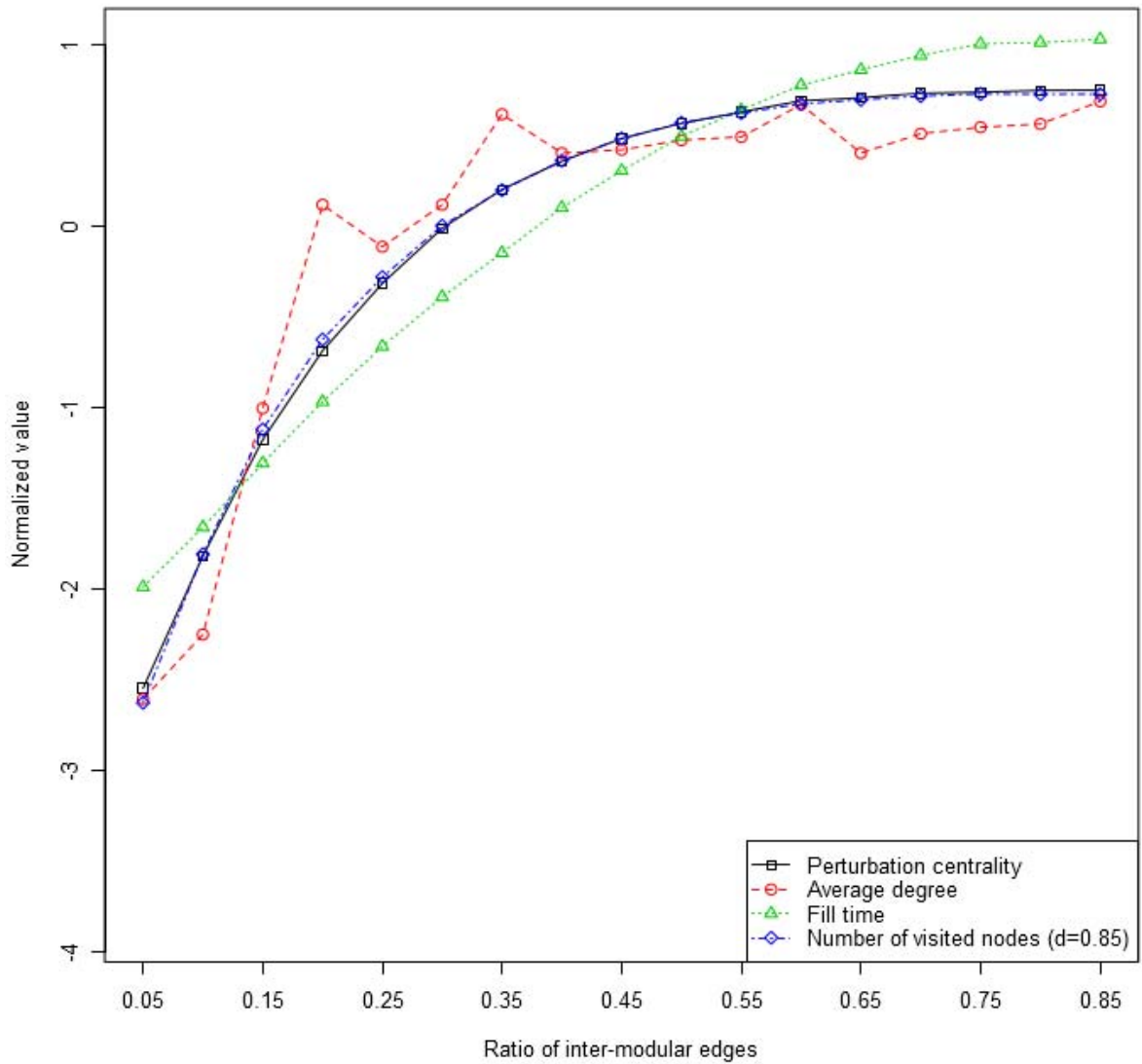

**Figure S1. Average perturbation centrality and "number of visited nodes" values plotted against the ratio of inter-modular edges.**

Scale-free, modular benchmark graphs [1] were generated as described in Supplementary Methods. Average perturbation centrality (black squares), average degree (red circles), the reciprocal of average fill time (green triangles, added perturbation: 10,000 units per step) and average "number of visited nodes" (blue diamonds) for a damping value of 0.85 were calculated from 3 randomly generated benchmark graphs with ratios of inter-modular edges ranging from 0.05 to 0.85 with steps of 0.05 as described in Methods of the main text for Turbine (with the change that a 95% threshold was used for the fill time – that is, 95% of the network had to have an energy value larger than 1 – since the benchmark networks were much more homogeneous than real-world networks) and in the Supplementary Methods for ITM-Probe [2]. Values were normalized using the scale function of the R package [3]. Note that the Spearman correlation between the various values was more than 0.95, except for the average degree.

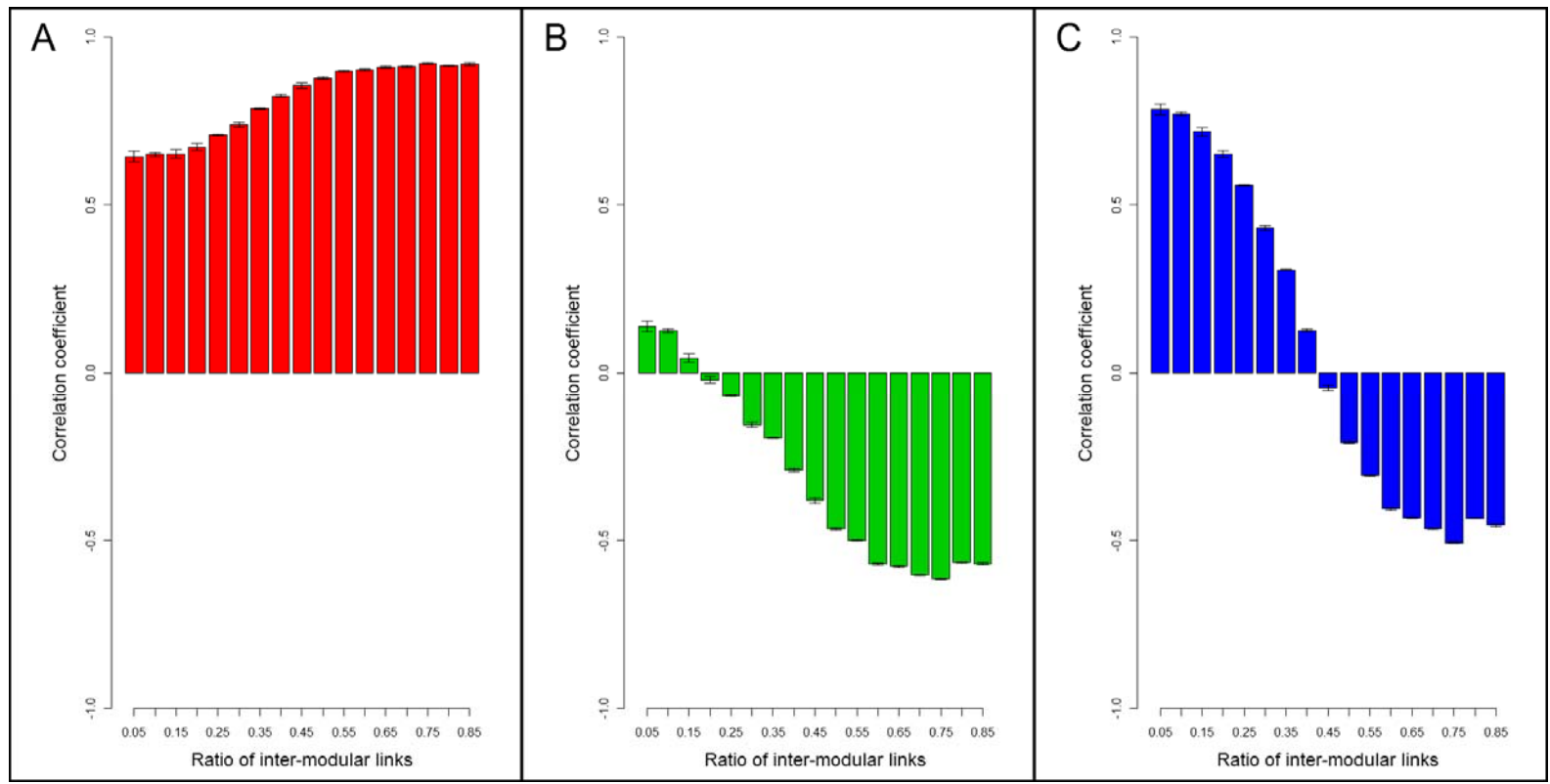

**Figure S2. Correlation of perturbation centrality, "number of visited nodes" and node degree.**
Scale-free, modular benchmark graphs [1] were generated as described in Supplementary Methods. Perturbation centrality and "number of visited nodes" measures were calculated from 3 sets of randomly generated benchmark graphs with ratios of inter-modular edges ranging from 0.05 to 0.85 with steps of 0.05 as described in Methods of the main text and in Supplementary Methods for ITM-Probe [2], respectively. Spearman correlations were calculated using the R package [3]. Panel A shows correlation of perturbation centrality versus node degree, Panel B shows the correlation of number of visited nodes versus the node degree, and Panel C shows the correlation of perturbation centrality and number of visited nodes. The data in Panel A reinforces the observation in the main text that the degree becomes more important in the determination of the silencing time as the modules become more and more fuzzy and overlapping. Interestingly, results from ITM-Probe behave in an exactly opposite way: as the communities become more overlapping, the number of visited nodes measure quickly becomes negatively correlated with the degree (possibly because random walks can "turn back"). These two effects taken together resulted in a large correlation between the perturbation centrality and number of visited nodes when there were pronounced modules, and a negative correlation when the modules became fuzzier, which is shown by the data in Panel C.

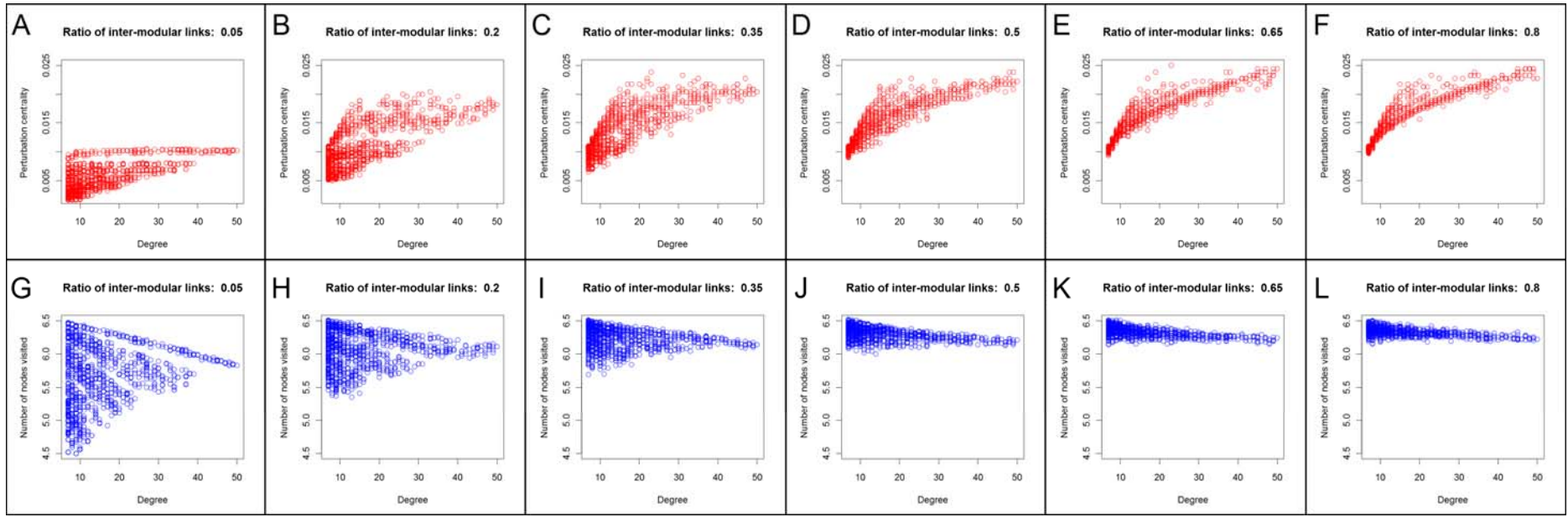

**Figure S1. Comparison of perturbation centrality with the "number of visited nodes" measure of ITM-Probe as a function of node degree with different ratios of inter-modular edges of benchmark graphs.**

Scale-free benchmark graphs [1] with overlapping modules were generated as described in Supplementary Methods. Perturbation centrality (red, Panels A through F) and "number of visited nodes" measures (blue, Panels G through L) were calculated as described in Methods of the main text and in Supplementary Methods for ITM-Probe [2], respectively. For the generation of the benchmark graphs with ratios of inter-modular edges 0.05, 0,2, 0.35, 0.5, 0.65 and 0.80 appearing on Panels A/G, B/H, C/I, D/J, E/K and F/L, respectively, a random seed of 87 was used. The results suggest that nodes in the networks with pronounced modules give similar results using Turbine and ITM-Probe (observe the same striping pattern showing the better perturbation propagation capability of nodes having inter-modular edges). On the contrary, in networks with fuzzy modules, the result is still correlated with the degree in Turbine, but ITM-Probe results do not seem to depend on the degree. These are the same results that can be obtained from Figure S2; this figure serves as an illustration of the possible underlying pattern behind the change in correlation: the number of nodes visited measure seems to have an upper saturation-like limit in ITM-Probe in networks with largely overlapping modules.

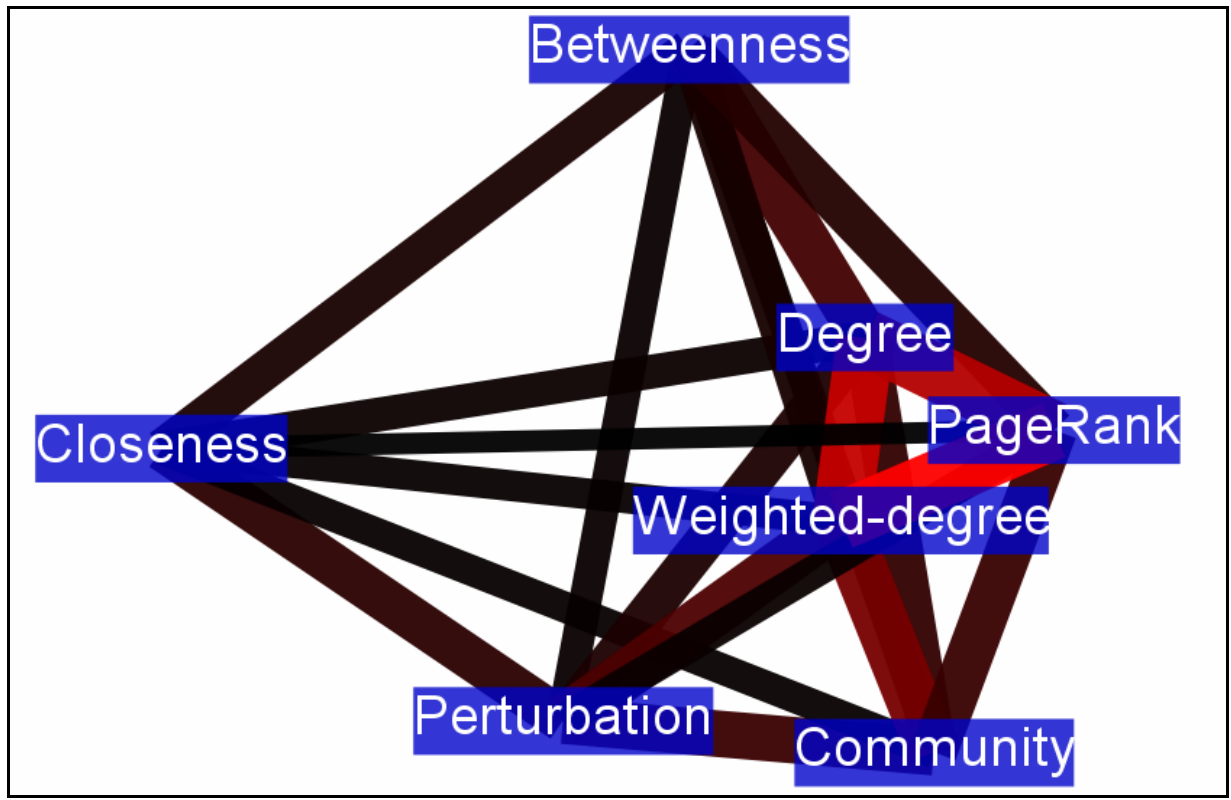

**Figure S4. A visual representation of the relation among different centrality measures.**

We calculated Spearman correlations between different centrality measures for the 10 benchmark and real-world networks shown in Table 2 of the main text. A 7-node graph was created from this data using the different centrality measures as nodes, and the average correlation between pairs of centralities as edge weights. The graph was thereafter laid out using the ForceAtlas 2 layout algorithm (which uses edge weights) of Gephi [4] with default settings except for the "Edge weight influence" option, which was set to 5.0. The layout generated this way can be a good approximation of the relations between different centralities, since the more correlated centrality measures are connected by more powerful "springs"; thereby their final position is closer to each other. It is visible on the figure that the perturbation centrality measure occupies a new position with largest correlations to closeness centrality, community centrality and weighted degree, just as the mean correlations of Table 2 in the main text show.

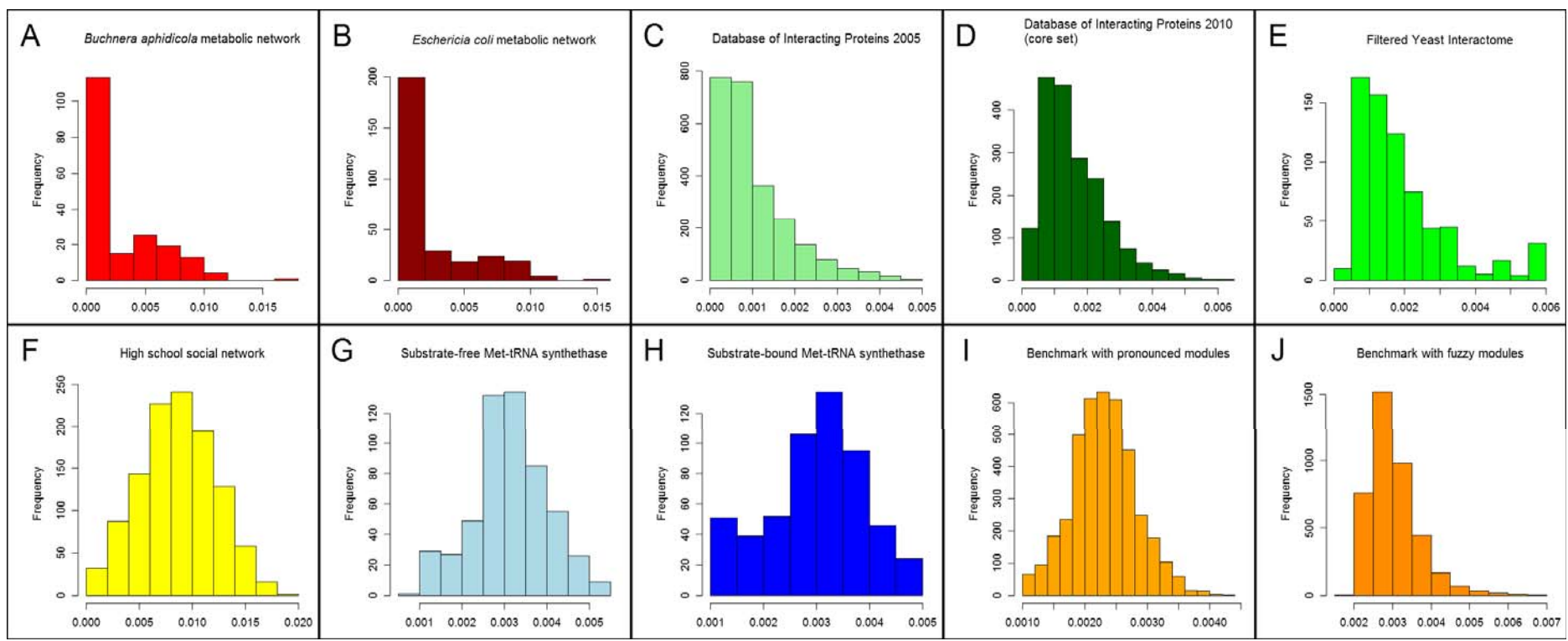

**Figure S5. Distribution of perturbation centralities in 10 benchmark and real-world networks.**
Histograms were generated using the same data set for 2 benchmark and 8 real-world networks that were used for **Table S1** and **S2**. Detailed descriptions of the networks are available in **Supplementary Methods**. Perturbation centralities were calculated according to **Methods** of the main text. Histograms from the perturbation centralities were generated using the "hist" command of R [3], with default settings. The tested social network (Panel F), the modular benchmark network [1] with pronounced modules (Panel I) and both conformations of Met-tRNA synthetase [5] (Panels G and H) had approximately normal distributions of perturbation centrality values. The Filtered Yeast Interactome [6] (Panel E) and the 2010 release of the Database of Interacting Proteins [7] (Panel D), as well as the modular benchmark network with fuzzy modules (Panel J) seemed to have approximately lognormal distributions for perturbation centrality. The histogram of the 2005 release of the Database of Interacting Proteins looked like a scale-free distribution (Panel C), and finally, the most skewed distributions were the perturbation centralities of the two metabolic networks [8,9], which looked like exponential distributions. However, all distributions failed the Shapiro-Wilk normality test ($p=5*10^{-16}$, $5*10^{-22}$, $10^{-44}$, $10^{-31}$, $3*10^{-26}$, 0.0003, 0.0094, $2*10^{-7}$, $3*10^{-9}$, $3*10^{-45}$, respectively), so the Wilcoxon rank-sum test had to be used for statistical significance analysis instead of a t-test.

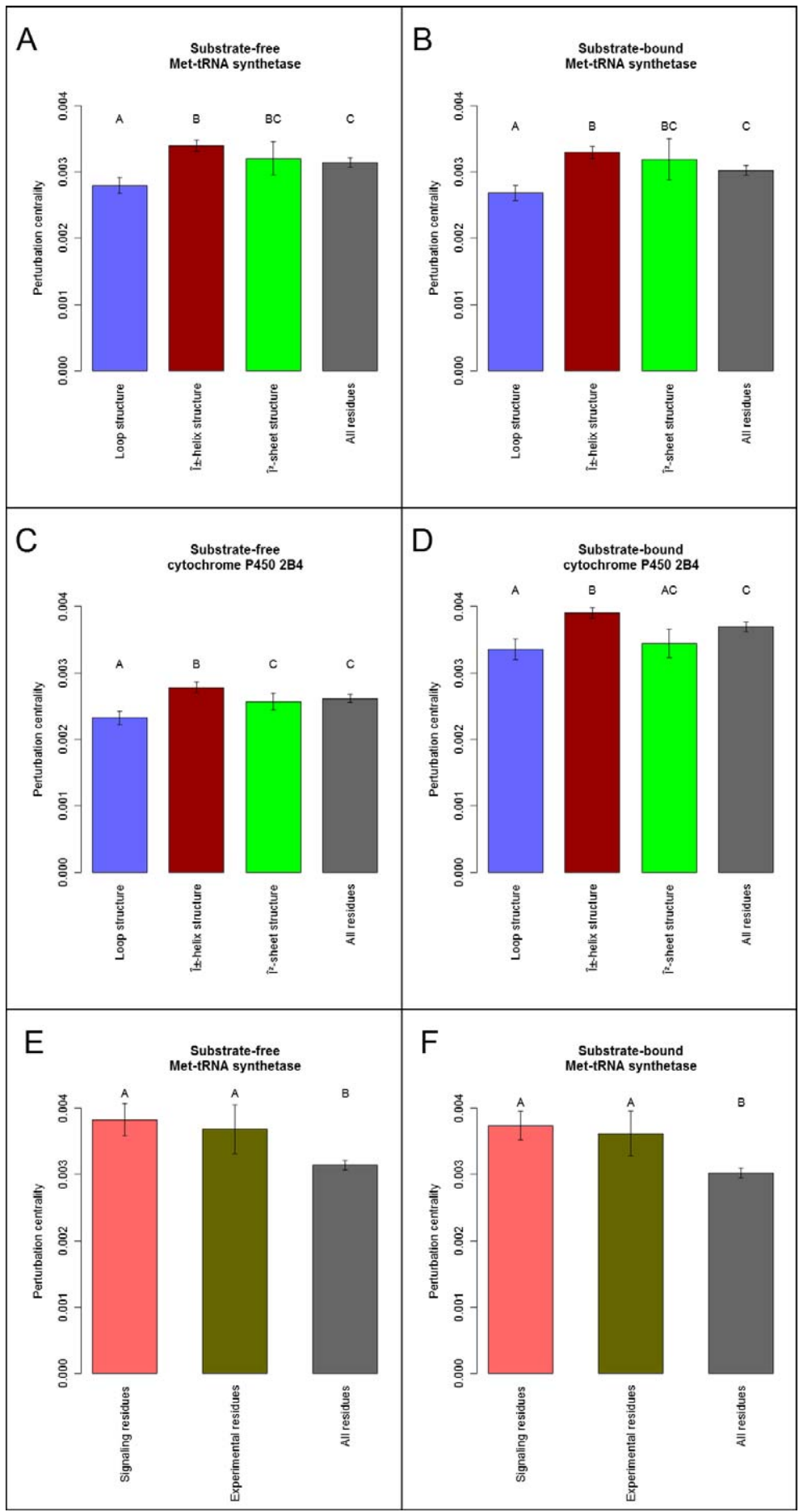

**Figure S6. Average perturbation centralities for different sets of residues in protein structure networks.**

Protein structure networks of the substrate-free and substrate-bound forms of the *E. coli* Met-tRNA synthetase and rabbit cytochrome P450 2B4 proteins were generated as described in **Methods** and **Supplementary Methods** of **Text S1**. Assignment of secondary structures for different amino acids was done by PyMOL. Error bars show standard error of the mean. Different letters on top of the bars mean significantly different groups ($\alpha=0.01$, Wilcoxon rank-sum test). Panels A and B show data calculated for the substrate-free and substrate (Met-AMP/tRNA$^{Met}$)-bound form of Met-tRNA synthetase, respectively. Panels C and D show data calculated for the substrate-free and substrate (imidazole)-bound form of cytochrome P450 2B4. In all cases amino acids of loops had significantly ($p=3.2*10^{-6}$, $9.5*10^{-6}$, $2.3*10^{-6}$, $0.0001$ for the free and bound conformations of Met-tRNA synthetase and cytochrome P450, respectively; Wilcoxon rank-sum test, $\alpha=0.00625$ adjusted with Bonferroni correction) lower perturbation centrality than average, while α-helices had significantly higher ($p=0.00023$, $0.00015$, $0.00083$, $0.0014$ for the free and bound conformations of Met-tRNA synthetase and cytochrome P450, respectively; Wilcoxon rank-sum test, $\alpha=0.00625$ adjusted with Bonferroni correction) than average perturbation centrality. Panels E and F show the average perturbation centralities of amino acids belonging to intra-protein communication pathways predicted by Ghosh and Vishveshwara [5] ("Signaling residues"), as well as amino acids with experimentally verified importance [5] ("Experimental residues"). Light blue, dark red and green bars show average perturbation centralities of amino acids in loop, α-helical and β-sheet structures, while gray bars show the global average perturbation centrality calculated for the whole protein. Pink and brown bars of Panels E and F show average perturbation centralities of the Signaling and Experimental residues, respectively. Note that both betweenness and closeness centralities were less successful than perturbation centrality in differentiating between the above amino acid sets (cf. the current Figure with **Figures S7** and **S8** of **Text S1**).

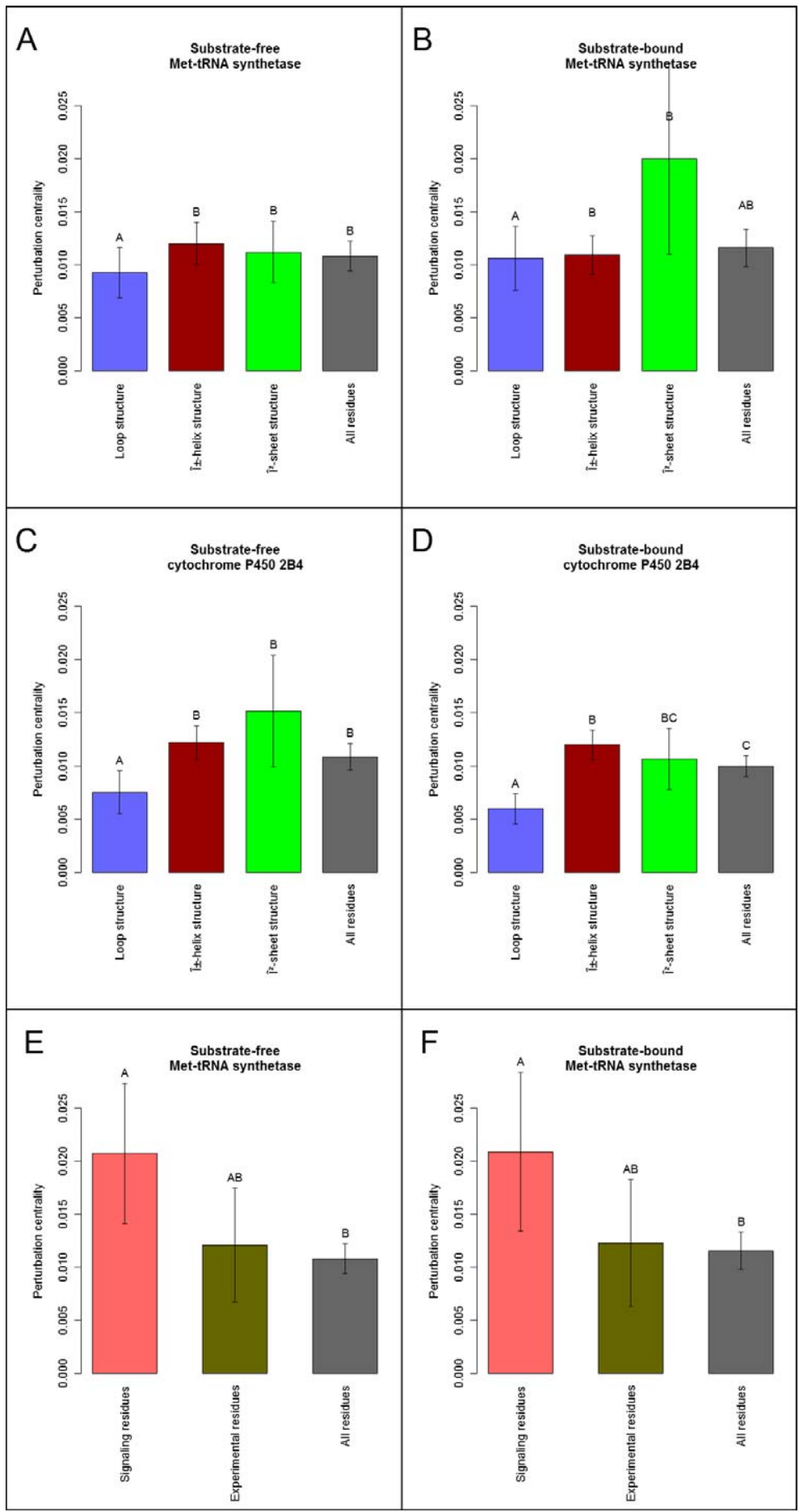

**Figure S7. Average betweenness centralities of different residue groups in Met-tRNA synthetase and cytochrome P450 enzymes.**

This figure is a direct parallel to **Figure S6** and **Figure S8**, here using *betweenness* centrality instead of perturbation and closeness centralities, respectively. Protein structure networks of the substrate-free and substrate-bound forms of the *E. coli* Met-tRNA synthetase and rabbit cytochrome P450 2B4 proteins were generated as described in **Methods** of the main text and **Supplementary Methods**. Assignment of secondary structures for different amino acids was performed by PyMOL. Error bars show standard error of the mean. Different letters mean significantly different groups ($\alpha=0.01$, Wilcoxon rank-sum test). Panels A and B show data calculated for the substrate-free and substrate-bound form of Met-tRNA synthetase, respectively. Panels C and D show data calculated for the substrate-free and imidazole-bound forms of cytochrome P450 2B4. Panels E and F show the average perturbation centralities of predicted communication pathways as described by Ghosh and Vishveshwara [5] ("Signaling residues") and other residues with experimentally verified importance [5] ("Experimental residues"). Light blue, dark red and green bars on Panels A through D show average perturbation centralities for residues in loops, α-helices and β-sheets, while gray bars show the global average perturbation centrality calculated for the whole protein. Pink bars on Panels E and F show average perturbation centralities of the Signaling residues, and brown bars show the means of the Experimental residues of Met-tRNA synthetase. Betweenness centrality returned by far the largest deviations of the three tested centralities (i.e. closeness, betweenness and perturbation centralities). Loops still had significantly ($p=0.0016$, $0.011$, $4.3*10^{-5}$, $6.8*10^{-7}$ for the free and bound conformations of Met-tRNA synthetase and cytochrome P450, respectively; Wilcoxon rank-sum test, $\alpha=0.00625$ adjusted with Bonferroni correction) lower mean centrality values than the global average in all networks (but the substrate-bound Met-tRNA synthetase). Using betweenness centrality α-helices can no longer be distinguished from the global mean (except for the substrate-bound form of cytochrome P450). Signaling residues could still be distinguished from the global mean (Panels E and F), but the differences in centralities for the Experimental residues were no longer significant.

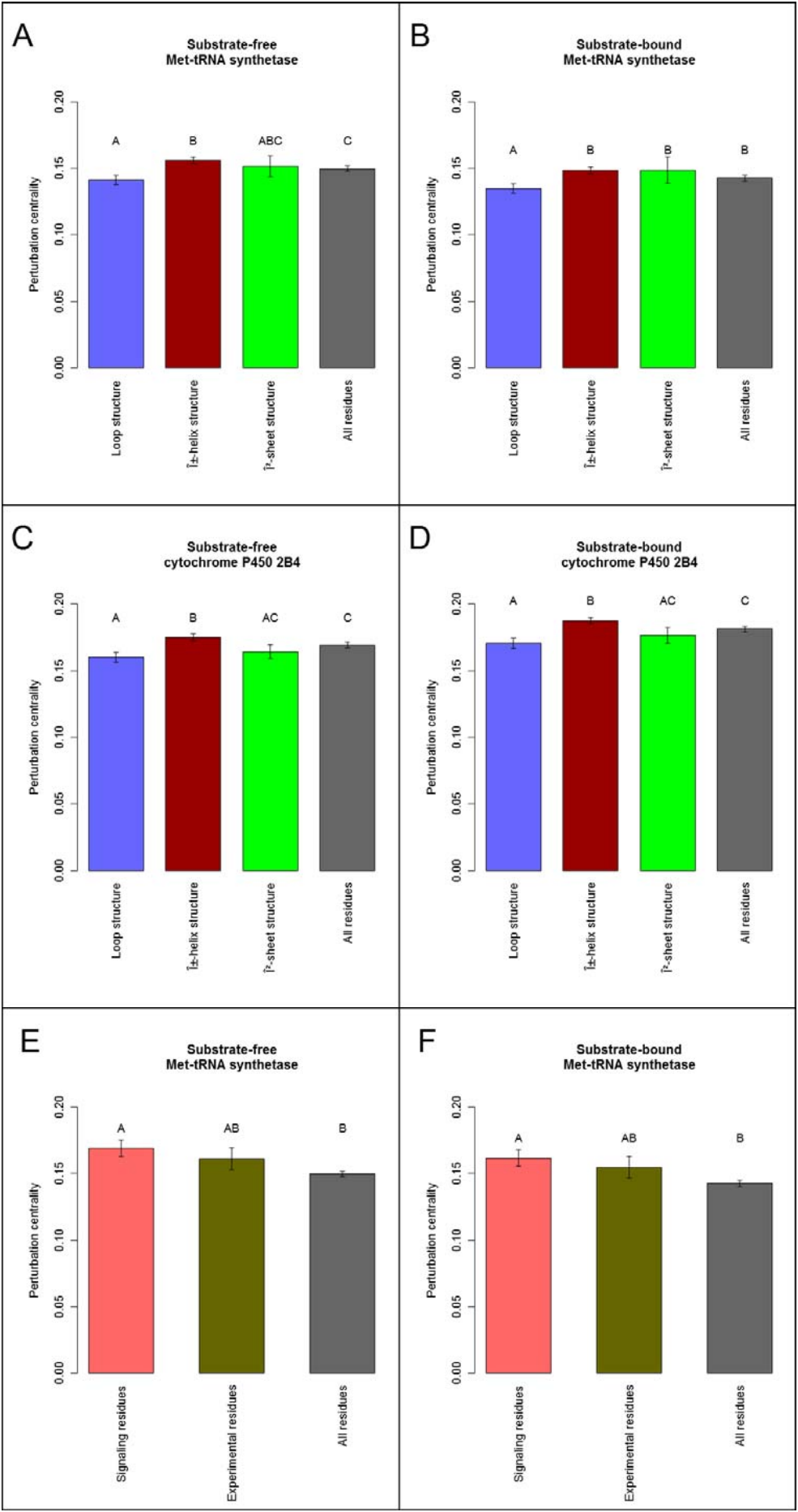

**Figure S8. Average closeness centralities of different residue groups in Met-tRNA synthetase and cytochrome P450 enzymes**.

This figure is a direct parallel to **Figure S6** and **Figure S7**, here using *closeness* centrality instead of perturbation or betweenness centralities, respectively. Protein structure networks of the substrate-free and substrate-bound forms of the *E. coli* Met-tRNA synthetase and rabbit cytochrome P450 2B4 proteins were generated as described in **Methods** of the main text and **Supplementary Methods**. Assignment of secondary structures for different amino acids was performed by PyMOL. Error bars show standard error of the mean. Different letters mean significantly different groups ($\alpha=0.01$, Wilcoxon rank-sum test). Panels A and B show data calculated for the substrate-free and substrate-bound form of Met-tRNA synthetase, respectively. Panels C and D show data calculated for the substrate-free and imidazole-bound form of cytochrome P450 2B4. Panels E and F show the average perturbation centralities of predicted communication pathways as described by Ghosh and Vishveshwara [5] ("Signaling residues") and other residues with experimentally verified importance [5] ("Experimental residues"). Light blue, dark red and green bars on Panels A through D show average perturbation centralities for residues in loops, $\alpha$-helices and $\beta$-sheets, while gray bars show the global average perturbation centrality calculated for the whole protein. Pink bars on Panels E and F show average perturbation centralities of the Signaling residues, and brown bars show the means of the Experimental residues of Met-tRNA synthetase. Closeness centrality returned smaller deviations than perturbation centrality. Interestingly, the distinction power of closeness centrality was slightly lower in Met-tRNA synthetase, and exactly the same in cytochrome P450 as the distinction power of perturbation centrality. In particular, the Experimental residues could no longer be distinguished from the global mean (Panels E and F, $p=0.034$, $p=0.04$, respectively; 0.007 and 0.006 with perturbation centrality), $\alpha$-helices did not have significantly higher mean closeness centrality than the global average in the substrate-bound Met-tRNA synthetase (Panel B, $p=0.014$ vs. 0.00015 with perturbation centrality), and $\beta$-sheets were no longer distinguishable from loops in the substrate-free Met-tRNA synthetase (Panel A, $p=0.035$ vs. 0.0034 with perturbation centrality). P-values were calculated using the Wilcoxon rank-sum test, $\alpha=0.0125$ adjusted with Bonferroni correction for a FWER of 0.05.

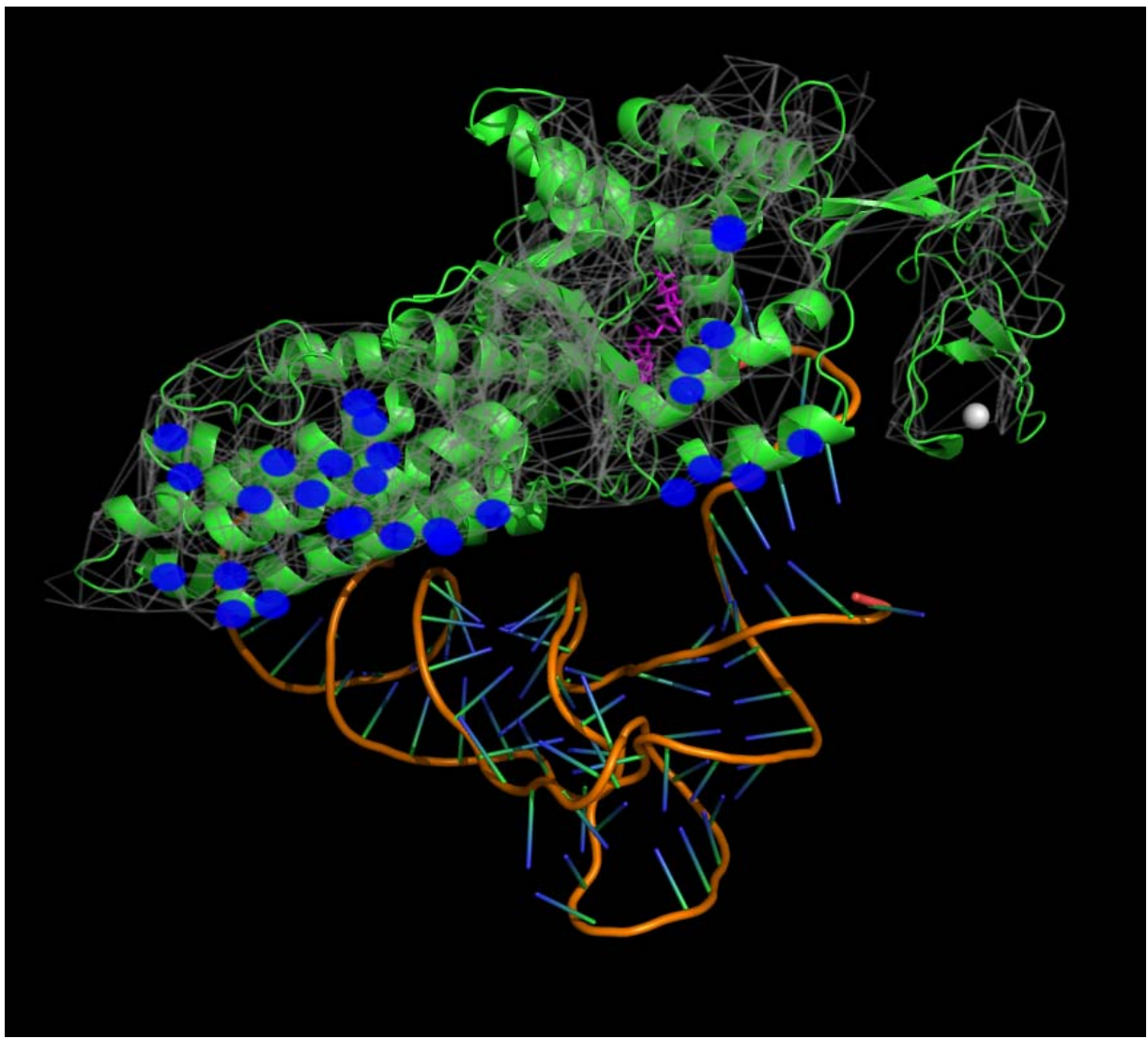

**Figure S9. Amino acids of Met-tRNA synthetase directly bound to tRNA$^{Met}$.**
The underlying protein structure network of Met-tRNA synthetase was calculated and visualized by the Turbine program as described in **Methods**, and was overlaid on the 3D image of the substrate-bound form of the protein (and its tRNA$^{Met}$ complex) generated with PyMOL [10] using ray-tracing. The bottom of the image shows the structure of tRNA$^{Met}$. The purple molecule in the middle of the protein structure is the substrate Met-AMP marking the active site of the enzyme, the white sphere on the right is the Zn$^{2+}$ ion. Blue circles mark those amino acids, which are directly bound to the tRNA$^{Met}$, evidenced by an atomic distance of less than 4.5Å between any atom of the residue and the tRNA$^{Met}$, excluding hydrogen atoms.

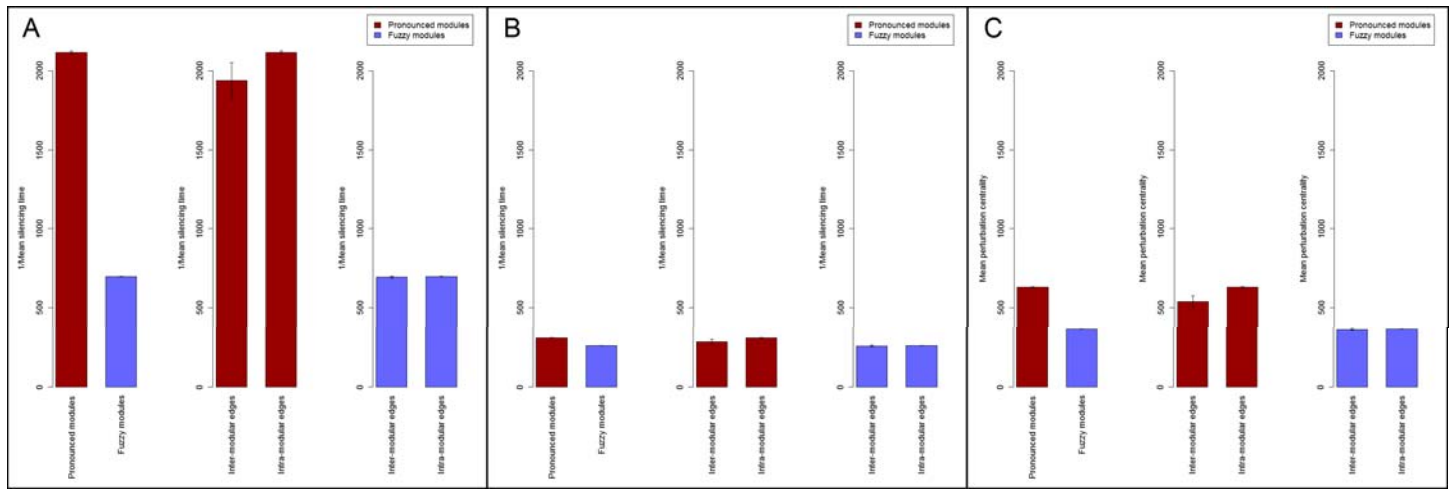

**Figure S10. Results for edgetic perturbations.**

We also tested the effects of edge-based perturbations as their importance have previously been stated in the literature [11]. Silencing time for an edge was calculated by propagating a given amount of energy (10,000, 40,000 and 1,000,000 units) from both end-nodes of an edge, simultaneously. We calculated silencing times for all edges for two modular benchmark networks [1], one with pronounced modules (only 5% of the links were inter-modular), and one with fuzzy modules (40% inter-modular links), both generated with the random seed 10. Silencing times were measured with a dissipation value of 1, and silencing threshold was also set to 1. Further definition of the silencing time is available in Methods of the main text. A link was termed inter-modular, if it was connecting two different communities. Dark red bars show data obtained on the network with pronounced modules, while light blue bars display data from the network with fuzzy modules. We could observe the very same effects for edge-based perturbations that we have observed for node-based perturbations. Panel A shows the mean perturbation centralities calculated for a large starting perturbation (1,000,000 units on both endpoints of an edge). There was a major (p=0, Wilcoxon rank-sum test) difference between the mean silencing times between fuzzy and pronounced modules in this case, and inter-modular edges were significantly (p=0.0094, α=0.025 with Bonferroni correction, Wilcoxon rank-sum test) better spreaders of perturbation in the network with pronounced modules. The disparity between the edgetic perturbations of networks with fuzzy versus pronounced modules was nearly eliminated, if a starting perturbation of 10,000 units was applied (Panel B). These effects were the same as what we have demonstrated with node-based perturbations on Figures 1 and 2 of the main text. Panel C shows silencing times calculated for 40,000 units of starting energy corresponding to the definition of perturbation centrality measure. (40,000 units of starting energy was used, since the benchmark networks contained 4000 nodes, and perturbation centrality was defined as the reciprocal of the silencing time resulting from applying a perturbation of $10*n$ units of energy, where $n$ is the number of nodes in the network, see **Methods** of the main text.) Data of Panel C verifies that the choice of $n*10$ units as a starting perturbation is a nice compromise between weighing the nodes' mesoscopic position, that is, the modular location (which can be detected using a large starting perturbation) and their local position, that is, the weighted degree of the node and its near neighbors (which can be detected using a small starting perturbation).

# Supplementary Tables

## Table S1. Correlation between the reciprocal of fill time and closeness centrality

| Networks[a] | Correlation[b] | Average fill time[c] |
|---|---|---|
| Benchmark graphs with fuzzy modules | 0.974 | 52 |
| Benchmark graphs with pronounced modules | 0.930 | 151 |
| Substrate-free Met-tRNA synthetase protein structure network | 0.957 | 76 |
| Substrate-bound Met-tRNA synthetase protein structure network | 0.941 | 94 |
| Filtered Yeast Interactome | 0.856 | 1344 |
| Database of Interacting Proteins yeast interactome (release 2005) | 0.857 | 215 |
| Database of Interacting Proteins yeast interactome (release 2010) | 0.941 | 168 |
| *B. aphidicola* metabolic network | 0.824 | 1225 |
| *E. coli* metabolic network | 0.746 | 1928 |
| School friendship network | 0.921 | 28 |
| **Mean and standard error** | **0.895 (0.023)** | **528 (219.9)** |

[a]Network descriptions are given in **Supplementary Methods** of **Text S1**.
[b]Correlation data show the Spearman's rho correlation of the reciprocal of the fill time versus the standard closeness centrality measure. Correlations between the reciprocal of fill time and closeness centrality are much stronger than those between perturbation centrality and closeness centrality (mean is 0.895 compared to 0.67 in **Table 1**, p=0.000487, Wilcoxon rank-sum test (correlations with closeness centrality in **Table 1** failed the Shapiro normality test with p=0.0019).
[c]Fill time was calculated for each node in each network by applying a perturbation of size 10,000 to the selected node in each time step until more than 80% of the network nodes had an energy level larger than 1. Dissipation was set to 0.

**Table S2. Correlation of silencing times calculated for three different-sized starting perturbations in 8 real-world and two benchmark networks**

| Network | Correlation between low- and medium-intensity perturbations | Correlation between medium- and high-intensity perturbations | Correlation between low- and high-intensity perturbations |
|---|---|---|---|
| Benchmark graph with pronounced modules | 0.41 | **0.90** | 0.22 |
| Benchmark graph with fuzzy modules | **0.94** | **0.95** | **0.80** |
| Substrate-free Met-tRNA synthetase protein structure network | **0.84** | 0.78 | 0.52 |
| Substrate-bound Met-tRNA synthetase protein structure network | **0.84** | 0.78 | 0.56 |
| Filtered Yeast Interactome | **0.88** | **0.81** | 0.57 |
| Database of Interacting Proteins yeast interactome (release 2005) | **0.94** | **0.93** | **0.80** |
| Database of Interacting Proteins yeast interactome (release 2010) | **0.85** | **0.89** | 0.63 |
| *E. coli* metabolic network | **0.99** | **0.97** | **0.96** |
| *B. aphidicola* metabolic network | **0.99** | **0.96** | **0.95** |
| School-friendship network | **0.87** | **0.81** | 0.66 |
| **Mean and standard error** | **0.853 (0.053)** | **0.877 (0.024)** | **0.668 (0.070)** |

Parameters of the different networks are available in the **Methods** section of the main article. Only the two benchmark graphs were model networks, generated using the benchmark graph generator tool of Lancichinetti and Fortunato [1]. The other graphs originated from different real-world scenarios. Low-intensity perturbation corresponds to an $n$ unit large Dirac-delta starting perturbation, medium-intensity perturbation corresponds to $10*n$ units, and high-intensity means that a $100*n$ unit-sized starting perturbation was applied to a single node when calculating its silencing time, $n$ being the number of nodes in the network. Silencing time was calculated for each node in the network as described in the **Methods** section of the main article. The columns show Spearman's correlation between silencing times calculated for nodes in the same network with different-sized starting perturbations. Correlations above 0.8 were marked with bold letters. It can be noticed that there are no abrupt changes in the importance in the perturbation dissipation capability of different nodes as the perturbation grows larger except for the benchmark network with disjunct modules, which means that real-world networks may display behavior closer to the benchmark graph with fuzzy modules. The table also underlines the choice of choosing $n*10$ as the size of the perturbation when calculating perturbation centrality, since it already seems to display module entrapment effects evidenced by the high correlation observed between medium- and high-intensity perturbations in the benchmark network with disjunct modules compared to the substantially lower correlation observed in the same network between low- and medium-intensity perturbations.

**Table S3. Statistically significantly enriched terms in the top 100 protein set of the DIP yeast interactome (release 2005)** [7] **containing proteins with *largest absolute* perturbation centrality in differently stressed cases.**

| A. Significantly enriched terms in the top 100 set of largest perturbation centrality in the *unstressed* DIP (2005) interactome | | |
|---|---|---|
| P-value | Term | Proteins |
| 3.98e-06 | non-homologous end-joining | YGL090W, YOR005C, YMR224C, YCR014C, YMR284W |
| 0.000139 | cell cycle | YOR026W, YGR188C, YOR353C, YOR005C, YKR031C, YBR057C, YJR140C, YMR224C, YER018C, YEL061C, YLR254C, YHR184W, YIL132C, YOR368W, YLR045C, YLR190W, YPL253C, YER147C, YGL075C, YGL249W, YJL090C, YGL251C, YLR383W, YCR063W, YML049C, YDR253C, YHL024W, YMR198W, YJL006C |
| 0.00189 | cellular response to stimulus | YLR240W, YOR353C, YGL090W, YOR005C, YBR077C, YKR031C, YMR224C, YBR128C, YGL155W, YCR027C, YCR014C, YIL132C, YLR007W, YOR368W, YBR020W, YML112W, YNL145W, YHR134W, YOL043C, YER147C, YIL128W, YJL090C, YHR079C, YLR383W, YMR284W, YOR120W, YLR094C, YJL006C, YPL046C, YGL220W, YDR098C, YDL138W |
| 0.00484 | M phase | YOR026W, YGR188C, YKR031C, YBR057C, YMR224C, YER018C, YEL061C, YLR254C, YHR184W, YIL132C, YOR368W, YLR045C, YPL253C, YER147C, YGL075C, YGL249W, YGL251C, YHL024W, YMR198W |
| 0.00637 | cell cycle phase | YOR026W, YGR188C, YKR031C, YBR057C, YJR140C, YMR224C, YER018C, YEL061C, YLR254C, YHR184W, YIL132C, YOR368W, YLR045C, YPL253C, YER147C, YGL075C, YGL249W, YJL090C, YGL251C, YHL024W, YMR198W |
| 0.00886 | response to DNA damage stimulus | YGL090W, YOR005C, YMR224C, YCR014C, YIL132C, YLR007W, YOR368W, YML112W, YHR134W, YOL043C, YER147C, YIL128W, YJL090C, YLR383W, YMR284W, YJL006C, YPL046C |
| 0.0165 | cell cycle process | YOR026W, YGR188C, YKR031C, YBR057C, YJR140C, YMR224C, YER018C, YEL061C, YLR254C, YHR184W, YIL132C, YOR368W, YLR045C, YPL253C, YER147C, YGL075C, YGL249W, YJL090C, YGL251C, YLR383W, YCR063W, YHL024W, YMR198W |
| 0.0212 | double-strand break repair | YGL090W, YOR005C, YMR224C, YCR014C, YOR368W, YER147C, YJL090C, YLR383W, YMR284W |
| 0.023 | DNA metabolic process | YGL090W, YOR005C, YBR057C, YJR140C, YMR224C, YCR014C, YIL132C, YLR007W, YOR368W, YHR134W, YOL043C, YER147C, YGL249W, YIL128W, YJL090C, YGL251C, YLR383W, YMR284W, YLR010C, YDR082W, YHL024W, YPL046C |
| 0.0256 | DNA repair | YGL090W, YOR005C, YMR224C, YCR014C, YIL132C, YLR007W, YOR368W, YHR134W, YOL043C, YER147C, YIL128W, YJL090C, YLR383W, YMR284W, YPL046C |
| 0.048 | double-strand break repair via non-homologous end joining | YGL090W, YOR005C, YMR224C, YCR014C, YMR284W |
| B. Significantly enriched terms in the top 100 set of largest perturbation centrality in the *heat-shocked* DIP (2005) interactome | | |
| P-value | Term | Proteins |
| 2.6e-05 | condensed chromosome | YOR026W, YGR188C, YBR156C, YER018C, YHR014W, YIL072W, YEL061C, YOL034W, YLR007W, YOR368W, YLR045C, YPL194W, YHR079C-A |
| 0.000126 | double-strand break repair via nonhomologous end joining | YGL090W, YOR005C, YMR224C, YHR056C, YCR014C, YJL092W, YDR369C |
| 0.000414 | double-strand break repair | YGL090W, YOR005C, YMR224C, YHR056C, YCR014C, YDR004W, YJL092W, YOR368W, YHR079C-A, YDR369C, YPR135W |
| 0.000638 | non-recombinational repair | YGL090W, YOR005C, YMR224C, YHR056C, YCR014C, YJL092W, YDR369C |
| 0.00304 | condensed nuclear chromosome | YOR026W, YGR188C, YER018C, YHR014W, YIL072W, YEL061C, YOR368W, YLR045C, YPL194W, YHR079C-A |

| P-value | Term | Proteins |
|---|---|---|
| 0.0063 | recombinational repair | YMR224C, YDR004W, YIL132C, YOL034W, YPL194W, YHR079C-A, YDR369C, YPR135W |
| 0.00869 | DNA repair | YGL090W, YOR005C, YMR224C, YHR056C, YCR014C, YDR004W, YJL092W, YIL132C, YOL034W, YLR007W, YOR368W, YPL194W, YHR079C-A, YHR134W, YDR369C, YPR135W |
| 0.0112 | cell cycle | YOR026W, YGR188C, YOR353C, YOR005C, YKR031C, YJR140C, YMR224C, YKL092C, YPL255W, YER018C, YHR014W, YKR072C, YDR004W, YIL072W, YEL061C, YLR254C, YHR184W, YIL132C, YOL034W, YOR368W, YLR045C, YLR190W, YPL194W, YHR079C-A, YDR369C, YPR135W |
| 0.0127 | nuclear part | YDR312W, YOR026W, YGR188C, YJR002W, YNL075W, YGL090W, YOR005C, YNL182C, YMR033W, YJR140C, YNR054C, YMR224C, YHR056C, YKR092C, YDL148C, YMR025W, YPR112C, YBL018C, YIR009W, YER018C, YHR014W, YJL039C, YBL014C, YDR004W, YIL072W, YEL061C, YOR368W, YLR045C, YPL211W, YOR064C, YML112W, YOR191W, YHR004C, YPL194W, YHR079C-A, YHR134W, YDR369C, YPR135W |
| 0.0138 | response to DNA damage stimulus | YGL090W, YOR005C, YMR224C, YHR056C, YCR014C, YDR004W, YJL092W, YIL132C, YOL034W, YLR007W, YOR368W, YML112W, YPL194W, YHR079C-A, YHR134W, YDR369C, YPR135W |
| 0.0304 | M phase | YOR026W, YGR188C, YKR031C, YMR224C, YER018C, YHR014W, YDR004W, YIL072W, YEL061C, YLR254C, YHR184W, YIL132C, YOR368W, YLR045C, YPL194W, YHR079C-A, YDR369C, YPR135W |
| 0.0322 | cellular response to stimulus | YOR353C, YGL090W, YOR005C, YBR077C, YOL067C, YKR031C, YMR224C, YBR128C, YHR056C, YGL155W, YNL242W, YMR025W, YKL092C, YCR027C, YDL166C, YCR014C, YDR004W, YJL092W, YIL132C, YOL034W, YLR007W, YOR368W, YBR020W, YML112W, YNL145W, YPL194W, YHR079C-A, YHR134W, YDR369C, YPR135W |
| **C.** | **Significantly enriched terms in the top 100 set of largest perturbation centrality in the *oxidatively stressed* DIP (2005) interactome** | |
| P-value | Term | Proteins |
| 1.7e-05 | cell cycle process | YOR026W, YGR188C, YKR031C, YLR254C, YPL253C, YGL075C, YGL251C, YLR383W, YCR063W, YHL024W, YMR198W, YNL152W, YDR439W, YKL049C, YIL132C, YLR045C, YJL090C, YKR010C, YOR368W, YER149C, YER132C, YEL061C, YJR140C, YER018C, YKL092C, YJL013C, YER147C, YGL174W |
| 2.94e-05 | cell cycle | YOR026W, YGR188C, YKR031C, YLR254C, YPL253C, YGL075C, YGL251C, YLR383W, YCR063W, YHL024W, YMR198W, YJL006C, YNL152W, YDR439W, YKL049C, YIL132C, YLR045C, YJL090C, YKR010C, YOR368W, YER149C, YER132C, YEL061C, YJR140C, YDR253C, YER018C, YKL092C, YJL013C, YER147C, YGL174W |
| 0.000225 | M phase | YOR026W, YGR188C, YKR031C, YLR254C, YPL253C, YGL075C, YGL251C, YHL024W, YMR198W, YNL152W, YDR439W, YKL049C, YIL132C, YLR045C, YKR010C, YOR368W, YER132C, YEL061C, YER018C, YJL013C, YER147C |
| 0.000284 | microtubule motor activity | YPL253C, YMR198W, YKL079W, YDR488C, YEL061C |
| 0.000354 | cell cycle phase | YOR026W, YGR188C, YKR031C, YLR254C, YPL253C, YGL075C, YGL251C, YHL024W, YMR198W, YNL152W, YDR439W, YKL049C, YIL132C, YLR045C, YJL090C, YKR010C, YOR368W, YER132C, YEL061C, YJR140C, YER018C, YJL013C, YER147C |
| 0.00104 | organelle fission | YOR026W, YGR188C, YLR254C, YPL253C, YGL075C, YMR198W, YNL152W, YDR439W, YKL049C, YKR010C, YIL065C, YEL061C, YER018C, YJL013C, YER147C |
| 0.00294 | mitosis | YOR026W, YGR188C, YLR254C, YPL253C, YGL075C, YMR198W, YNL152W, YDR439W, YKL049C, YKR010C, YEL061C, YER018C, YJL013C, YER147C |
| 0.00331 | nuclear division | YOR026W, YGR188C, YLR254C, YPL253C, YGL075C, YMR198W, YNL152W, YDR439W, YKL049C, YKR010C, YEL061C, YER018C, YJL013C, YER147C |
| 0.00386 | chromosome segregation | YGR188C, YPL253C, YLR383W, YMR198W, YDR439W, YKL049C, YIL132C, YKR010C, YEL061C, YER018C, YJL013C, YER147C |
| 0.00531 | nucleus | YOR026W, YGR188C, YKR031C, YKR092C, YBL014C, YCR014C, YLR254C, YHR134W, YPL253C, YDR020C, YJR119C, YGL075C, YHR079C, YGL251C, YLR383W, YCR063W, YKR022C, YMR198W, YJL006C, YPL046C, YGR278W, YDR439W, YKL049C, YFL049W, YGR006W, YBL010C, YIL132C, YGL131C, YLR045C, YPR112C, YHR004C, YJL090C, YNR011C, YDR082W, YKR010C, YDL080C, YOR368W, YLR010C, YLR094C, YIL128W, YHL006C, YAL051W, YEL061C, YJR140C, YDR253C, YGL220W, YMR284W, YER018C, YDR098C, YLR007W, YJL013C, YER147C, YML112W, YGL174W, YHR167W, YPR034W |

| P-value | Term | Proteins |
|---|---|---|
| 0.00595 | mitotic cell cycle | YOR026W, YGR188C, YLR254C, YPL253C, YGL075C, YMR198W, YNL152W, YDR439W, YKL049C, YLR045C, YJL090C, YKR010C, YEL061C, YJR140C, YDR253C, YER018C, YJL013C, YER147C |
| 0.00851 | M phase of mitotic cell cycle | YOR026W, YGR188C, YLR254C, YPL253C, YGL075C, YMR198W, YNL152W, YDR439W, YKL049C, YKR010C, YEL061C, YER018C, YJL013C, YER147C |
| 0.0107 | condensed chromosome | YOR026W, YGR188C, YLR383W, YDR439W, YKL049C, YLR045C, YOR368W, YEL061C, YER018C, YLR007W |
| 0.0125 | microtubule-based process | YGR188C, YLR254C, YPL253C, YGL075C, YMR198W, YKL079W, YDR488C, YLR045C, YEL061C, YER018C |
| 0.0153 | cellular response to stimulus | YKR031C, YCR027C, YCR014C, YNL145W, YHR134W, YHR079C, YLR383W, YJL006C, YPL046C, YIL132C, YJL090C, YBR128C, YBR077C, YOR368W, YER149C, YLR094C, YIL128W, YHL006C, YAL051W, YGL220W, YMR284W, YDL138W, YDR098C, YPL002C, YKL092C, YLR007W, YER147C, YML112W, YBR020W, YLR240W |
| 0.0201 | motor activity | YPL253C, YMR198W, YKL079W, YDR488C, YEL061C |
| 0.0256 | mitotic sister chromatid segregation | YGR188C, YPL253C, YMR198W, YDR439W, YKL049C, YKR010C, YEL061C, YER147C |
| 0.0257 | microtubule-based movement | YLR254C, YMR198W, YKL079W, YDR488C, YEL061C |
| 0.0348 | chromosome organization | YGR188C, YPL253C, YJR119C, YGL251C, YMR198W, YNL152W, YDR439W, YKL049C, YFL049W, YIL132C, YDR082W, YKR010C, YLR010C, YHL006C, YEL061C, YJR140C, YMR284W, YER147C, YPR034W |
| 0.0413 | sister chromatid segregation | YGR188C, YPL253C, YMR198W, YDR439W, YKL049C, YKR010C, YEL061C, YER147C |
| 0.0443 | condensed chromosome kinetochore | YOR026W, YGR188C, YDR439W, YKL049C, YLR045C, YEL061C, YER018C |
| 0.0479 | chromosomal part | YOR026W, YGR188C, YLR383W, YDR439W, YKL049C, YGL131C, YLR045C, YJL090C, YDR082W, YOR368W, YLR010C, YEL061C, YMR284W, YER018C, YLR007W, YER147C |
| **D. Significantly enriched terms in the top 100 set of largest perturbation centrality in the *osmotically stressed* DIP (2005) interactome** | | |
| **P-value** | **Term** | **Proteins** |
| 0.000101 | cell cycle process | YGL075C, YGL251C, YLR383W, YCR063W, YDR439W, YPL253C, YGR188C, YKL049C, YHL024W, YOR026W, YJL090C, YLR045C, YMR198W, YJR140C, YHR184W, YER132C, YKR031C, YEL061C, YGL174W, YHL023C, YJL013C, YMR224C, YGR262C, YPR135W, YOR368W, YER018C, YER147C |
| 0.000167 | cell cycle | YGL075C, YGL251C, YLR383W, YCR063W, YDR439W, YPL253C, YGR188C, YKL049C, YHL024W, YJL006C, YOR026W, YOR005C, YJL090C, YLR045C, YMR198W, YJR140C, YHR184W, YER132C, YKR031C, YEL061C, YGL174W, YHL023C, YJL013C, YMR224C, YGR262C, YPR135W, YOR368W, YER018C, YER147C |
| 0.00027 | microtubule motor activity | YDR488C, YPL253C, YKL079W, YMR198W, YEL061C |
| 0.00028 | M phase | YGL075C, YGL251C, YDR439W, YPL253C, YGR188C, YKL049C, YHL024W, YOR026W, YLR045C, YMR198W, YHR184W, YER132C, YKR031C, YEL061C, YHL023C, YJL013C, YMR224C, YPR135W, YOR368W, YER018C, YER147C |
| 0.00028 | non-homologous end-joining | YCR014C, YOR005C, YGL090W, YMR224C |
| 0.000453 | cell cycle phase | YGL075C, YGL251C, YDR439W, YPL253C, YGR188C, YKL049C, YHL024W, YOR026W, YJL090C, YLR045C, YMR198W, YJR140C, YHR184W, YER132C, YKR031C, YEL061C, YHL023C, YJL013C, YMR224C, YPR135W, YOR368W, YER018C, YER147C |
| 0.00389 | combined immunodeficiency | YGR188C, YOR005C, YJL013C |
| 0.00769 | nuclear part | YGL075C, YCR063W, YGR278W, YDR439W, YKR092C, YLR010C, YGR188C, YDR082W, YKL049C, YKR022C, YGL131C, YMR219W, YJL006C, YOR026W, YFL049W, YOR005C, YGR006W, YJL090C, YBL014C, YLR045C, YJR140C, YPL046C, YNL286W, YLR051C, YEL061C, YGL174W, YML112W, YGL090W, YHL006C, YJL013C, YMR224C, YKR086W, YPR034W, YPR135W, YOR368W, YER018C, YER147C, YOR064C |
| 0.0109 | nucleus | YJR119C, YGL075C, YGL251C, YLR383W, YCR063W, YGR278W, YDR439W, YBL010C, YCR014C, YPL253C, YKR092C, YLR010C, YGR188C, YDR082W, YKL049C, YHR079C, YKR022C, YGL131C, YMR219W, YJL006C, YOR026W, YLR007W, YFL049W, YOR005C, YGR006W, YJL090C, YBL014C, YLR045C, YMR198W, YGL220W, YJR140C, YDR098C, YPL046C, |
| | | YNL286W, YLR051C, YKR031C, YLR014C, YEL061C, YGL174W, YIL128W, YML112W, YGL090W, YDR020C, YNR011C, YHL006C, YJL013C, YMR224C, YKR086W, YGR262C, YPR034W, YOL043C, YPR135W, YOR368W, YER018C, YER147C, YOR064C |
| 0.011 | condensed chromosome | YLR383W, YDR439W, YGR188C, YKL049C, YOR026W, YLR007W, YLR045C, YEL061C, YOR368W, YER018C |
| 0.0128 | upslanted palpebral fissure | YJR119C, YGR188C, YOR005C, YEL061C, YJL013C |
| 0.0133 | chromosomal part | YLR383W, YDR439W, YLR010C, YGR188C, YDR082W, YKL049C, YGL131C, YOR026W, YLR007W, YJL090C, YLR045C, YEL061C, YGR262C, YPR135W, YOR368W, YER018C, YER147C |
| 0.019 | motor activity | YDR488C, YPL253C, YKL079W, YMR198W, YEL061C |
| 0.0216 | double-strand break repair | YLR383W, YCR014C, YOR005C, YJL090C, YGL090W, YMR224C, YPR135W, YOR368W, YER147C |
| 0.0252 | chromosome segregation | YLR383W, YDR439W, YPL253C, YGR188C, YKL049C, YMR198W, YEL061C, YJL013C, YPR135W, YER018C, YER147C |
| 0.0253 | macromolecular complex | YLR383W, YCR063W, YGR278W, YDR439W, YDR488C, YPL253C, YLR010C, YBR128C, YGR188C, YDR082W, YKL049C, YKR022C, YGL131C, YLR439W, YJL006C, YOR026W, YLR007W, YFL049W, YOR005C, YNL014W, YKL079W, YGR006W, YNR049C, YJL090C, YBL014C, YLR045C, YMR198W, YLR185W, YJR140C, YPL046C, YNL286W, YEL061C, YGL174W, YML112W, YGL090W, YHL023C, YLR240W, YPL002C, YHL006C, YJL013C, YMR224C, YKR086W, YGR262C, YBR077C, YPR034W, YPR135W, YIL068C, YOR368W, YER018C, YER147C, YOR064C |
| 0.0255 | mitotic sister chromatid segregation | YDR439W, YPL253C, YGR188C, YKL049C, YMR198W, YEL061C, YPR135W, YER147C |
| 0.0256 | DNA metabolic process | YGL251C, YLR383W, YDR439W, YCR014C, YLR010C, YDR082W, YKL049C, YHL024W, YLR007W, YOR005C, YJL090C, YJR140C, YPL046C, YIL128W, YGL090W, YHL006C, YMR224C, YGR262C, YOL043C, YPR135W, YOR368W, YER147C |
| 0.0289 | cafe-au-lait spot | YJR119C, YGR188C, YOR005C, YJL013C |
| 0.0322 | organelle fission | YGL075C, YDR439W, YPL253C, YGR188C, YKL049C, YOR026W, YMR198W, YIL065C, YEL061C, YJL013C, YPR135W, YER018C, YER147C |
| 0.0369 | protein complex | YLR383W, YDR439W, YDR488C, YPL253C, YBR128C, YGR188C, YKL049C, YGL131C, YJL006C, YOR026W, YLR007W, YFL049W, YOR005C, YKL079W, YNR049C, YJL090C, YBL014C, YLR045C, YMR198W, YJR140C, YPL046C, YEL061C, YGL174W, YML112W, YGL090W, YHL023C, YLR240W, YPL002C, YHL006C, YJL013C, YMR224C, YGR262C, YBR077C, YPR034W, YPR135W, YIL068C, YOR368W, YER018C, YER147C, YOR064C |
| 0.0395 | response to DNA damage stimulus | YLR383W, YCR014C, YJL006C, YLR007W, YOR005C, YJL090C, YPL046C, YIL128W, YML112W, YGL090W, YHL006C, YMR224C, YOL043C, YPR135W, YOR368W, YER147C |
| 0.0411 | sister chromatid segregation | YDR439W, YPL253C, YGR188C, YKL049C, YMR198W, YEL061C, YPR135W, YER147C |
| 0.0433 | condensed chromosome kinetochore | YDR439W, YGR188C, YKL049C, YOR026W, YLR045C, YEL061C, YER018C |

Perturbation centralities were calculated with the Turbine software as described in **Methods** of the main text. Term enrichment analysis was performed with the R plug-in of g:Profiler [12], which returns both the enriched terms, and the proteins connected with the term. A term was stated as statistically significant, if the resulting p-value was strictly less than 0.05 after applying Bonferroni correction. Results show the high importance of cell cycle maintenance and DNA repair in both stressed and unstressed cases.

**Table S4. Statistically significantly enriched terms in the top 100 protein set of the DIP yeast interactome (release 2005)** [7] **containing proteins with *largest increase* of perturbation centrality in differently stressed cases.**

| A. Significantly enriched terms in the top 100 set of proteins having largest *increase* of perturbation centrality on *heat shock* (compared to the unstressed case) | | |
|---|---|---|
| **P-value** | **Term** | **Proteins** |
| 5.18e-08 | cellular carbohydrate metabolic process | YLR273C, YJL137C, YIL045W, YNL216W, YPR160W, YMR261C, YML100W, YDR074W, YCR005C, YLR258W, YPL201C, YNR001C, YBR149W, YIL053W, YER062C, YJL089W, YMR105C, YPL180W |
| 6.15e-07 | response to stimulus | YDL190C, YDL059C, YNL076W, YBR169C, YNL103W, YNL216W, YER103W, YMR261C, YEL060C, YML100W, YDR074W, YFL021W, YDL124W, YBL075C, YNL314W, YIR023W, YIL113W, YLR178C, YPL022W, YLR019W, YOL128C, YGR088W, YPL154C, YFL016C, YPR054W, YIL053W, YDR168W, YDR214W, YER062C, YOR363C, YKL109W, YDR200C, YOR120W, YGL163C, YLR259C, YDL017W, YMR250W, YJL089W, YHR186C, YLL021W, YPL180W, YNR007C, YBR274W, YJR032W |
| 1.25e-06 | cellular carbohydrate biosynthetic process | YLR273C, YJL137C, YIL045W, YMR261C, YML100W, YDR074W, YLR258W, YPL201C, YIL053W, YER062C, YMR105C |
| 2.31e-05 | carbohydrate biosynthetic process | YLR273C, YJL137C, YIL045W, YMR261C, YML100W, YDR074W, YLR258W, YPL201C, YIL053W, YER062C, YJL089W, YMR105C |
| 0.000188 | response to stress | YDL190C, YDL059C, YBR169C, YNL216W, YER103W, YMR261C, YEL060C, YML100W, YDR074W, YDL124W, YBL075C, YPL022W, YLR019W, YOL128C, YGR088W, YPL154C, YFL016C, YIL053W, YDR168W, YDR214W, YER062C, YOR120W, YGL163C, YLR259C, YDL017W, YMR250W, YHR186C, YPL180W, YNR007C, YBR274W, YJR032W |
| 0.000323 | organic substance catabolic process | YPL065W, YJL172W, YIR032C, YDL190C, YLR270W, YCL008C, YCL052C, YNL216W, YOR173W, YPR160W, YPR111W, YNL230C, YEL060C, YER143W, YNL314W, YCR005C, YIR023W, YPL022W, YNR001C, YPL154C, YFL016C, YKL148C, YDR214W, YOR363C, YOR120W, YGR058W, YLL041C, YNL311C, YMR250W, YMR105C, YMR287C |
| 0.000354 | catalytic activity | YJL137C, YML016C, YJL172W, YIR032C, YMR315W, YDL190C, YLR270W, YCL008C, YAL060W, YBR265W, YOR173W, YOR317W, YER103W, YPR160W, YLR096W, YIL177C, YMR261C, YPR111W, YPL074W, YEL060C, YML100W, YDR074W, YDL124W, YBL075C, YER143W, YCR005C, YLR258W, YMR104C, YIL113W, YER081W, YML065W, YPL022W, YLR019W, YOL128C, YGR088W, YNR001C, YBR149W, YPL154C, YIL108W, YPR054W, YIL053W, YBL045C, YNL092W, YKL148C, YPR191W, YER062C, YOR120W, YGL163C, YLR259C, YDL017W, YLL041C, YMR250W, YMR105C, YKL210W, YNR007C, YMR287C, YBR274W, YJR032W |
| 0.00046 | catabolic process | YPL065W, YJL172W, YIR032C, YDL190C, YLR270W, YCL008C, YCL052C, YNL216W, YOR173W, YPR160W, YPR111W, YNL230C, YEL060C, YER143W, YNL314W, YCR005C, YIR023W, YPL022W, YGR088W, YNR001C, YPL154C, YFL016C, YKL148C, YDR214W, YOR363C, YOR120W, YGR058W, YLL041C, YNL311C, YMR250W, YMR105C, YNR007C, YMR287C |
| 0.000616 | cellular catabolic process | YPL065W, YJL172W, YIR032C, YDL190C, YLR270W, YCL008C, YCL052C, YNL216W, YOR173W, YPR111W, YNL230C, YEL060C, YER143W, YNL314W, YCR005C, YIR023W, YPL022W, YGR088W, YNR001C, YPL154C, YFL016C, YKL148C, YDR214W, YOR363C, YGR058W, YLL041C, YNL311C, YMR250W, YNR007C, YMR287C |
| 0.00102 | energy derivation by oxidation of organic compounds | YLR273C, YJL137C, YIL045W, YPR160W, YCR005C, YLR258W, YNR001C, YBL045C, YKL148C, YPR191W, YKL109W, YLL041C, YMR105C |
| 0.00105 | oxidation-reduction process | YLR273C, YJL137C, YMR315W, YIL045W, YAL060W, YBR265W, YPR160W, YDL124W, YCR005C, YLR258W, YER081W, YGR088W, YNR001C, YBR149W, YBL045C, YKL148C, YPR191W, YOR363C, YKL109W, YOR120W, YLL041C, YMR105C |
| 0.00116 | carbohydrate metabolic process | YLR273C, YJL137C, YIL045W, YNL216W, YPR160W, YMR261C, YML100W, YDR074W, YCR005C, YLR258W, YPL201C, YNR001C, YBR149W, YIL053W, YER062C, YOR120W, YJL089W, YMR105C, YPL180W |
| 0.00276 | glycoside biosynthetic process | YMR261C, YML100W, YDR074W, YMR105C |
| 0.00276 | oligosaccharide biosynthetic | YMR261C, YML100W, YDR074W, YMR105C |

| P-value | Term | Proteins |
|---|---|---|
| | process | |
| 0.00276 | disaccharide biosynthetic process | YMR261C, YML100W, YDR074W, YMR105C |
| 0.00276 | trehalose biosynthetic process | YMR261C, YML100W, YDR074W, YMR105C |
| 0.00547 | alditol biosynthetic process | YPL201C, YIL053W, YER062C |
| 0.00547 | glycerol biosynthetic process | YPL201C, YIL053W, YER062C |
| 0.00873 | generation of precursor metabolites and energy | YLR273C, YJL137C, YIL045W, YNL216W, YPR160W, YCR005C, YLR258W, YNR001C, YBL045C, YKL148C, YPR191W, YKL109W, YLL041C, YMR105C |
| 0.0139 | glycogen metabolic process | YLR273C, YJL137C, YIL045W, YPR160W, YLR258W, YMR105C |
| 0.0156 | glycogen biosynthetic process | YLR273C, YJL137C, YIL045W, YLR258W, YMR105C |
| 0.0188 | glycogen breakdown (glycogenolysis) | YJL137C, YPR160W, YMR105C |
| 0.0191 | organic substance metabolic process | YLR273C, YDR034C, YJL137C, YPL065W, YML016C, YJL172W, YIR032C, YMR315W, YDL190C, YLR270W, YDL059C, YNL076W, YCL008C, YIL045W, YBR169C, YAL060W, YNL103W, YBR265W, YCL052C, YNL216W, YOR173W, YOR317W, YER103W, YPR160W, YLR096W, YMR261C, YPR111W, YNL230C, YEL060C, YML100W, YDR074W, YFL021W, YDL124W, YBL075C, YLR453C, YER143W, YNL314W, YCR005C, YLR258W, YIR023W, YMR104C, YIL113W, YER081W, YLR178C, YPL201C, YML065W, YPL022W, YOL128C, YNR001C, YBR149W, YPL154C, YIL108W, YFL016C, YPR054W, YIL053W, YBL045C, YDR168W, YKL148C, YPR191W, YDR214W, YER062C, YOR363C, YKL109W, YOR120W, YGL163C, YLR259C, YGR058W, YDL017W, YLL041C, YNL311C, YMR250W, YJL089W, YMR105C, YKL210W, YPL180W, YNR007C, YMR287C, YDR515W, YBR274W, YJR032W |
| 0.0199 | energy reserve metabolic process | YLR273C, YJL137C, YIL045W, YPR160W, YLR258W, YMR105C |
| 0.0217 | alpha,alpha-trehalose-phosphate synthase complex (UDP-forming) | YMR261C, YML100W, YDR074W |
| 0.0248 | trehalose metabolic process | YMR261C, YML100W, YDR074W, YMR105C |
| 0.0368 | fungal-type vacuole lumen | YJL172W, YEL060C, YLR178C, YPL154C |
| 0.0437 | single-organism metabolic process | YLR273C, YDR034C, YJL137C, YPL065W, YML016C, YJL172W, YIR032C, YMR315W, YDL190C, YLR270W, YDL059C, YNL076W, YCL008C, YIL045W, YBR169C, YAL060W, YNL103W, YBR265W, YCL052C, YNL216W, YOR173W, YOR317W, YER103W, YPR160W, YLR096W, YMR261C, YPR111W, YNL230C, YEL060C, YML100W, YDR074W, YFL021W, YDL124W, YBL075C, YLR453C, YER143W, YNL314W, YCR005C, YLR258W, YIR023W, YMR104C, YIL113W, YER081W, YLR178C, YPL201C, YML065W, YPL022W, YOL128C, YGR088W, YNR001C, YBR149W, YPL154C, YIL108W, YFL016C, YPR054W, YIL053W, YBL045C, YDR168W, YKL148C, YPR191W, YDR214W, YER062C, YOR363C, YKL109W, YOR120W, YGL163C, YLR259C, YGR058W, YDL017W, YLL041C, YNL311C, YMR250W, YJL089W, YMR105C, YKL210W, YPL180W, YNR007C, YMR287C, YDR515W, YBR274W, YJR032W |
| **B.** | **Significantly enriched terms in the top 100 set of proteins having largest *increase* of perturbation centrality in *oxidative stress* (compared to the unstressed case)** | |
| **P-value** | **Term** | **Proteins** |
| 4.24e-07 | cellular carbohydrate metabolic process | YLR273C, YJL137C, YIL045W, YML100W, YMR261C, YDR074W, YLR258W, YGR143W, YPR160W, YBR149W, YGR166W, YCR005C, YOR178C, YNR001C, YJL089W, YDR001C, YBL058W |
| 2.2e-05 | cellular carbohydrate biosynthetic process | YLR273C, YJL137C, YIL045W, YML100W, YMR261C, YDR074W, YLR258W, YGR143W, YGR166W, YOR178C |
| 8.53e-05 | glucan metabolic process | YLR273C, YJL137C, YIL045W, YLR258W, YGR143W, YPR160W, YGR166W, YOR178C, YBL058W |
| 8.53e-05 | cellular glucan metabolic process | YLR273C, YJL137C, YIL045W, YLR258W, YGR143W, YPR160W, YGR166W, YOR178C, YBL058W |
| 0.000254 | carbohydrate biosynthetic process | YLR273C, YJL137C, YIL045W, YML100W, YMR261C, YDR074W, YLR258W, YGR143W, YGR166W, YOR178C, YJL089W |
| 0.000585 | glucan biosynthetic process | YLR273C, YJL137C, YIL045W, YLR258W, YGR143W, YGR166W, YOR178C |

| P-value | Term | Proteins |
|---|---|---|
| 0.000596 | cellular polysaccharide metabolic process | YLR273C, YJL137C, YIL045W, YLR258W, YGR143W, YPR160W, YGR166W, YOR178C, YBL058W |
| 0.000649 | TRAPP complex | YOR115C, YDR246W, YGR166W, YMR218C, YDR407C |
| 0.000928 | glycogen metabolic process | YLR273C, YJL137C, YIL045W, YLR258W, YPR160W, YOR178C, YBL058W |
| 0.00102 | carbohydrate metabolic process | YLR273C, YJL137C, YIL045W, YML100W, YMR261C, YDR074W, YLR258W, YGR143W, YPR160W, YBR149W, YBR229C, YGR166W, YCR005C, YOR178C, YMR200W, YNR001C, YJL089W, YDR001C, YBL058W |
| 0.00143 | energy reserve metabolic process | YLR273C, YJL137C, YIL045W, YLR258W, YPR160W, YOR178C, YBL058W |
| 0.00194 | response to stimulus | YLR309C, YJL173C, YNL076W, YER103W, YPR019W, YML032C, YDL124W, YMR250W, YDR453C, YDR179C, YLR178C, YPL026C, YML100W, YMR261C, YDR074W, YBR202W, YPL154W, YHL035C, YDL059C, YOL128C, YBR216C, YFR014C, YOR018W, YMR038C, YOR178C, YDR049W, YMR200W, YMR218C, YJL201W, YKL213C, YJL089W, YDR168W, YDR001C, YDL216C, YDL101C, YBL058W, YDR407C |
| 0.00249 | polysaccharide metabolic process | YLR273C, YJL137C, YIL045W, YLR258W, YGR143W, YPR160W, YGR166W, YOR178C, YBL058W |
| 0.00852 | cellular polysaccharide biosynthetic process | YLR273C, YJL137C, YIL045W, YLR258W, YGR143W, YGR166W, YOR178C |
| 0.00993 | polysaccharide biosynthetic process | YLR273C, YJL137C, YIL045W, YLR258W, YGR143W, YGR166W, YOR178C |
| 0.0142 | oxidation-reduction process | YMR315W, YLR273C, YJL137C, YIL045W, YDL124W, YBR265W, YDR453C, YLR258W, YPR160W, YDR231C, YBR149W, YBR213W, YMR038C, YNL202W, YLL051C, YCR005C, YOR178C, YER023W, YNR001C, YBL058W |
| 0.0168 | acetyl-CoA + $H_2O$ + oxaloacetate => citrate + CoA | YCR005C, YNR001C |
| 0.0196 | glycogen biosynthetic process | YLR273C, YJL137C, YIL045W, YLR258W, YOR178C |
| 0.0196 | cis-Golgi network | YOR115C, YDR246W, YGR166W, YMR218C, YDR407C |
| 0.0289 | alpha,alpha-trehalose-phosphate synthase complex (UDP-forming) | YML100W, YMR261C, YDR074W |
| 0.0321 | trehalose metabolic process | YML100W, YMR261C, YDR074W, YDR001C |
| 0.0477 | protein phosphatase type 1 regulator activity | YLR273C, YIL045W, YOR178C, YBL058W |
| **C.** | **Significantly enriched terms in the top 100 set of proteins having largest *increase* of perturbation centrality in *osmotic stress* (compared to the unstressed case)** | |
| **P-value** | **Term** | **Proteins** |
| 4.42e-09 | cellular carbohydrate biosynthetic process | YLR273C, YJL137C, YIL045W, YPL201C, YIL053W, YER062C, YLR258W, YMR261C, YFR015C, YML100W, YDR074W, YMR105C, YOR178C |
| 2.66e-06 | carbohydrate biosynthetic process | YLR273C, YJL137C, YIL045W, YPL201C, YIL053W, YER062C, YLR258W, YMR261C, YFR015C, YML100W, YDR074W, YMR105C, YOR178C |
| 5.85e-06 | cellular carbohydrate metabolic process | YLR273C, YJL137C, YIL045W, YPL201C, YIL053W, YER062C, YLR258W, YMR261C, YFR015C, YIR031C, YML100W, YDR074W, YCR005C, YDR001C, YMR105C, YOR178C |
| 1.98e-05 | glycogen biosynthetic process | YLR273C, YJL137C, YIL045W, YLR258W, YFR015C, YMR105C, YOR178C |
| 0.000582 | glucan biosynthetic process | YLR273C, YJL137C, YIL045W, YLR258W, YFR015C, YMR105C, YOR178C |
| 0.000586 | trehalose metabolic process | YMR261C, YML100W, YDR074W, YDR001C, YMR105C |
| 6e-04 | protein phosphorylation | YJR059W, YKL116C, YKL048C, YDL025C, YLR096W, YFL033C, YGR052W, YPL203W, YLR210W, YMR104C, YDR460W, YFL029C, YER129W, YDR052C, YDR490C |
| 0.000619 | protein serine/threonine kinase activity | YJR059W, YKL116C, YKL048C, YDL025C, YLR096W, YFL033C, YGR052W, YPL203W, YMR104C, YFL029C, YER129W, YDR490C |
| 0.000922 | glycogen metabolic process | YLR273C, YJL137C, YIL045W, YLR258W, YFR015C, YMR105C, YOR178C |
| 0.00104 | protein kinase activity | YJR059W, YKL116C, YKL048C, YDL025C, YLR096W, YFL033C, YGR052W, YPL203W, YMR104C, YFL029C, YER129W, YDR490C |
| 0.00141 | energy reserve metabolic process | YLR273C, YJL137C, YIL045W, YLR258W, YFR015C, YMR105C, YOR178C |
| 0.00306 | glycoside biosynthetic process | YMR261C, YML100W, YDR074W, YMR105C |
| 0.00306 | oligosaccharide biosynthetic process | YMR261C, YML100W, YDR074W, YMR105C |
| 0.00306 | disaccharide biosynthetic process | YMR261C, YML100W, YDR074W, YMR105C |
| 0.00306 | trehalose biosynthetic process | YMR261C, YML100W, YDR074W, YMR105C |
| 0.00316 | phosphorylation | YJR059W, YKL116C, YKL048C, YKL067W, YDL025C, YLR096W, YHR033W, YFL033C, YGR052W, YPL203W, YLR210W, YMR104C, YDR460W, YFL029C, YER129W, YDR052C, YDR490C, YKL141W |
| 0.00372 | response to stress | YDR159W, YIL053W, YER103W, YBR169C, YER062C, YDR501W, YMR261C, YKL048C, YML100W, YDR074W, YDL124W, YKL067W, YCR065W, YOR141C, YDL190C, YJL173C, YDR217C, YJR032W, YFL033C, YDL020C, YDR001C, YDR460W, YDL059C, YGR088W, YBR066C, YDR113C, YGL163C, YLR183C, YOR178C |
| 0.00586 | response to stimulus | YDR159W, YIL053W, YER103W, YNL076W, YBR169C, YER062C, YDR379W, YDR501W, YMR261C, YKL116C, YKL048C, YML100W, YDR074W, YDL124W, YKL067W, YCR065W, YAL024C, YOR141C, YDL190C, YJL173C, YDR217C, YJR032W, YFL033C, YHL035C, YDL020C, YPL203W, YDR001C, YDR460W, YDL059C, YOR018W, YGR088W, YBR066C, YDR113C, YGL163C, YLR183C, YOR178C, YDR490C |
| 0.00595 | alditol biosynthetic process | YPL201C, YIL053W, YER062C |
| 0.00595 | glycerol biosynthetic process | YPL201C, YIL053W, YER062C |
| 0.007 | kinase activity | YJR059W, YKL116C, YKL048C, YKL067W, YDL025C, YLR096W, YHR033W, YFL033C, YGR052W, YPL203W, YMR104C, YFL029C, YER129W, YDR490C |
| 0.00716 | carbohydrate metabolic process | YLR273C, YJL137C, YIL045W, YPL201C, YIL053W, YER062C, YLR258W, YMR261C, YFR015C, YIR031C, YKL048C, YML100W, YDR074W, YCR005C, YDR001C, YMR105C, YER129W, YOR178C |
| 0.00839 | cellular polysaccharide biosynthetic process | YLR273C, YJL137C, YIL045W, YLR258W, YFR015C, YMR105C, YOR178C |
| 0.00958 | cell cycle phase | YDR159W, YJR059W, YDR285W, YOR058C, YPL155C, YGR238C, YOR195W, YGL216W, YCR065W, YAL024C, YOR177C, YJL173C, YDR217C, YFL033C, YLR210W, YNL309W, YFL029C, YNL068C, YDR052C, YDR113C, YOR178C |
| 0.00977 | polysaccharide biosynthetic process | YLR273C, YJL137C, YIL045W, YLR258W, YFR015C, YMR105C, YOR178C |
| 0.0127 | glycogen synthesis | YJL137C, YLR258W, YFR015C, YMR105C |
| 0.0133 | glycoside metabolic process | YMR261C, YML100W, YDR074W, YDR001C, YMR105C |
| 0.0156 | single-organism carbohydrate metabolic process | YLR273C, YJL137C, YIL045W, YPL201C, YIL053W, YER062C, YLR258W, YMR261C, YFR015C, YKL048C, YML100W, YDR074W, YDR001C, YMR105C, YER129W, YOR178C |
| 0.0173 | glucan metabolic process | YLR273C, YJL137C, YIL045W, YLR258W, YFR015C, YMR105C, YOR178C |
| 0.0173 | cellular glucan metabolic process | YLR273C, YJL137C, YIL045W, YLR258W, YFR015C, YMR105C, YOR178C |
| 0.0235 | alpha,alpha-trehalose-phosphate synthase complex (UDP-forming) | YMR261C, YML100W, YDR074W |
| 0.0252 | cell cycle process | YDR159W, YJR059W, YDR285W, YOR058C, YKL048C, YPL155C, YGR238C, YOR195W, YGL216W, YCR065W, YAL024C, YOR177C, YJL173C, YDR217C, YFL033C, YLR210W, YNL309W, YFL029C, YNL068C, YDR052C, YDR113C, YLR457C, YOR178C |
| 0.0323 | cell cycle | YDR159W, YJR059W, YDR285W, YOR058C, YKL048C, YPL155C, YGR238C, YOR195W, YGL216W, YCR065W, YAL024C, YOR177C, YJL173C, YDR217C, YFL033C, YLR210W, YDR460W, YNL309W, YFL029C, YNL068C, YDR052C, YDR113C, YNL007C, YLR457C, YOR178C |
| 0.0365 | mitotic cell cycle | YDR159W, YJR059W, YOR058C, YPL155C, YGR238C, YOR195W, YGL216W, YCR065W, YAL024C, YDR217C, YLR210W, |

| | | YNL309W, YFL029C, YNL068C, YDR052C, YDR113C, YOR178C |

Perturbation centralities were calculated with the Turbine software as described in **Methods** of the main text. Term enrichment analysis was performed with the R plug-in of g:Profiler [12], which returns both the enriched terms, and the proteins connected with the term. A term was stated as statistically significant, if the resulting p-value was strictly less than 0.05 after applying Bonferroni correction. Results clearly show the stress response as displayed by the significant enrichment of the terms "response to stimulus" and "response to stress". Furthermore, the strong up-regulation of carbohydrate metabolism is also obvious from the data, and it is a well-known stress response [13–15] successfully identified by the perturbation centrality measure.

**Table S5. Most significantly enriched terms in the top 100 protein set of the DIP yeast interactome (release 2005) [7] containing proteins with *largest decrease* of perturbation centrality in differently stressed cases.**

| A. Significantly enriched terms in the top 100 set of proteins having *largest decrease* of perturbation centrality in *heat shock* | | |
|---|---|---|
| P-value | Term | Proteins |
| 1.81e-59 | ribosome biogenesis | YCL054W, YLR129W, YOL010W, YPL093W, YPR137W, YDR324C, YDR449C, YLR276C, YMR239C, YHR088W, YLL008W, YGR159C, YGL171W, YJL033W, YNL124W, YOL077C, YLR186W, YPL217C, YFR001W, YDL166C, YOR310C, YKL099C, YDL031W, YHR072W-A, YKL009W, YOR206W, YHR170W, YGR128C, YHR148W, YBR267W, YHR066W, YJR002W, YDL148C, YHR197W, YKR060W, YPL211W, YKL021C, YNL075W, YJL109C, YEL026W, YGL111W, YBR247C, YGR245C, YDR087C, YGL078C, YLR221C, YHR081W, YMR093W, YJL069C, YOR243C, YIR026C, YGR081C, YPR144C, YNL308C, YER126C, YNL182C, YKL082C, YMR229C, YNR053C, YER006W, YKR081C, YOL142W, YHR052W, YCR072C, YMR128W, YER002W, YPL226W, YNL232W, YNR054C, YLR397C, YDL060W |
| 6.76e-57 | nucleolus | YCL054W, YLR129W, YOL010W, YPL093W, YPR137W, YDR324C, YDR449C, YLR276C, YMR239C, YHR088W, YLL008W, YGR159C, YNL248C, YGL171W, YJL033W, YOL077C, YLR186W, YPL217C, YFR001W, YOR310C, YKL099C, YDL031W, YHR072W-A, YKL009W, YOR206W, YGR128C, YHR148W, YHR066W, YJR002W, YDL148C, YOR340C, YHR143W-A, YKR060W, YPL211W, YKL021C, YNL075W, YJL109C, YEL026W, YGL111W, YBR247C, YGR245C, YDR087C, YGL078C, YLR221C, YMR093W, YJL069C, YJL076W, YGR081C, YPR144C, YNL308C, YER126C, YKL082C, YMR229C, YNR053C, YER006W, YKR081C, YOL142W, YHR052W, YCR072C, YMR128W, YER002W, YNL232W, YNR054C, YPL020C, YDL060W, YNL175C |
| 4.68e-56 | ribonucleoprotein complex biogenesis | YCL054W, YLR129W, YOL010W, YPL093W, YPR137W, YDR324C, YDR449C, YLR276C, YMR239C, YHR088W, YLL008W, YGR159C, YGL171W, YJL033W, YNL124W, YOL077C, YLR186W, YPL217C, YFR001W, YDL166C, YOR310C, YKL099C, YDL031W, YHR072W-A, YKL009W, YOR206W, YHR170W, YGR128C, YHR148W, YBR267W, YHR066W, YJR002W, YDL148C, YHR197W, YKR060W, YPL211W, YKL021C, YNL075W, YJL109C, YEL026W, YGL111W, YBR247C, YGR245C, YDR087C, YGL078C, YLR221C, YOR276W, YHR081W, YMR093W, YJL069C, YOR243C, YIR026C, YGR081C, YPR144C, YNL308C, YER126C, YNL182C, YKL082C, YMR229C, YNR053C, YER006W, YKR081C, YOL142W, YHR052W, YCR072C, YMR128W, YER002W, YPL226W, YNL232W, YNR054C, YLR397C, YDL060W |
| 7.16e-52 | preribosome | YCL054W, YLR129W, YOL010W, YPL093W, YPR137W, YDR324C, YDR449C, YHR088W, YLL008W, YGL171W, YOL077C, YLR186W, YPL217C, YFR001W, YOR310C, YKL099C, YDL031W, YKL009W, YOR206W, YGR128C, YHR148W, YJR002W, YDL148C, YKR060W, YPL211W, YNL075W, YJL109C, YEL026W, YGL111W, YBR247C, YDR087C, YGL078C, YLR221C, YMR093W, YJL069C, YGR081C, YPR144C, YER126C, YMR229C, YNR053C, YER006W, YKR081C, YHR052W, YMR128W, YER002W, YLR397C, YDL060W, YNL175C |
| 1.75e-49 | cellular component biogenesis at cellular level | YCL054W, YLR129W, YOL010W, YPL093W, YPR137W, YDR324C, YDR449C, YLR276C, YMR239C, YHR088W, YLL008W, YGR159C, YGL171W, YJL033W, YNL124W, YOL077C, YLR186W, YPL217C, YFR001W, YDL166C, YOR310C, YKL099C, YDL031W, YHR072W-A, YKL009W, YOR206W, YHR170W, YGR128C, YHR148W, YBR267W, YHR066W, YJR002W, YDL148C, YHR197W, YKR060W, YPL211W, YKL021C, YNL075W, YJL109C, YEL026W, YGL111W, YBR247C, YGR245C, YDR087C, YGL078C, YLR221C, YOR276W, YHR081W, YMR093W, YJL069C, YOR243C, YIR026C, YGR081C, YPR144C, YNL308C, YER126C, YNL182C, YKL082C, YMR229C, YNR053C, YER006W, YKR081C, YOL142W, YHR052W, YCR072C, YMR128W, YER002W, YPL226W, YNL232W, YNR054C, YLR397C, YDL060W |
| 3.56e-48 | nuclear lumen | YCL054W, YLR129W, YOL010W, YPL093W, YPR137W, YDR324C, YDR449C, YLR276C, YMR239C, YHR088W, YLL008W, YGR159C, YNL248C, YGL171W, YJL033W, YNL124W, YOL077C, YLR186W, YPL217C, YFR001W, YNL151C, YOR310C, YKL099C, YDL031W, YKL144C, YHR072W-A, YKL009W, YOR206W, YHR170W, YGR128C, YHR148W, YHR066W, YPR190C, YJR002W, YDL148C, YHR197W, YOR340C, YHR143W-A, YKR060W, YPL211W, YKL021C, YNL075W, YJL109C, YEL026W, YGL111W, YBR247C, YGR245C, YDR087C, YHR031C, YGL078C, YLR221C, YOR207C, YHR081W, YMR093W, YJL069C, YJL076W, YGR081C, YPR144C, YNL308C, YER126C, YNL182C, YKL082C, YMR229C, YNR053C, YDR045C, YER006W, YKR081C, YOL142W, YHR052W, YCR072C, YMR128W, YER002W, YNL232W, YNR054C, YPL233W, YPL020C, YDL060W, YNL175C |
| 1.91e-47 | rRNA processing | YCL054W, YLR129W, YOL010W, YPL093W, YPR137W, YDR324C, YDR449C, YLR276C, YMR239C, YHR088W, YLL008W, YGR159C, YGL171W, YJL033W, YNL124W, YLR186W, YPL217C, YFR001W, YDL166C, YOR310C, YKL099C, YDL031W, YHR072W-A, YKL009W, YGR128C, YHR148W, YJR002W, YDL148C, YHR197W, YKR060W, YPL211W, YKL021C, YNL075W, YJL109C, YEL026W, YGL111W, YBR247C, YDR087C, YGL078C, YHR081W, YMR093W, YJL069C, YOR243C, YGR081C, YPR144C, YNL308C, YER126C, YNL182C, YKL082C, YMR229C, YER006W, YKR081C, YOL142W, YMR128W, YER002W, YNL232W, YNR054C, YDL060W |
| 1.84e-46 | rRNA metabolic process | YCL054W, YLR129W, YOL010W, YPL093W, YPR137W, YDR324C, YDR449C, YLR276C, YMR239C, YHR088W, YLL008W, YGR159C, YGL171W, YJL033W, YNL124W, YLR186W, YPL217C, YFR001W, YDL166C, YOR310C, YKL099C, YDL031W, YHR072W-A, YKL009W, YGR128C, YHR148W, YJR002W, YDL148C, YHR197W, YKR060W, YPL211W, YKL021C, YNL075W, YJL109C, YEL026W, YGL111W, YBR247C, YDR087C, YGL078C, YHR081W, YMR093W, YJL069C, YOR243C, YGR081C, YPR144C, YNL308C, YER126C, YNL182C, YKL082C, YMR229C, YER006W, YKR081C, YOL142W, YMR128W, YER002W, YNL232W, YNR054C, YDL060W |
| 2.09e-41 | ncRNA processing | YCL054W, YLR129W, YOL010W, YPL093W, YPR137W, YDR324C, YDR449C, YLR276C, YMR239C, YHR088W, YLL008W, YGR159C, YGL171W, YJL033W, YNL124W, YLR186W, YPL217C, YFR001W, YDL166C, YOR310C, YKL099C, YDL031W, YHR072W-A, YKL009W, YGR128C, YHR148W, YJR002W, YDL148C, YHR197W, YKR060W, YPL211W, YKL021C, YNL075W, YJL109C, YEL026W, YGL111W, YBR247C, YDR087C, YGL078C, YNL062C, YHR081W, YMR093W, YJL069C, YOR243C, YGR081C, YPR144C, YNL308C, YER126C, YNL182C, YKL082C, YMR229C, YER006W, YKR081C, YOL142W, YMR128W, YER002W, YNL232W, YNR054C, YDL060W |
| 1.78e-39 | ncRNA metabolic process | YCL054W, YLR129W, YOL010W, YPL093W, YPR137W, YDR324C, YDR449C, YLR276C, YMR239C, YHR088W, YLL008W, YGR159C, YGL171W, YJL033W, YNL124W, YLR186W, YPL217C, YFR001W, YDL166C, YOR310C, YKL099C, YDL031W, YHR072W-A, YKL009W, YGR128C, YHR148W, YJR002W, YDL148C, YHR197W, YKR060W, YPL211W, YKL021C, YNL075W, YJL109C, YEL026W, YGL111W, YBR247C, YDR087C, YGL078C, YNL062C, YHR081W, YMR093W, YJL069C, YOR243C, YGR081C, YPR144C, YNL308C, YER126C, YNL182C, YKL082C, YMR229C, YER006W, YKR081C, YOL142W, YMR128W, YER002W, YNL232W, YNR054C, YDR037W, YDL060W |
| 7.58e-39 | organelle lumen | YCL054W, YLR129W, YOL010W, YPL093W, YPR137W, YDR324C, YDR449C, YLR276C, YMR239C, YHR088W, YLL008W, YGR159C, YNL248C, YGL171W, YJL033W, YNL124W, YOL077C, YLR186W, YPL217C, YFR001W, YNL151C, YOR310C, YKL099C, YDL031W, YKL144C, YHR072W-A, YKL009W, YOR206W, YHR170W, YGR128C, YHR148W, YHR066W, YPR190C, YJR002W, YDL148C, YHR197W, YOR340C, YHR143W-A, YKR060W, YPL211W, YKL021C, YNL075W, YJL109C, YEL026W, YGL111W, YBR247C, YGR245C, YDR087C, YHR031C, YGL078C, YLR221C, YOR207C, YHR081W, YMR093W, YJL069C, YJL076W, YGR081C, YPR144C, YNL308C, YER126C, YNL182C, YKL082C, YMR229C, YNR053C, YDR045C, YER006W, YKR081C, YOL142W, YHR052W, YCR072C, YMR128W, YER002W, YNL232W, YNR054C, YPL233W, YPL020C, YDL060W, YNL175C |
| 7.58e-39 | intracellular organelle lumen | YCL054W, YLR129W, YOL010W, YPL093W, YPR137W, YDR324C, YDR449C, YLR276C, YMR239C, YHR088W, YLL008W, YGR159C, YNL248C, YGL171W, YJL033W, YNL124W, YOL077C, YLR186W, YPL217C, YFR001W, YNL151C, YOR310C, YKL099C, YDL031W, YKL144C, YHR072W-A, YKL009W, YOR206W, YHR170W, YGR128C, YHR148W, YHR066W, YPR190C, YJR002W, YDL148C, YHR197W, YOR340C, YHR143W-A, YKR060W, YPL211W, YKL021C, YNL075W, YJL109C, YEL026W, YGL111W, YBR247C, YGR245C, YDR087C, YHR031C, YGL078C, YLR221C, YOR207C, YHR081W, YMR093W, YJL069C, YJL076W, YGR081C, YPR144C, YNL308C, YER126C, YNL182C, YKL082C, YMR229C, YNR053C, YDR045C, YER006W, YKR081C, YOL142W, YHR052W, YCR072C, YMR128W, YER002W, YNL232W, YNR054C, YPL233W, YPL020C, YDL060W, YNL175C |
| 1.14e-38 | membrane-enclosed lumen | YCL054W, YLR129W, YOL010W, YPL093W, YPR137W, YDR324C, YDR449C, YLR276C, YMR239C, YHR088W, YLL008W, YGR159C, YNL248C, YGL171W, YJL033W, YNL124W, YOL077C, YLR186W, YPL217C, YFR001W, YNL151C, YOR310C, YKL099C, YDL031W, YKL144C, YHR072W-A, YKL009W, YOR206W, YHR170W, YGR128C, YHR148W, YHR066W, YPR190C, YJR002W, YDL148C, YHR197W, YOR340C, YHR143W-A, YKR060W, YPL211W, YKL021C, YNL075W, YJL109C, YEL026W, YGL111W, YBR247C, YGR245C, YDR087C, YHR031C, YGL078C, YLR221C, YOR207C, YHR081W, YMR093W, YJL069C, YJL076W, YGR081C, YPR144C, YNL308C, YER126C, YFR011C, YNL182C, YKL082C, YMR229C, YNR053C, YDR045C, YER006W, YKR081C, YOL142W, YHR052W, YCR072C, YMR128W, YER002W, YNL232W, YNR054C, YPL233W, YPL020C, YDL060W, YNL175C |
| 2.48e-34 | nuclear part | YCL054W, YLR129W, YOL010W, YPL093W, YPR137W, YDR324C, YDR449C, YLR276C, YMR239C, YHR088W, YLL008W, YGR159C, YNL248C, YGL171W, YJL033W, YNL124W, YOL077C, YLR186W, YPL217C, YFR001W, YNL151C, YOR310C, YKL099C, YDL031W, YKL144C, |
| | | YHR072W-A, YKL009W, YOR206W, YHR170W, YGR128C, YHR148W, YHR066W, YPR190C, YJR002W, YDL148C, YHR197W, YOR340C, YHR143W-A, YKR060W, YPL211W, YKL021C, YNL075W, YJL109C, YEL026W, YGL111W, YBR247C, YGR245C, YDR087C, YHR031C, YGL078C, YLR221C, YOR207C, YHR081W, YMR093W, YJL069C, YJL076W, YGR081C, YPR144C, YNL308C, YER126C, YNL182C, YKL082C, YMR229C, YNR053C, YDR045C, YER006W, YKR081C, YOL142W, YHR052W, YCR072C, YMR128W, YER002W, YNL232W, YNR054C, YPL233W, YPL020C, YDL060W, YNL175C |
| 6.61e-32 | RNA processing | YCL054W, YLR129W, YOL010W, YPL093W, YPR137W, YDR324C, YDR449C, YLR276C, YMR239C, YHR088W, YLL008W, YGR159C, YGL171W, YJL033W, YNL124W, YLR186W, YPL217C, YFR001W, YDL166C, YOR310C, YKL099C, YDL031W, YHR072W-A, YGR128C, YHR148W, YJR002W, YDL148C, YHR197W, YKR060W, YPL211W, YKL021C, YNL075W, YJL109C, YEL026W, YGL111W, YBR247C, YDR087C, YGL078C, YNL062C, YHR081W, YMR093W, YJL069C, YOR243C, YGR081C, YPR144C, YNL308C, YER126C, YNL182C, YKL082C, YMR229C, YER006W, YKR081C, YOL142W, YMR128W, YER002W, YNL232W, YNR054C, YDL060W |
| 1.67e-31 | cellular component biogenesis | YCL054W, YLR129W, YOL010W, YPL093W, YPR137W, YDR324C, YDR449C, YLR276C, YMR239C, YHR088W, YLL008W, YGR159C, YGL171W, YJL033W, YNL124W, YOL077C, YLR186W, YPL217C, YFR001W, YDL166C, YOR310C, YKL099C, YDL031W, YHR072W-A, YKL009W, YOR206W, YHR170W, YGR128C, YHR148W, YBR267W, YHR066W, YJR002W, YDL148C, YHR197W, YKR060W, YPL211W, YKL021C, YNL075W, YJL109C, YEL026W, YGL111W, YBR247C, YGR245C, YDR087C, YGL078C, YLR221C, YOR276W, YHR081W, YMR093W, YJL069C, YOR243C, YIR026C, YGR081C, YPR144C, YNL308C, YER126C, YNL182C, YKL082C, YMR229C, YNR053C, YER006W, YKR081C, YOL142W, YHR052W, YCR072C, YMR128W, YER002W, YPL226W, YNL232W, YNR054C, YLR397C, YDL060W |
| 1.47e-28 | ribosomal large subunit biogenesis | YCL054W, YPL093W, YLR276C, YHR088W, YLL008W, YOL077C, YFR001W, YDL031W, YKL009W, YBR267W, YHR066W, YHR197W, YPL211W, YKL021C, YGL111W, YGR245C, YGL078C, YLR221C, YIR026C, YER126C, YNL182C, YKL082C, YMR229C, YKR081C, YHR052W, YCR072C, YER002W, YLR397C |
| 4.33e-27 | maturation of 5.8S rRNA from tricistronic rRNA transcript (SSU-rRNA, 5.8S rRNA, LSU-rRNA) | YCL054W, YLR129W, YOL010W, YDR449C, YHR088W, YGL171W, YLR186W, YFR001W, YOR310C, YKL099C, YDL031W, YJR002W, YDL148C, YKL021C, YJL109C, YBR247C, YGL078C, YHR081W, YJL069C, YPR144C, YNL308C, YER126C, YMR229C, YKR081C, YOL142W, YNL232W, YNR054C |
| 6.41e-27 | maturation of 5.8S rRNA | YCL054W, YLR129W, YOL010W, YDR449C, YHR088W, YGL171W, YLR186W, YFR001W, YOR310C, YKL099C, YDL031W, YJR002W, YDL148C, YKL021C, YJL109C, YBR247C, YGL078C, YHR081W, YJL069C, YPR144C, YNL308C, YER126C, YMR229C, YKR081C, YOL142W, YNL232W, YNR054C |
| 2.17e-25 | non-membrane-bounded organelle | YCL054W, YLR129W, YOL010W, YPL093W, YPR137W, YDR324C, YDR449C, YLR276C, YMR239C, YFR031C-A, YHR088W, YLL008W, YGR159C, YNL248C, YGL171W, YJL033W, YOL077C, YLR186W, YPL217C, YFR001W, YOR310C, YKL099C, YDL031W, YHR072W-A, YKL009W, YOR206W, YGR128C, YHR148W, YHR066W, YJR002W, YDL148C, YOR340C, YHR143W-A, YKR060W, YPL211W, YKL021C, YNL075W, YJL109C, YEL026W, YGL111W, YBR247C, YGR245C, YDR087C, YHR031C, YGL078C, YLR221C, YMR093W, YJL069C, YJL076W, YGR081C, YPR144C, YNL308C, YER126C, YKL082C, YMR229C, YNR053C, YER006W, YKR081C, YOL142W, YHR052W, YCR072C, YMR128W, YER002W, YNL232W, YNR054C, YPL233W, YPL020C, YDL060W, YNL175C |
| 2.17e-25 | intracellular non-membrane-bounded organelle | YCL054W, YLR129W, YOL010W, YPL093W, YPR137W, YDR324C, YDR449C, YLR276C, YMR239C, YFR031C-A, YHR088W, YLL008W, YGR159C, YNL248C, YGL171W, YJL033W, YOL077C, YLR186W, YPL217C, YFR001W, YOR310C, YKL099C, YDL031W, YHR072W-A, YKL009W, YOR206W, YGR128C, YHR148W, YHR066W, YJR002W, YDL148C, YOR340C, YHR143W-A, YKR060W, YPL211W, YKL021C, YNL075W, YJL109C, YEL026W, YGL111W, YBR247C, YGR245C, YDR087C, YHR031C, YGL078C, YLR221C, YMR093W, YJL069C, YJL076W, YGR081C, YPR144C, YNL308C, YER126C, YKL082C, YMR229C, YNR053C, YER006W, YKR081C, YOL142W, YHR052W, YCR072C, YMR128W, YER002W, YNL232W, YNR054C, YPL233W, YPL020C, YDL060W, YNL175C |
| 5.69e-25 | ribonucleoprotein complex | YCL054W, YLR129W, YOL010W, YPL093W, YPR137W, YDR324C, YDR449C, YFR031C-A, YHR088W, YLL008W, YGL171W, YNL124W, YOL077C, YLR186W, YPL217C, YFR001W, YOR310C, YKL099C, YDL031W, YHR072W-A, YKL009W, YOR206W, YGR128C, YHR148W, YJR002W, YDL148C, YKR060W, YPL211W, YNL075W, YJL109C, YEL026W, YGL111W, YBR247C, YDR087C, YGL078C, YLR221C, |

| P-value | Term | Protein |
|---|---|---|
| | | YMR093W, YJL069C, YGR081C, YPR144C, YER126C, YMR229C, YCL037C, YNR053C, YER006W, YKR081C, YHR052W, YMR128W, YER002W, YPL226W, YLR397C, YDL060W, YAL036C, YNL175C |
| 2.39e-24 | preribosome, large subunit precursor | YCL054W, YPL093W, YHR088W, YLL008W, YOL077C, YFR001W, YDL031W, YKL009W, YOR206W, YPL211W, YGL111W, YDR087C, YGL078C, YLR221C, YER126C, YNR053C, YER006W, YKR081C, YHR052W, YER002W, YLR397C |
| 5.36e-24 | nucleus | YCL054W, YLR129W, YOL010W, YPL093W, YPR137W, YDR324C, YDR449C, YLR276C, YMR239C, YHR088W, YLL008W, YGR159C, YNL248C, YGL171W, YJL033W, YNL124W, YOL077C, YLR186W, YPL217C, YFR001W, YDL166C, YNL151C, YOR310C, YKL099C, YDL031W, YKL144C, YDR399W, YHR072W-A, YKL009W, YOR206W, YHR170W, YGR128C, YHR148W, YHR066W, YPR190C, YJR002W, YDL148C, YHR197W, YOR340C, YHR143W-A, YKR060W, YPL211W, YKL021C, YNL075W, YJL109C, YEL026W, YGL111W, YBR247C, YGR245C, YDR087C, YHR031C, YGL078C, YNL062C, YLR221C, YOR207C, YHR081W, YMR093W, YJL069C, YJL076W, YOR243C, YIR026C, YGR081C, YPR144C, YNL308C, YER126C, YNL182C, YKL082C, YMR229C, YNR053C, YDR045C, YMR310C, YER006W, YML021C, YKR081C, YOL142W, YHR052W, YCR072C, YMR128W, YER002W, YPL226W, YNL232W, YNR054C, YPL233W, YPL020C, YPR189W, YDL060W, YNL175C |
| 2.08e-22 | ribosomal small subunit biogenesis | YLR129W, YOL010W, YDR324C, YDR449C, YGR159C, YGL171W, YLR186W, YFR001W, YDL166C, YOR310C, YKL099C, YGR128C, YJR002W, YDL148C, YKR060W, YJL109C, YEL026W, YBR247C, YMR093W, YJL069C, YGR081C, YPR144C, YNL308C, YKL082C, YMR229C, YMR128W, YPL226W, YNR054C |
| 3.22e-22 | RNA metabolic process | YCL054W, YLR129W, YOL010W, YPL093W, YPR137W, YDR324C, YDR449C, YLR276C, YMR239C, YHR088W, YLL008W, YGR159C, YNL248C, YGL171W, YJL033W, YNL124W, YLR186W, YPL217C, YFR001W, YDL166C, YNL151C, YOR310C, YKL099C, YDL031W, YKL144C, YHR072W-A, YKL009W, YGR128C, YHR148W, YPR190C, YJR002W, YDL148C, YHR197W, YOR340C, YHR143W-A, YKR060W, YPL211W, YKL021C, YNL075W, YJL109C, YEL026W, YGL111W, YBR247C, YDR087C, YGL078C, YNL062C, YOR207C, YHR081W, YMR093W, YJL069C, YJL076W, YOR243C, YGR081C, YPR144C, YNL308C, YER126C, YNL182C, YKL082C, YMR229C, YDR045C, YER006W, YKR081C, YOL142W, YMR128W, YER002W, YNL232W, YNR054C, YDR037W, YPR189W, YDL060W |
| 3.68e-22 | 90S preribosome | YLR129W, YOL010W, YPR137W, YDR324C, YDR449C, YGL171W, YLR186W, YPL217C, YOR310C, YOR206W, YGR128C, YHR148W, YJR002W, YDL148C, YKR060W, YNL075W, YJL109C, YBR247C, YMR093W, YJL069C, YGR081C, YPR144C, YMR229C, YMR128W, YDL060W |
| 6.62e-21 | cleavage involved in rRNA processing | YLR129W, YOL010W, YDR449C, YGL171W, YLR186W, YFR001W, YOR310C, YKL099C, YHR072W-A, YJR002W, YDL148C, YJL109C, YBR247C, YGL078C, YHR081W, YJL069C, YPR144C, YNL308C, YMR229C, YOL142W, YNL232W, YNR054C |
| 9.54e-21 | RNA phosphodiester bond hydrolysis | YLR129W, YOL010W, YDR449C, YGL171W, YLR186W, YFR001W, YOR310C, YKL099C, YHR072W-A, YJR002W, YDL148C, YJL109C, YBR247C, YGL078C, YHR081W, YJL069C, YPR144C, YNL308C, YMR229C, YOL142W, YNL232W, YNR054C |
| 4.42e-20 | maturation of SSU-rRNA from tricistronic rRNA transcript (SSU-rRNA, 5.8S rRNA, LSU-rRNA) | YLR129W, YOL010W, YDR324C, YDR449C, YGL171W, YLR186W, YFR001W, YDL166C, YOR310C, YKL099C, YGR128C, YJR002W, YDL148C, YJL109C, YEL026W, YBR247C, YMR093W, YJL069C, YGR081C, YPR144C, YNL308C, YMR229C, YMR128W, YNR054C |
| **B.** | **Significantly enriched terms in the top 100 set of proteins having *largest decrease* of perturbation centrality in *oxidative stress*** | |
| **P-value** | **Term** | **Protein** |
| 6.62e-38 | ribosome biogenesis | YMR239C, YKL099C, YPL217C, YDL148C, YGL171W, YDR324C, YNL308C, YOL077C, YLR186W, YDR299W, YKR060W, YPR137W, YGL111W, YPL211W, YHR088W, YGL078C, YMR229C, YKL082C, YDL166C, YGR128C, YPL226W, YLR129W, YOR206W, YPR144C, YLR002C, YBR267W, YFR001W, YHR066W, YHR148W, YNL182C, YLR409C, YLR221C, YJR002W, YMR093W, YLL008W, YOR078W, YNL124W, YNR054C, YHR196W, YLR276C, YFL002C, YER002W, YJL069C, YHR170W, YOR243C, YDR398W, YPR112C, YNL075W, YHR081W, YKL021C, YGL099W, YHR197W, YBR247C, YKR024C, YHR052W, YNL224C |
| 5.98e-37 | preribosome | YKL099C, YPL217C, YDL148C, YGL171W, YDR324C, YOL077C, YLR186W, YDR299W, YKR060W, YPR137W, YGL111W, YPL211W, YHR088W, YGL078C, YMR229C, YGR128C, YLR129W, YOR206W, YPR144C, YLR002C, YFR001W, YHR148W, YLR409C, YLR221C, YJR002W, YMR093W, YLL008W, YOR078W, YHR196W, YFL002C, YER002W, YJL069C, YNL175C, YDR398W, YPR112C, YNL075W, YBR247C, YHR052W, YNL224C |
| 1.99e-35 | ribonucleoprotein complex biogenesis | YMR239C, YKL099C, YPL217C, YDL148C, YGL171W, YDR324C, YNL308C, YOL077C, YLR186W, YDR299W, YKR060W, YPR137W, YGL111W, YPL211W, YHR088W, YGL078C, YMR229C, YKL082C, YDL166C, YGR128C, YPL226W, YLR129W, YOR206W, YPR144C, YLR002C, YBR267W, YFR001W, YHR066W, YHR148W, YNL182C, YLR409C, YLR221C, YJR002W, YMR093W, YLL008W, YOR078W, YNL124W, YNR054C, YHR196W, YLR276C, YFL002C, YER002W, YJL069C, YHR170W, YOR243C, YDR398W, YPR112C, YNL075W, YIR005W, YHR081W, YKL021C, YGL099W, YHR197W, YBR247C, YKR024C, YHR052W, YNL224C |
| 4.58e-34 | nucleolus | YMR239C, YKL099C, YPL217C, YDL148C, YGL171W, YNL248C, YDR324C, YNL308C, YOL077C, YLR186W, YDR299W, YKR060W, YPR137W, YGL111W, YPL211W, YHR088W, YGL078C, YMR229C, YKL082C, YGR128C, YLR129W, YOR206W, YPR144C, YLR002C, YEL055C, YFR001W, YHR066W, YHR148W, YLR409C, YLR221C, YOR340C, YJR002W, YMR093W, YLL008W, YOR078W, YNR054C, YHR196W, YLR276C, YFL002C, YER002W, YJL069C, YNL175C, YDR398W, YPR112C, YNL075W, YKL021C, YBR247C, YKR024C, YNL113W, YHR052W |
| 2.38e-32 | rRNA processing | YMR239C, YKL099C, YPL217C, YDL148C, YGL171W, YDR324C, YNL308C, YLR186W, YDR299W, YKR060W, YPR137W, YGL111W, YPL211W, YHR088W, YGL078C, YMR229C, YKL082C, YDL166C, YGR128C, YLR129W, YPR144C, YLR002C, YFR001W, YHR148W, YNL182C, YLR409C, YJR002W, YMR093W, YLL008W, YOR078W, YNL124W, YNR054C, YHR196W, YLR276C, YFL002C, YER002W, YJL069C, YOR243C, YDR398W, YPR112C, YNL075W, YHR081W, YKL021C, YHR197W, YBR247C, YKR024C, YNL224C |
| 1.39e-31 | rRNA metabolic process | YMR239C, YKL099C, YPL217C, YDL148C, YGL171W, YDR324C, YNL308C, YLR186W, YDR299W, YKR060W, YPR137W, YGL111W, YPL211W, YHR088W, YGL078C, YMR229C, YKL082C, YDL166C, YGR128C, YLR129W, YPR144C, YLR002C, YFR001W, YHR148W, YNL182C, YLR409C, YJR002W, YMR093W, YLL008W, YOR078W, YNL124W, YNR054C, YHR196W, YLR276C, YFL002C, YER002W, YJL069C, YOR243C, YDR398W, YPR112C, YNL075W, YHR081W, YKL021C, YHR197W, YBR247C, YKR024C, YNL224C |
| 1.82e-30 | cellular component biogenesis at cellular level | YMR239C, YKL099C, YPL217C, YDL148C, YGL171W, YDR324C, YNL308C, YOL077C, YLR186W, YDR299W, YKR060W, YPR137W, YGL111W, YPL211W, YHR088W, YGL078C, YMR229C, YKL082C, YDL166C, YGR128C, YPL226W, YLR129W, YOR206W, YPR144C, YLR002C, YBR267W, YFR001W, YHR066W, YHR148W, YNL182C, YLR409C, YLR221C, YJR002W, YMR093W, YLL008W, YOR078W, YNL124W, YNR054C, YHR196W, YLR276C, YFL002C, YER002W, YJL069C, YHR170W, YOR243C, YDR398W, YPR112C, YNL075W, YIR005W, YHR081W, YKL021C, YGL099W, YHR197W, YBR247C, YKR024C, YHR052W, YNL224C |
| 9.66e-30 | nuclear lumen | YMR239C, YKL099C, YPL217C, YDL148C, YGL171W, YNL248C, YDR324C, YNL308C, YOL077C, YLR186W, YNL151C, YPR190C, YDR299W, YKR060W, YPR137W, YGL111W, YPL211W, YHR088W, YGL078C, YMR229C, YKL082C, YGR128C, YLR129W, YKL144C, YDR227W, YOR206W, YPR144C, YLR002C, YEL055C, YFR001W, YHR066W, YHR148W, YNL182C, YLR409C, YLR221C, YOR340C, YJR002W, YMR093W, YLL008W, YOR078W, YNL124W, YNR054C, YHR196W, YLR276C, YFL002C, YER002W, YMR176W, YJL069C, YHR170W, YNL175C, YDR398W, YPR112C, YNL075W, YHR081W, YKL021C, YBR289W, YHR197W, YBR247C, YKR024C, YNL113W, YOL012C, YHR052W, YOR207C |
| 1.34e-26 | ncRNA processing | YMR239C, YKL099C, YPL217C, YDL148C, YGL171W, YDR324C, YNL308C, YLR186W, YDR299W, YKR060W, YPR137W, YGL111W, YPL211W, YHR088W, YGL078C, YMR229C, YKL082C, YDL166C, YGR128C, YLR129W, YPR144C, YLR002C, YFR001W, YHR148W, YNL182C, YLR409C, YJR002W, YMR093W, YLL008W, YOR078W, YNL124W, YNR054C, YHR196W, YLR276C, YFL002C, YER002W, YJL069C, YOR243C, YDR398W, YPR112C, YNL075W, YHR081W, YKL021C, YHR197W, YBR247C, YKR024C, YNL224C |
| 3.12e-25 | ncRNA metabolic process | YMR239C, YKL099C, YPL217C, YDL148C, YGL171W, YDR324C, YNL308C, YLR186W, YDR299W, YKR060W, YPR137W, YGL111W, YPL211W, YHR088W, YOL097C, YGL078C, YMR229C, YKL082C, YDL166C, YGR128C, YLR129W, YPR144C, YLR002C, YFR001W, YHR148W, YNL182C, YLR409C, YJR002W, YMR093W, YLL008W, YOR078W, YNL124W, YNR054C, YHR196W, YLR276C, YFL002C, YER002W, YJL069C, YOR243C, YDR398W, YPR112C, YNL075W, YHR081W, YKL021C, YHR197W, YBR247C, YKR024C, YNL224C |
| 1.01e-23 | 90S preribosome | YPL217C, YDL148C, YGL171W, YDR324C, YLR186W, YDR299W, YKR060W, YPR137W, YMR229C, YGR128C, YLR129W, YOR206W, YPR144C, YHR148W, YLR409C, YJR002W, YMR093W, YOR078W, YHR196W, YFL002C, YJL069C, YDR398W, YPR112C, YNL075W, YBR247C, YNL224C |
| 1.2e-22 | organelle lumen | YMR239C, YKL099C, YPL217C, YDL148C, YGL171W, YNL248C, YDR324C, YNL308C, YOL077C, YLR186W, YNL151C, YPR190C, YDR299W, YKR060W, YPR137W, YGL111W, YPL211W, YHR088W, YGL078C, YMR229C, YKL082C, YGR128C, YLR129W, YKL144C, YDR227W, YOR206W, YPR144C, YLR002C, YEL055C, YFR001W, YHR066W, YHR148W, YNL182C, YLR409C, YLR221C, YOR340C, YJR002W, |

| P-value | Term | Proteins |
|---|---|---|
| | | YMR093W, YLL008W, YOR078W, YNL124W, YNR054C, YHR196W, YLR276C, YFL002C, YER002W, YMR176W, YJL069C, YHR170W, YNL175C, YDR398W, YPR112C, YNL075W, YHR081W, YKL021C, YBR289W, YHR197W, YBR247C, YKR024C, YNL113W, YOL012C, YHR052W, YOR207C |
| 1.2e-22 | intracellular organelle lumen | YMR239C, YKL099C, YPL217C, YDL148C, YGL171W, YNL248C, YDR324C, YNL308C, YOL077C, YLR186W, YNL151C, YPR190C, YDR299W, YKR060W, YPR137W, YGL111W, YPL211W, YHR088W, YGL078C, YMR229C, YKL082C, YGR128C, YLR129W, YKL144C, YDR227W, YOR206W, YPR144C, YLR002C, YEL055C, YFR001W, YHR066W, YHR148W, YNL182C, YLR409C, YLR221C, YOR340C, YJR002W, YMR093W, YLL008W, YOR078W, YNL124W, YNR054C, YHR196W, YLR276C, YFL002C, YER002W, YMR176W, YJL069C, YHR170W, YNL175C, YDR398W, YPR112C, YNL075W, YHR081W, YKL021C, YBR289W, YHR197W, YBR247C, YKR024C, YNL113W, YOL012C, YHR052W, YOR207C |
| 1.64e-22 | membrane-enclosed lumen | YMR239C, YKL099C, YPL217C, YDL148C, YGL171W, YNL248C, YDR324C, YNL308C, YOL077C, YLR186W, YNL151C, YPR190C, YDR299W, YKR060W, YPR137W, YGL111W, YPL211W, YHR088W, YGL078C, YMR229C, YKL082C, YGR128C, YLR129W, YKL144C, YDR227W, YOR206W, YPR144C, YLR002C, YEL055C, YFR001W, YHR066W, YHR148W, YNL182C, YLR409C, YLR221C, YOR340C, YJR002W, YMR093W, YLL008W, YOR078W, YNL124W, YNR054C, YHR196W, YLR276C, YFL002C, YER002W, YMR176W, YJL069C, YHR170W, YNL175C, YDR398W, YPR112C, YNL075W, YFR011C, YHR081W, YKL021C, YBR289W, YHR197W, YBR247C, YKR024C, YNL113W, YOL012C, YHR052W, YOR207C |
| 4.57e-21 | nuclear part | YMR239C, YKL099C, YPL217C, YDL148C, YGL171W, YNL248C, YDR324C, YNL308C, YOL077C, YLR186W, YNL151C, YPR190C, YDR299W, YKR060W, YPR137W, YGL111W, YPL211W, YHR088W, YGL078C, YMR229C, YKL082C, YGR128C, YLR129W, YKL144C, YDR227W, YOR206W, YPR144C, YLR002C, YEL055C, YFR001W, YHR066W, YHR148W, YNL182C, YLR409C, YLR221C, YOR340C, YJR002W, YMR093W, YLL008W, YOR078W, YNL124W, YNR054C, YHR196W, YLR276C, YFL002C, YER002W, YMR176W, YJL069C, YHR170W, YNL175C, YDR398W, YPR112C, YDR303C, YNL075W, YIR005W, YHR081W, YKL021C, YBR289W, YHR197W, YBR247C, YKR024C, YNL113W, YOL012C, YHR052W, YOR207C |
| 2.28e-20 | RNA processing | YMR239C, YKL099C, YPL217C, YDL148C, YGL171W, YDR324C, YNL308C, YLR186W, YDR299W, YKR060W, YPR137W, YGL111W, YPL211W, YHR088W, YGL078C, YMR229C, YKL082C, YDL166C, YGR128C, YLR129W, YPR144C, YLR002C, YFR001W, YHR148W, YNL182C, YLR409C, YJR002W, YMR093W, YLL008W, YOR078W, YNL124W, YNR054C, YHR196W, YLR276C, YFL002C, YER002W, YJL069C, YOR243C, YDR398W, YPR112C, YNL075W, YIR005W, YHR081W, YKL021C, YHR197W, YBR247C, YKR024C, YNL224C |
| **C.** | **Significantly enriched terms in the top 100 set of proteins having *largest decrease* of perturbation centrality in *osmotic stress*** | |
| **P-value** | **Term** | **Proteins** |
| 1.29e-40 | ribosome biogenesis | YGL111W, YNL308C, YMR239C, YGL078C, YHR066W, YOL077C, YFL002C, YOR206W, YKL172W, YLR276C, YLR002C, YDL166C, YMR229C, YIR026C, YDL148C, YDR299W, YDR324C, YKL009W, YKL099C, YCL054W, YNL182C, YGR103W, YBR267W, YLR221C, YLR129W, YOR243C, YNL124W, YGR128C, YHR170W, YGR081C, YPL226W, YOR294W, YPL211W, YHR052W, YPR102C, YHR088W, YPR137W, YKL021C, YHR197W, YKR081C, YHR072W-A, YFR001W, YDR091C, YHR148W, YNL061W, YDR087C, YGL099W, YLR409C, YPR112C, YNL002C, YOL010W, YNL207W, YPR144C, YNL075W, YPL126W, YLR186W, YNL224C, YGR159C |
| 2.56e-39 | ribonucleoprotein complex biogenesis | YGL111W, YNL308C, YMR239C, YGL078C, YHR066W, YOL077C, YFL002C, YOR206W, YKL172W, YLR276C, YLR002C, YDL166C, YMR229C, YIR026C, YDL148C, YDR299W, YDR324C, YKL009W, YKL099C, YCL054W, YNL182C, YGR103W, YBR267W, YLR221C, YIR005W, YLR129W, YOR243C, YNL124W, YGR128C, YHR170W, YGR081C, YPL226W, YOR294W, YPL211W, YHR052W, YPR102C, YHR088W, YPR137W, YKL021C, YHR197W, YKR081C, YHR072W-A, YFR001W, YDR091C, YHR148W, YNL061W, YDR087C, YGL099W, YGR178C, YLR409C, YPR112C, YNL002C, YOL010W, YNL207W, YPR144C, YNL075W, YPL126W, YLR186W, YNL224C, YGR159C |
| 5.54e-37 | preribosome | YGL111W, YGL078C, YOL077C, YFL002C, YOR206W, YKL172W, YLR002C, YMR229C, YDL148C, YDR299W, YDR324C, YKL009W, YKL099C, YCL054W, YNL175C, YGR103W, YLR221C, YLR129W, YGR128C, YGR081C, YOR294W, YPL211W, YHR052W, YHR088W, YPR137W, YKR081C, YFR001W, YHR148W, YNL061W, YDR087C, YLR409C, YPR112C, YNL002C, YOL010W, YPR144C, YNL075W, YPL126W, YLR186W, YNL224C |
| 2.1e-35 | nucleolus | YGL111W, YNL308C, YMR239C, YGL078C, YNL248C, YHR066W, YOL077C, YFL002C, YOR206W, YKL172W, YEL055C, YLR276C, YLR002C, YMR229C, YDL148C, YDR299W, YDR324C, YKL009W, YKL099C, YCL054W, YNL175C, YGR103W, YLR221C, YOR340C, YLR129W, |
| | | YGR128C, YGR081C, YOR294W, YPL211W, YHR052W, YHR088W, YPR137W, YKL021C, YKR081C, YHR072W-A, YFR001W, YHR143W-A, YHR148W, YNL061W, YDR087C, YPL020C, YLR409C, YPR112C, YNL002C, YOL010W, YPR144C, YNL075W, YPL126W, YLR186W, YGR159C, YNL113W |
| 4.88e-34 | cellular component biogenesis at cellular level | YGL111W, YNL308C, YMR239C, YGL078C, YHR066W, YOL077C, YFL002C, YOR206W, YKL172W, YLR276C, YLR002C, YDL166C, YMR229C, YIR026C, YDL148C, YDR299W, YDR324C, YKL009W, YKL099C, YCL054W, YNL182C, YGR103W, YBR267W, YLR221C, YIR005W, YLR129W, YOR243C, YNL124W, YGR128C, YHR170W, YGR081C, YPL226W, YOR294W, YPL211W, YHR052W, YPR102C, YHR088W, YPR137W, YKL021C, YHR197W, YKR081C, YHR072W-A, YFR001W, YDR091C, YHR148W, YNL061W, YDR087C, YGL099W, YGR178C, YLR409C, YPR112C, YNL002C, YOL010W, YNL207W, YPR144C, YNL075W, YPL126W, YLR186W, YNL224C, YGR159C |
| 2.21e-32 | rRNA processing | YGL111W, YNL308C, YMR239C, YGL078C, YFL002C, YKL172W, YLR276C, YLR002C, YDL166C, YMR229C, YDL148C, YDR299W, YDR324C, YKL009W, YKL099C, YCL054W, YNL182C, YGR103W, YLR129W, YOR243C, YNL124W, YGR128C, YGR081C, YOR294W, YPL211W, YHR088W, YPR137W, YKL021C, YHR197W, YKR081C, YHR072W-A, YFR001W, YDR091C, YHR148W, YNL061W, YDR087C, YLR409C, YPR112C, YNL002C, YOL010W, YNL207W, YPR144C, YNL075W, YPL126W, YLR186W, YNL224C, YGR159C |
| 1.29e-31 | rRNA metabolic process | YGL111W, YNL308C, YMR239C, YGL078C, YFL002C, YKL172W, YLR276C, YLR002C, YDL166C, YMR229C, YDL148C, YDR299W, YDR324C, YKL009W, YKL099C, YCL054W, YNL182C, YGR103W, YLR129W, YOR243C, YNL124W, YGR128C, YGR081C, YOR294W, YPL211W, YHR088W, YPR137W, YKL021C, YHR197W, YKR081C, YHR072W-A, YFR001W, YDR091C, YHR148W, YNL061W, YDR087C, YLR409C, YPR112C, YNL002C, YOL010W, YNL207W, YPR144C, YNL075W, YPL126W, YLR186W, YNL224C, YGR159C |
| 1.67e-27 | ncRNA metabolic process | YGL111W, YNL308C, YMR239C, YGL078C, YFL002C, YKL172W, YLR276C, YLR002C, YDL166C, YMR229C, YOL097C, YDL148C, YDR299W, YDR324C, YKL009W, YKL099C, YCL054W, YNL182C, YGR102W, YGR103W, YLR129W, YOR243C, YNL124W, YGR128C, YGR081C, YOR294W, YPL211W, YHR088W, YPR137W, YKL021C, YHR197W, YKR081C, YHR072W-A, YFR001W, YDR091C, YHR148W, YNL061W, YDR087C, YLR409C, YPR112C, YNL002C, YOL010W, YNL207W, YPR144C, YNL075W, YPL126W, YLR186W, YNL224C, YHR019C, YGR159C |
| 1.24e-26 | ncRNA processing | YGL111W, YNL308C, YMR239C, YGL078C, YFL002C, YKL172W, YLR276C, YLR002C, YDL166C, YMR229C, YDL148C, YDR299W, YDR324C, YKL009W, YKL099C, YCL054W, YNL182C, YGR103W, YLR129W, YOR243C, YNL124W, YGR128C, YGR081C, YOR294W, YPL211W, YHR088W, YPR137W, YKL021C, YHR197W, YKR081C, YHR072W-A, YFR001W, YDR091C, YHR148W, YNL061W, YDR087C, YLR409C, YPR112C, YNL002C, YOL010W, YNL207W, YPR144C, YNL075W, YPL126W, YLR186W, YNL224C, YGR159C |
| 1.69e-25 | nuclear lumen | YGL111W, YNL308C, YMR239C, YGL078C, YNL248C, YHR066W, YOL077C, YFL002C, YOR206W, YKL172W, YEL055C, YLR276C, YLR002C, YMR229C, YDL148C, YDR299W, YDR324C, YKL009W, YKL099C, YCL054W, YNL182C, YNL175C, YGR103W, YLR221C, YOR340C, YLR129W, YNL124W, YKL144C, YGR128C, YHR170W, YGR081C, YOR294W, YPL211W, YHR052W, YHR088W, YPR137W, YKL021C, YHR197W, YKR081C, YHR072W-A, YFR001W, YHR143W-A, YHR148W, YNL061W, YDR087C, YPL020C, YCR092C, YLR409C, YPR112C, YNL002C, YOL010W, YPR144C, YNL075W, YPL126W, YLR186W, YMR167W, YHR031C, YGR159C, YNL113W |
| 2.12e-25 | ribosomal large subunit biogenesis | YGL111W, YGL078C, YHR066W, YOL077C, YFL002C, YLR276C, YMR229C, YIR026C, YKL009W, YCL054W, YNL182C, YGR103W, YBR267W, YLR221C, YOR294W, YPL211W, YHR052W, YPR102C, YHR088W, YKL021C, YHR197W, YKR081C, YFR001W, YDR091C, YGL099W, YNL002C |
| 1.41e-22 | preribosome, large subunit precursor | YGL111W, YGL078C, YOL077C, YOR206W, YKL172W, YLR002C, YKL009W, YCL054W, YGR103W, YLR221C, YOR294W, YPL211W, YHR052W, YHR088W, YKR081C, YFR001W, YNL061W, YDR087C, YNL002C, YNL224C |
| 2.19e-21 | RNA processing | YGL111W, YNL308C, YMR239C, YGL078C, YFL002C, YKL172W, YLR276C, YLR002C, YDL166C, YMR229C, YDL148C, YDR299W, YDR324C, YKL009W, YKL099C, YCL054W, YNL182C, YGR103W, YIR005W, YLR129W, YOR243C, YNL124W, YGR128C, YGR081C, YOR294W, YPL211W, YHR088W, YPR137W, YKL021C, YHR197W, YKR081C, YHR072W-A, YFR001W, YDR091C, YHR148W, YNL061W, YDR087C, YGR178C, YLR409C, YPR112C, YNL002C, YOL010W, YNL207W, YPR144C, YNL075W, YPL126W, YLR186W, YNL224C, YGR159C |
| 3.37e-20 | ribonucleoprotein complex | YGL111W, YGL078C, YOL077C, YFL002C, YOR206W, YKL172W, YLR002C, YMR229C, YDL148C, YDR299W, YDR324C, YKL009W, YKL099C, YCL054W, YNL175C, YGR103W, YLR221C, YIR005W, YLR129W, YNL124W, YGR128C, YGR081C, YPL226W, YGR054W, YOR294W, YPL211W, YHR052W, YPR102C, YHR088W, YPR137W, YKR081C, YFR031C-A, YHR072W-A, YFR001W, YDR091C, YAL035W, YHR148W, YNL061W, YDR087C, YGR178C, YLR409C, YPR112C, YNL002C, YOL010W, YPR144C, YNL075W, YPL126W, YLR186W, YNL224C |
| 9.28e-20 | cellular component | YGL111W, YNL308C, YMR239C, YGL078C, YHR066W, YOL077C, YFL002C, YOR206W, YKL172W, YLR276C, YLR002C, YDL166C, |

| | biogenesis | YMR229C, YIR026C, YDL148C, YDR299W, YDR324C, YKL009W, YKL099C, YCL054W, YNL182C, YGR103W, YBR267W, YLR221C, YIR005W, YLR129W, YOR243C, YNL124W, YGR128C, YHR170W, YGR081C, YPL226W, YOR294W, YPL211W, YHR052W, YPR102C, YHR088W, YPR137W, YKL021C, YHR197W, YKR081C, YHR072W-A, YFR001W, YDR091C, YHR148W, YNL061W, YDR087C, YGL099W, YGR178C, YLR409C, YPR112C, YNL002C, YOL010W, YNL207W, YPR144C, YNL075W, YPL126W, YLR186W, YNL224C, YGR159C |
|---|---|---|

Perturbation centralities were calculated with the Turbine software as described in **Methods** of the main text. Term enrichment analysis was performed with the R plug-in of g:Profiler [12], which returns both the enriched terms, and the proteins connected with the term. A term was stated as statistically significant, if the resulting p-value was strictly less than 0.05 after applying Bonferroni correction. The list was cut at $p=10^{-20}$ for brevity of the table. The full list can be obtained by running the „stressed_profile.R" script available at the Turbine web-site http://turbine.linkgroup.hu. The data suggests a very strong down-regulation of ribosome biogenesis and protein translation in all types of stress, which is a well-studied change in stress [16–18] successfully identified by the perturbation centrality measure.

**Table S6. Run-time and memory usage of the Turbine program during simulation of different synthetic networks**

| Network[1] | Number of nodes | Number of links | Run-time[2] | Peak memory usage | Disk space usage | Data-on-disk[3] |
|---|---|---|---|---|---|---|
| BA-100/2 | 100 | 200 | 0.03 s | 4,452 kB | 796 kB | no |
| BA-1K/2 | 1,000 | 2,000 | 0.33 s | 11,892 kB | 7,908 kB | no |
| BA-10K/2 | 10,000 | 20,000 | 4.16 s | 86,008 kB | 78,980 kB | no |
| BA-100K/2 | 100,000 | 200,000 | 50.24 s | 828,292 kB | 771 MB | no |
| BA-100K/2 | 100,000 | 200,000 | 51.36 s | 45,476 kB | 771 MB | yes |
| BA-1M/2 | 1,000,000 | 2,000,000 | 8 m 32 s | 421,640 kB | 7713 MB | yes |
| BA-1M/6 | 1,000,000 | 6,000,000 | 14 m 3 s | 921,668 kB | 7797 MB | yes |

[1]Networks were generated with the "grown" plug-in of the *netgen* tool of the Turbine toolkit. This plug-in creates Barabási–Albert-type scale-free networks given the total requested number of nodes and the degree of newly added nodes.

[2]Run-times are reported as measured by Turbine. This is only the raw simulation time, without accounting for loading and saving the file to/from the hard disk in the cases, where data-on-disk[3] mode is switched off. 1000 time steps were simulated with the communicating vessels model, on a server with a Xeon 3.6 GHz processor, using only one CPU. Column-major matrix ordering was used for better performance. The run-time is linear function of both the number of nodes and the simulation time. Note that the calculation of the perturbation centrality measure for all nodes in a network requires a separate simulation for each node, thus having a quadratic computational complexity with respect to the number of nodes.

[3]Data-on-disk mode is a special feature of Turbine enabling long simulations that would otherwise not fit in the computer memory. In this mode, data files are accessed directly on the disk without consuming any memory, at the price of lower access speed. The performance penalty for using data-on-disk mode is actually surprisingly small according to the data in the Table. Using row-major ordering with data-on-disk mode leads to lower performance, because sequential writes in each time step will become random writes on the disk.

**Table S7. Details of the used realizations of modular benchmark graphs**

| A. Networks used for testing fuzzy versus pronounced modules (all except the ITM-Probe series) | | | | | |
|---|---|---|---|---|---|
| Pre-set ratio of inter-modular links | Random seed | Number of links | Number of intra-modular links | Number of inter-modular links | Overlap[a] |
| 5% | 10 | 13491 | 12819 | 672 | No |
| 5% | 10 | 13695 | 13326 | 369 | Yes |
| 5% | 19 | 13583 | 12872 | 711 | No |
| 5% | 19 | 13755 | 13614 | 141 | Yes |
| 5% | 20 | 13938 | 13237 | 701 | No |
| 5% | 20 | 13583 | 13392 | 191 | Yes |
| 5% | 42 | 13966 | 13263 | 703 | No |
| 5% | 42 | 14206 | 13942 | 264 | Yes |
| 5% | 85 | 13619 | 12919 | 700 | No |
| 5% | 85 | 13908 | 13830 | 78 | Yes |
| 5% | 87 | 13772 | 13086 | 686 | No |
| 5% | 87 | 13884 | 13595 | 289 | Yes |
| 5% | 88 | 13818 | 13134 | 684 | No |
| 5% | 88 | 13772 | 13445 | 327 | Yes |
| 40% | 10 | 13491 | 8111 | 5380 | No |
| 40% | 10 | 13695 | 9590 | 4105 | Yes |
| 40% | 19 | 13583 | 8140 | 5443 | No |
| 40% | 19 | 13755 | 10444 | 3311 | Yes |
| 40% | 20 | 13938 | 8361 | 5577 | No |
| 40% | 20 | 13583 | 10052 | 3531 | Yes |
| 40% | 42 | 13966 | 8383 | 5583 | No |
| 40% | 42 | 14206 | 10331 | 3875 | Yes |
| 40% | 85 | 13619 | 8170 | 5449 | No |
| 40% | 85 | 13908 | 10746 | 3162 | Yes |
| 40% | 87 | 13772 | 8269 | 5503 | No |
| 40% | 87 | 13887 | 9913 | 3974 | Yes |
| 40% | 88 | 13818 | 8292 | 5526 | No |
| 40% | 88 | 13772 | 9748 | 4024 | Yes |
| B. Networks used in comparing ITM-Probe and Turbine with different fuzziness values | | | | | |
| Pre-set ratio of inter-modular links | Random seed | Number of links | Number of intra-modular links | Number of inter-modular links | Overlap |
| 5% | 59 | 7717 | 7319 | 398 | No |
| 5% | 87 | 7824 | 7429 | 395 | No |
| 5% | 88 | 7788 | 7395 | 393 | No |
| 10% | 59 | 7692 | 6916 | 776 | No |
| 10% | 87 | 7837 | 7054 | 783 | No |
| 10% | 88 | 7820 | 7035 | 785 | No |
| 15% | 59 | 7755 | 6581 | 1174 | No |
| 15% | 87 | 7837 | 6653 | 1184 | No |
| 15% | 88 | 7827 | 6653 | 1174 | No |
| 20% | 59 | 7780 | 6213 | 1567 | No |
| 20% | 87 | 7852 | 6279 | 1573 | No |
| 20% | 88 | 7850 | 6269 | 1581 | No |
| 25% | 59 | 7795 | 5844 | 1951 | No |
| 25% | 87 | 7850 | 5891 | 1959 | No |
| 25% | 88 | 7824 | 5861 | 1963 | No |
| 30% | 59 | 7776 | 5438 | 2338 | No |
| 30% | 87 | 7849 | 5485 | 2364 | No |
| 30% | 88 | 7857 | 5499 | 2358 | No |
| 35% | 59 | 7803 | 5059 | 2744 | No |
| 35% | 87 | 7850 | 5103 | 2747 | No |
| 35% | 88 | 7857 | 5102 | 2755 | No |
| 40% | 59 | 7792 | 4670 | 3122 | No |
| 40% | 87 | 7848 | 4710 | 3138 | No |
| 40% | 88 | 7858 | 4710 | 3148 | No |
| 45% | 59 | 7805 | 4292 | 3513 | No |
| 45% | 87 | 7845 | 4309 | 3536 | No |
| 45% | 88 | 7849 | 4305 | 3544 | No |
| 50% | 59 | 7794 | 3885 | 3909 | No |
| 50% | 87 | 7850 | 3923 | 3927 | No |
| 50% | 88 | 7858 | 3937 | 3921 | No |
| 55% | 59 | 7805 | 3509 | 4296 | No |
| 55% | 87 | 7848 | 3528 | 4320 | No |
| 55% | 88 | 7850 | 3524 | 4326 | No |
| 60% | 59 | 7805 | 3124 | 4681 | No |
| 60% | 87 | 7850 | 3127 | 4723 | No |
| 60% | 88 | 7858 | 3147 | 4711 | No |
| 65% | 59 | 7797 | 2716 | 5081 | No |
| 65% | 87 | 7849 | 2750 | 5099 | No |
| 65% | 88 | 7852 | 2743 | 5109 | No |
| 70% | 59 | 7803 | 2340 | 5463 | No |
| 70% | 87 | 7850 | 2359 | 5491 | No |
| 70% | 88 | 7851 | 2340 | 5511 | No |
| 75% | 59 | 7801 | 1939 | 5862 | No |
| 75% | 87 | 7852 | 1961 | 5891 | No |
| 75% | 88 | 7853 | 1963 | 5890 | No |
| 80% | 59 | 7803 | 1556 | 6247 | No |
| 80% | 87 | 7849 | 1572 | 6277 | No |
| 80% | 88 | 7855 | 1560 | 6295 | No |
| 85% | 59 | 7805 | 1166 | 6639 | No |
| 85% | 87 | 7852 | 1169 | 6683 | No |
| 85% | 88 | 7857 | 1180 | 6677 | No |

The used networks were generated with the benchmark network generator tool of Lancichinetti and Fortunato [1] with the random seed ("seed.dat" file) set to the value in the second column. The exact commands used for generation were the following:
```
benchmark -N 4000 -k 6 -maxk 100 -mu %fuzziness% -t1 3 -t2 0.2 -minc 20 -maxc 100 -on 200 -om 2
benchmark -N 4000 -k 6 -maxk 100 -mu %fuzziness% -t1 3 -t2 0.2 -minc 20 -maxc 100 -on 0 -om 0
```
for networks in part A with and without overlapping communities, respectively. Networks in part B were generated with the command
```
benchmark -N 1000 -k 15 -maxk 50 -mu %fuzziness% -t1 2 -t2 1
```
The variable %fuzziness% was substituted with the corresponding values of the first column in all commands.

[a]"Yes" indicates that the network was generated with overlapping modules, that is, 200 nodes were selected as "overlapping nodes" having two parent communities and receiving intra-modular links from both of them. "No" indicates that no such nodes were present in the network.

**Table S8: Average perturbation centrality of inter-modular nodes and inter-modular hubs as compared to the average perturbation centrality of intra-modular, non-hub nodes**

| Networks[a] | Relative perturbation centrality of inter-modular non-hubs[b] | Relative perturbation centrality of intra-modular hubs[b] |
|---|---|---|
| Benchmark graphs with pronounced modules[e] | 126% | 115% |
| Benchmark graphs with fuzzy modules | 102% | 149% |
| Substrate-bound Met-tRNA synthetase protein structure network | 118% | 126% |
| Substrate-free Met-tRNA synthetase protein structure network | 113% | 125% |
| Filtered Yeast Interactome | 96% | 273% |
| Database of Interacting Proteins yeast interactome (release 2010) | 129% | 193% |
| Database of Interacting Proteins yeast interactome (release 2005) | 153% | 256% |
| *E. coli* metabolic network | 90% | 479% |
| *B. aphidicola* metabolic network | 112% | 281% |
| School friendship network | 107% | 149% |
| **Mean and standard error** | **115% (5.81%)** | **215% (35.6%)** |

Perturbation centralities were calculated as described in **Methods** for each node of the network, and their average was taken. Hubs were nodes with a degree in the top 10%. Modularization was performed using the ModuLand Cytoscape plug-in [19]. Inter-modular nodes were defined as nodes having more than 40% inter-modular edges.
[a]Network descriptions are given in **Supplementary Methods** of **Text S1**.
[b]Percentages reported are the percentage of the average perturbation centrality of the node set marked in the description of the column header compared to the average perturbation centrality of intra-modular, non-hub nodes.

**Table S9. Signaling residues and residues of experimentally verified importance in Met-tRNA synthetase protein structure network**

| A. List of individual perturbation properties of residues with experimentally verified importance [5] | | | | |
|---|---|---|---|---|
| **Residue** | High importance in the substrate-free conformation | High importance in the substrate-bound conformation | High increase of importance on binding | High decrease of importance on binding |
| ASP-456 | | | Y | |
| ASN-452 | | | | |
| ASP-353 | Y | Y | | |
| ALA-352 | Y | Y | | Y |
| TYR-357 | Y | Y | | |
| LEU-355 | | | | |
| TYR-359 | | | | |
| ARG-356 | Y | Y | | |
| ARG-395 | | | | |
| ASN-348 | | | | |
| PHE-350 | | | | |
| HIS-349 | | | | |
| ASP-351 | Y | Y | | |
| **TRP-461** | Y | | Y | |
| ASP-449 | | | | |
| TYR-358 | | | | Y |
| THR-360 | Y | Y | | |
| ALA-361 | Y | Y | | |
| LYS-362 | Y | Y | | |
| SER-354 | Y | Y | | |
| B. List of individual perturbation properties of residues predicted to participate in intra-protein signaling [5] | | | | |
| **Residue** | High importance in the substrate-free conformation | High importance in the substrate-bound conformation | High increase of importance on binding | High decrease of importance on binding |
| GLN-538 | | | | Y |
| VAL-543 | | | | |
| PHE-484 | | | | |
| MET-488 | | | | |
| PHE-437 | Y | Y | | |
| ASP-456 | | | Y | |
| ASN-452 | | | | |
| LEU-495 | Y | Y | | |
| TRP-432 | Y | Y | | |
| THR-499 | Y | | | Y |
| HIS-21 | | | | |
| ARG-36 | Y | Y | | |
| ASP-32 | Y | Y | | |
| **LEU-13** | Y | Y | | |
| PHE-377 | | | | Y |
| VAL-381 | | | | |
| PHE-84 | Y | Y | | |
| ASN-391 | | | | |
| LEU-392 | | | | |
| PHE-87 | Y | Y | | |
| ILE-89 | **Y** | **Y** | | |
| ASP-384 | Y | Y | | |
| MET-25 | **Y** | Y | | |
| LYS-388 | | Y | | |
| TYR-357 | Y | **Y** | | |
| ILE-385 | Y | Y | | |
| HIS-323 | Y | Y | | |
| LEU-26 | | Y | Y | |
| HIS-28 | **Y** | Y | | |
| LEU-355 | | | | |
| TYR-531 | | | | |
| VAL-326 | | | Y | |
| TYR-359 | | | | |
| TRP-346 | | | | |
| LEU-498 | | | | Y |
| MET-333 | | | **Y** | Y |
| ARG-395 | | | | |
| PHE-350 | | | | |
| TYR-37 | Y | Y | | |
| **TRP-461** | Y | | Y | |
| LYS-492 | Y | Y | | |
| LEU-363 | **Y** | **Y** | | |

The list of amino acids of *E. coli* Met-tRNA synthetase predicted to participate in the transmission of conformational changes was taken according to the text and Figure 5 of the paper of Ghosh and Vishveshwara [5]. The list of amino acids with experimentally verified importance was taken from the same article [5] listing earlier findings. Protein structure networks of the substrate-free and substrate-bound forms of *E. coli* Met-tRNA synthetase were constructed as described in **Supplementary Methods**. Perturbation centralities were calculated as described in **Methods** of the main text. Differences of perturbation centralities were obtained by subtracting the *rank* of the perturbation centrality of the substrate-bound conformation from the *rank* of the perturbation centrality of the substrate-free conformation. Residues in the top 20% of largest perturbation centrality in either the substrate-free or the substrate-bound conformation, and residues having an increase or decrease of perturbation importance in the top 20% are shown in the respective column marked by letter "Y". Boldface Y-s signify a top 10% level of importance. Starting and ending residues of the predicted communication pathways, Leu-13 and Trp-461 were also marked with bold letters.

# Supplementary Results

**Connection of the communicating vessels model with perturbation dissipation using shortest paths and having an exponential decay**

In the communicating vessels model the change of the energy of a node is given by the following differential equation:

$$\frac{dS}{dt} = -\sum_{i=0}^{l}\left(\frac{S - S_i}{2} w_i\right) - D_0 \quad (1)$$

where $S$ is the energy of the current node, $l$ is the number of edges of the current node, $w_i$ is the weight of the $i^{th}$ edge, $S_i$ is the current energy of the node on the other end of the $i^{th}$ edge, and $D_0$ is a parameter defining the amount of energy dissipated by a node in a given time step.

Let us define the case of unobstructed propagation as $S \gg S_i$. The consequence of the above definition is that at any given time step, the amount of perturbation transmitted to neighbors is approximately $S\frac{w_i}{2}$. The amount transferred to the $n^{th}$ neighbor is approximately $S\left(\frac{w_i}{2}\right)^n$ using the same logic, which results in an exponential decline with distance. If $w_i \geq 1$, transfer will stop after a second or less in simulation with energy levels evening out to S/2, S/2 for the two affected nodes, which results in the same exponential decline, if the condition $S \gg S_i$ holds for the other nodes as well. If there is a connection between nodes at the same distance from the origin, no energy will be transmitted, since $S = S_i$ if the edge weight from the origin is the same for both nodes. Generally, if there are no large differences between edge weights, nodes with the same distance from the origin will tend to have similar values. If there are multiple edges at a given network shell (i.e. distance from the node where perturbation started) to the next one, this only changes the amount of energy propagated by a proportional amount. According to this logic, we can hypothesize that these perturbations tend to travel on shortest paths.

The dissipation of energy in time will be exponential in the unobstructed case. This comes from integrating equation (1) with the condition $S \gg S_i$, which yields the result

$$S(t) = \frac{S_0 W_d}{2} e^{-\frac{W_d t}{2}} - \frac{2 D_0}{W_d} \quad (2),$$

which describes an exponential decay, if the weighted out-degree, $W_d = \sum_{i=0}^{l} w_i > 0$.

If the flow of perturbation is totally obstructed (all nodes in the neighborhood have the same energy level) then $S = S_i$, thus $S(t) = S_0 - D_0 t$, which is a simple linear dissipation with time, resulting in no energy transfer with distance. Most real world cases are somewhere between the above two extremes.

Module entrapment occurs when the energy level of a module is too high for the number and weight of inter-modular edges to propagate the perturbation outwards the module, so the energy level inside the module is raised to $S_j = S_i$ for every *i* and *j* node inside the module.

**Comparison of the communicating vessels model with the random walk model of ITM-Probe**

ITM-Probe is an algorithm modeling information propagation in complex networks using a random walk model [2]. For the comparison of the communicating vessels model used by Turbine with the ITM-Probe method, perturbation centralities obtained on networks with different ratios of inter-modular edges were compared with the "number of visited nodes" measure of the emitting model of the ITM-Probe method as described in the Supplementary Methods section.

The "number of visited nodes" measure used by the emitting model of the ITM-Probe method is theoretically quite similar to the perturbation centrality of the communicating vessels model used by Turbine, since in the ITM-Probe model the random walk has a defined probability (with a default value of 15%) to end at every node visited. This means that the average distance of the last node of the random walk is proportional with the distance from the originating node, so in the end the nodes closer to the origin tend to dissipate more walks in ITM-Probe [2] in the same manner as nodes closer to the origin dissipate more energy in our communicating vessels model.

In the comparison of Turbine with ITM-Probe a total of 51 benchmark graphs [1] were tested, with a ratio of inter-modular edges increasing from 0.05 to 0.85 with steps of 0.05. Three networks were generated with each ratio of inter-modular edges value, with three different random seeds to make the results more robust as described in **Supplementary Methods**. Perturbation centrality and the "number of visited nodes measure" were calculated for each node (1,000 in each network).

In the first simulation perturbation centrality and the "number of visited nodes" measure were averaged for all the 3 randomly generated networks for each inter-modular edge ratio. Spearman correlation of perturbation centrality and the "number of visited nodes" measure was calculated with the R program package [3]. Results are shown in **Figure S1**. Spearman correlation of the average number of visited nodes with the average perturbation centrality calculated for low-intensity perturbations was a perfect 1 meaning that the limitation of perturbation propagation imposed by modular boundaries decreased in exactly the same manner, when assessed by either the Turbine or the ITM-Probe method as the modular structure was coalescing by the gradual increase of the number of inter-modular edges.

Despite the above effect, large differences were observed between the tested measures of ITM-Probe and Turbine. Correlations of degrees, number of visited nodes and perturbation centrality were much lower in different simulations depending on the level of modularity as shown on **Figure S2**. The correlation between perturbation centrality and the "number of visited nodes" measure (**Figure S2C**) was high when the modularity of the network was noticeable (the ratio of inter-modular edges was smaller than 0.3). However, when the modules started to merge, the correlation diminished, and even turned to negative at the highest ratios of inter-modular edges. An explanation of this effect lies in the fact that in the ITM-Probe method, the degree of a node is inversely correlated with the "number of visited nodes" measures at ratios of inter-modular edges higher than 0.2 (**Figure S2B**). The diminished importance of hubs in information spread is rather counter-intuitive. For low ratios of inter-modular edges, the correlation between the "number of visited nodes" measure and node degree was positive, which shows that high-degree inter-modular nodes are the most important information spreaders in highly modular networks, and in the scale-free benchmark graphs nodes having a higher degree also have a higher chance of gaining inter-modular edges. Correlation of perturbation centrality with node degree varied with a much lower effect size ($\Delta r=0.25$), as shown on **Figure S2A**. The differences observed can be explained by the limiting effect of modular boundaries on perturbation propagation described in the main text.

**Figure S3** illustrates the individual perturbation centralities and "number of visited nodes" measures for a single scale-free benchmark graph generated using the random seed value of 87. Perturbation centrality was positively correlated with the node degree in all ratios of inter-modular edges analyzed, but the limiting effect of modular boundaries on the propagation of perturbations was noticeable at low ratios of inter-modular edges. This effect was evident by the segregation of perturbation centrality data to stripes at the ratio of inter-modular edges of 0.05. At this benchmark graph configuration nodes had only one or two inter-modular edges, if any. The top segregated layer of perturbation centrality data corresponds to nodes having two inter-modular edges, while the middle layer of perturbation centrality values corresponds to nodes having one inter-modular edge. The highest "number of visited nodes" values did not change with growing ratios of inter-modular edges (cf. Panels G through L of **Figure S3**), only the lowest "number of visited nodes" values – and thus the average – got higher as ratio of inter-modular nodes increased, resulting in a saturation-like effect for the "number of visited nodes" measure possibly explaining the inverse correlation with degree in networks with highly overlapping modules.

**Dissipation-free propagation of perturbation is characterized by fill time, which reciprocally correlates with closeness centrality**

Fill time was defined as a measure assessing the propagation of a continuous perturbation in the communicating vessels model without dissipation as stated in the main text. The fill time of node $i$ is the time needed to raise the energy level of 80% of the nodes above 1 unit (0.01% of initial perturbation) in a simulation, where a perturbation of 10,000 units was added to node $i$ <u>in each time step</u>, and was propagated without dissipation. Fill time was calculated for all nodes in several benchmark graphs and in real-world networks. **Table S1** shows that the reciprocal of the fill time of node $i$ strongly correlated with the closeness centrality of the same node ($\bar{r} = 0.895; 95\% CI = 0.843 - 0.946; p = 2.1 \cdot 10^{-11}$, one-sample t-test, Shapiro–Wilk normality test successful with p=0.178). We note that the two metabolic networks, where the correlation with the fill time was the lowest, correspond to a different class of networks according to Guimerà *et al.* [20] Since the closeness centrality of node $i$ is defined as the mean geodesic distance (mean shortest path) between node $i$ and all other nodes, the high correlation between the reciprocal of fill time and closeness centrality meant that the shortest paths determined most of the dissipation-free propagation of perturbations. This agrees with the expectations [21], and validates the use of the communication vessels model, which has this property as shown in the **Supplementary Results** of **Text S1**.

**Comparing the effect of degree and modular position on perturbation centrality in real-world networks**

In the experiment comparing the silencing times of the Lancichinetti [1] benchmark networks, the ratio of perturbation dissipation efficiencies of two node categories out of the 4, namely: inter-modular non-hubs and intra-modular hubs, had interesting changes with the modular structure. As we noted earlier, modules of real world networks seem to be more overlapping than the pronounced modules of our benchmark graphs. Starting from these notions we compared the mean perturbation centrality of intra-modular hubs against inter-modular non-hubs as the percentage of the mean perturbation centrality of intra-modular non-hubs in multiple real-world networks (**Table S8**). Hubs were nodes with the top 10% degree, and inter-modular non-hubs were those with at least 40% inter-modular edges – just as in the previous calculations. Inter-modular non-hubs had only a 15% larger perturbation centrality, while hubs had a 115% larger perturbation centrality than intra-modular non-hubs. The large (87%) difference between the effect of hubs versus the effect of inter-modular non-hubs suggest that from a perturbation perspective real-world networks resemble the benchmark graphs with fuzzy modules more, than the benchmark graphs with pronounced modules. (Note that the same observation was obtained when we compared the low-intensity and high-intensity silencing times – see **Table S2**.)

**Amino acids participating in intra-protein signaling have a high perturbation centrality in the protein structure network**

We have assessed the perturbation centrality of amino acids forming α-helices, β-sheets and loops in two pairs of protein structure networks corresponding to the substrate-free and substrate-bound conformations of *E. coli* Met-tRNA synthetase and rabbit cytochrome P450 2B4, respectively. The Wilcoxon rank-sum test ($\alpha=0.00625$ adjusted with Bonferroni correction) indicated significantly (p=0.00023, 0.00015, 0.00083, 0.0014 for the free and bound conformations of Met-tRNA synthetase and cytochrome P450, respectively) larger perturbation centralities of α-helices compared to the global mean, while the same test indicated significantly (p=$3.2*10^{-6}$, $9.5*10^{-6}$, $2.3*10^{-6}$, 0.0001 for the free and bound conformations of Met-tRNA synthetase and cytochrome P450, respectively) smaller perturbation centralities for loops compared to the global mean. (**Figure S6A through S6D**; $\alpha=0.001$). The average perturbation centrality for β-sheets showed a larger variation. The same data calculated for betweenness and closeness centralities is shown on **Figures S7** and **S8** of **Text S1**. Betweenness centrality had a much larger deviation than perturbation centrality, and both closeness and betweenness centralities could differentiate less between amino acids in different secondary structures than perturbation centrality.

We continued the analysis of protein structure networks by assessing the perturbation centralities of *E. coli* Met-tRNA synthetase amino acids participating in intra-protein signaling. We selected Met-tRNA synthetase, since an earlier molecular dynamics study [5] identified key amino acids involved in the transmission of conformational changes upon substrate binding (termed as "Signaling residues" on **Figures S6E** and **S6F**). We have also checked a set of residues, whose importance have already been experimentally established [5] termed as "Experimental residues" on **Figures S6E** and **S6F**). **Figures S6E** and **S6F** show that the perturbation centrality was significantly (Wilcoxon rank-sum test, $\alpha=0.0125$ adjusted with Bonferroni correction for a FWER of 0.05) higher for both the Signaling (p=$1.7*10^{-6}$, $1.1*10^{-6}$ for the free and Met-tRNA-bound conformations, respectively) and Experimental residues (p=0.0072, 0.061 for the free and Met-tRNA-bound conformations, respectively) than average. Data in **Figures S7** and **S8** of **Text S1** show that both closeness and betweenness centralities could differentiate Signaling residues, but neither closeness nor betweenness centralities could differentiate the Experimental residues from the average centrality of the whole protein. 67 or 60% of Signaling or Experimental residues, respectively, were in the top 20% of amino acids having the largest perturbation centrality and/or the largest change of perturbation centrality upon substrate binding (**Table S9** of **Text S1**).

**Testing the effects of edgetic perturbations**

The definition of perturbation centrality described in the main text: "the reciprocal of silencing time retrieved by using a Dirac delta type starting perturbation of $10*n$ units, where $n$ is the number of nodes in the network, using a dissipation value of 1 can be extended to describe the perturbation centrality of an edge, where the perturbation centrality of the edge connecting nodes $i$ and $j$ is the reciprocal of the silencing time obtained when the same perturbation was started from nodes $i$ and $j$ at the same time. Testing these edgetic perturbations on benchmark networks revealed that edgetic perturbations show the same basic properties as (single) node-based perturbations (cf. **Figure 1** and **Figure S10** of **Text S1**).

# Supplementary Methods

### Description of the networks used

### Benchmark graphs

Scale-free, modular benchmark graphs were generated using the unweighted, undirected benchmark graph generating tool of Lancichinetti and Fortunato [1]. Double edges were removed from the generated networks. Random seed values of 59, 87 and 88 were used to generate three sets of networks. When 7 sets of networks were generated, the additional seeds 19, 20, 42 and 85 were used. These benchmark graphs had 4,000 nodes and 13,785 ± 421 edges. The ratio of inter-modular edges was set to 0.05 in the case of the networks termed as "networks with pronounced modules", and 0.4 for the networks termed as "networks with fuzzy modules". Networks with no overlapping nodes were used in all cases except for the testing of inter-modular nodes (**Figure 2** of the main text), where 200 overlapping nodes were generated, each belonging to two separate modules. A total of 28 networks of this type were created. For the ITM-Probe simulations in this Supplementary Information the number of nodes was 1,000 and the number of edges was 7,825 ± 133 with the ratio of inter-modular edges varying between 0.05 and 0.85 in steps of 0.05 for a total of 51 networks. None of the networks generated for the ITM-Probe comparison contained overlapping nodes. Detailed information about the generated benchmark graphs, as well as the exact commands used for generation is available in **Table S7**. Benchmark graph data can be downloaded from our web-site: http://turbine.linkgroup.hu.

### Protein structure networks

The protein structure networks of *Escherichia coli* Met-tRNA synthetase were generated from the starting and equilibrated state of the molecular simulation of the *E. coli* Met-tRNA synthetase/tRNA$^{Met}$/Met-AMP complex corresponding to the substrate-free and substrate-bound forms of the enzyme, respectively. The structure for the substrate-free form of *E. coli* Met-tRNA synthetase is available (pdb ID: 1QQT) [22]. However, the substrate-bound conformation containing tRNA$^{Met}$ and Met-ATP was not experimentally solved yet, but was created using molecular simulation software using the known substrate-bound structure of *Aquifex aeolicus* Met-tRNA synthase as a template, as described and kindly shared by Ghosh and Vishveshwara [5]. The protein structure network was obtained from the PDB data with the help of the RINerator software [23], which uses the Probe program [24] for calculating interaction strengths. Probe returns negative values for repulsive interactions; we have taken the absolute value of the interaction strength for all interactions, since both repulsive and attractive interactions transmit perturbations. In contrast with the previous article of Szalay-Bekő et al [19], where multiple edges adversely affect calculations, in Turbine, multiple edges are neutral, so the full network was used with every edge and self-loop intact instead of averaging the strengths of multiple links into a single one. The rationale behind this is to retain as much information as possible from the original file. The protein structure network had 547 nodes, since the first 3 N-terminal amino acids were not participating in the network and all ligands and cofactors, including the tRNA were removed. The final weighted, undirected protein structure networks of the substrate-free and substrate-bound enzyme contained 6,901 and 6,744 edges, respectively.

Protein structure networks of the substrate-free and substrate-bound forms of rabbit cytochrome P450 2B4 protein were also created with the RINerator software [23] by using the 1PO5 and 1SUO pdb files of the Protein Data Bank, and taking the absolute value of the resulting interaction strengths. The surrounding water molecules, the ligands and the cofactors were removed from the networks, resulting in 465 nodes with 6,278 edges in the substrate-free (open) conformation and 465 nodes with 6,409 edges in the substrate-bound (closed) conformation, both undirected, weighted networks. Protein structure network data can be downloaded from our web-site: http://turbine.linkgroup.hu.

**Yeast protein-protein interaction networks**

The Filtered Yeast Interactome (FYI) is a high-fidelity yeast protein-protein interaction dataset containing data consistently obtained using several different methods [6]. The downloadable data set in the Supplement of the article contained 1,302 proteins (nodes) having 2,312 interactions (edges). The giant component of this network had 695 nodes and 1,614 edges.

The "Database of Interacting Proteins yeast interactome (release 2005)" network is the giant component of the unweighted and undirected yeast protein-protein interaction network assembled by Ekman et al. [25] using the 2005 March compilation of the Database of Interacting Proteins [26] consisting of 2,444 nodes and 6,271 edges covering approximately half the proteins of yeast genome. Besides the rather high confidence of its data, we choose this network, because it was used in the identification of party and date hubs, an interesting dynamic feature of protein-protein interaction networks [25] and its properties were assessed in our earlier publications [19].

The "Database of Interacting Proteins yeast interactome (release 2010)" network was created from the 2010 October compilation of the Database of Interacting Proteins [26]. Only high-fidelity interactions marked as "core" were included in the network, yielding a giant component of 1,884 nodes and 4,234 edges.

Interaction weights of yeast proteins were obtained from the yeast whole-genome mRNA expression dataset of Holstege et al. [27] containing data of 6,180 yeast genes (5461 with expression levels, and one gene with two different expression levels). Missing data (719 nodes total, less than 12%) were substituted as the ln-transformed average expression level of all other proteins, 0.9205, taking into account that the distribution of expression data is approximately lognormal [28]. For calculating the changes of expression levels in different types of stress, the datasets of Gasch *et al.* [29] were used, which describe the relative changes of expression levels based on a set of microarray data. Thus, the stressed expression levels were calculated by multiplying the Holstege baseline with the Gasch relative changes according to the previous article [30]. The particular datasets used for the different stresses were the following: "Heat Shock 15 minutes hs-1" (25°C to 37°C heat shock for 15 minutes) for the heat-shocked network; "constant 0.32 mM $H_2O_2$ (30 min) redo" (0.32 mM hydrogen-peroxide treatment for 30 minutes) for the oxidative stress, and "1 M sorbitol - 15 min" (hyper-osmotic shock using 1 M sorbitol for 15 minutes) for the osmotic stress. Detailed experimental data is available in the article describing the dataset [29]. Edge-weights of non-stressed and stressed protein-protein interaction networks were generated from the expression data by multiplying the abundances of the two connected proteins as described earlier [30], since larger concentrations of involved proteins make their interaction more likely. Thus all final interactomes were weighted and undirected.

The use of even larger interactomes such as BioGRID or STRING were also considered, but was dropped due to computational constraints, since there were 6 magnitudes of difference between weighted out-degrees. Adding the fact that the weight distribution was approximately lognormal (so most of the nodes had low weighted outdegrees), and the required constraint that the maximum weighted outdegree should be no more than 1 ($-1 \leq \Delta t \sum_{i=0}^{l} w_i \leq 1$) meant that unrealistically high analysis times would have been required to attenuate the perturbations, since enforcing an upper limit for the maximum weighted outdegree results in median weighted outdegrees becoming extremely low, which in turn makes propagation speed disproportionately slow. Protein-protein interaction network data can be downloaded from our web-site: http://turbine.linkgroup.hu.

**Metabolic networks**

Metabolic networks were created by Balázs Szappanos (Biological Research Centre, Hungarian Academy of Sciences, Szeged, Hungary), and were the same as used in the papers of Mihalik and Csermely [30] and Szalay-Bekő *et al.* [19]. *Escherichia coli* metabolic network contained 249 metabolites (nodes) and their 730 transformations (edges), while the *Buchnera aphidicola* metabolic network contained 190 nodes and 563 edges. These networks were constructed based on the primary data of Feist *et al.* [9] and Thomas *et al.* [8], respectively. Frequent cofactors were deleted from the networks, except of those metabolic reactions, where cofactors were considered as main components. For the better comparison of networks, metabolic reactions were taken irreversible, and flux balance analyses (FBA) were performed resulting in weighted networks. All flux quantities were minimized, whereas reactions non-affecting the biomass production were considered having zero flux. Weights were generated as the mean of the appropriate flux quantities in absolute value, except the case when one of the fluxes was zero that automatically resulted in a zero weight [30]. For *E. coli*, data from rich medium was used to make the metabolic network more similar to the network of *B. aphidicola*. Final networks were thus weighted and undirected. Metabolic network data can be downloaded from our web-site: http://turbine.linkgroup.hu.

**School friendship network**

The social network was community-44 from the Add Health survey[2] as described by James Moody [31] and Mark Newman [32]. This network has an approximately equal number of black and white students and 4 well-developed, rather dense communities. The network contains 1,147 students with 6,189 directed edges between them. In our current study directed parallel edges were merged into a single undirected edge with a weight equal to the sum of the original weights, and only the giant component of the network was used. This process resulted in a weighted undirected network consisting of 1,127 nodes and 5,096 edges with weights between 1 and 12. Network data can be downloaded from our web-site: http://turbine.linkgroup.hu.

---

[2]This research uses data from Add Health, a program project designed by J. Richard Udry, Peter S. Bearman, and Kathleen Mullan Harris, and funded by a grant P01-HD31921 from the National Institute of Child Health and Human Development, with cooperative funding from 17 other agencies. Special acknowledgment is due Ronald R. Rindfuss and Barbara Entwisle for assistance in the original design. Persons interested in obtaining data files from Add Health should contact Add Health, Carolina Population Center, 123 W. Franklin Street, Chapel Hill, NC 27516-2524 (addhealth@unc.edu).

**Network measures and their calculation**

The degree of a node is defined as the number of edges of the node. More precisely for the perturbation simulations we have to consider the number of outgoing edges (i.e. the out-degree). However, all networks we used were undirected, so the degree of a node was equivalent to its out-degree in all cases. Weighted degree was calculated by summarizing the edge weights of all edges for a given node. Degree and weighted degree were calculated using the built-in algorithm of Turbine.

PageRank is a random-walk based measure [33], where the outbound edges of a node increase the centrality of those nodes for which the given edge is an inbound edge. The exact value of increase is proportional to the current PageRank value of the edge's outbound node. Multiple iterations of this method yielded a converging result [33]. PageRank values were calculated using the algorithm of the igraph [34] package.

Closeness centrality [35] is defined as the mean geodesic distance (mean shortest path) between a given node and all other nodes reachable from it. Betweenness centrality [36] gives the (relative) number of shortest paths between every two nodes in the network, which include the examined node. Closeness centrality and betweenness centrality were calculated using the Pajek [37] package.

Community centrality is a measure coined for the ModuLand overlapping community structure determination program [19], which measures the centrality of a node separately for different communities. Community centrality was calculated by the ModuLand Cytoscape plug-in [19].

**Network modularization**

Community structures of networks were determined by the ModuLand Cytoscape plug-in [19] with basic settings. A threshold of 0.9 was used for merging highly correlated modules in all networks. For the school friendship network, modules of the second hierarchical level were used (where meta-nodes of the level represent modules of the original network and meta-edges of the level represent the overlap of the modules of the original network as described in [19]) to obtain the 4 densely connected communities of the network [32,38].

**Determination of the "number of visited nodes" measure of the ITM-Probe method**

To compare the "number of visited nodes" measure of the ITM-Probe method [39] with the perturbation centrality obtained by using the Turbine program, the recently released standalone version of the ITM-Probe method [39] was used, since calculating the ITM-Probe results for all nodes on a network would take a tremendous amount of time using the web-based interface. A plug-in was written for Turbine that converts any complex network in Turbine format (needed for the fast-calculation of the Turbine program) to the format required by ITM-Probe (JSON). This plug-in can be downloaded from the Turbine web-site at http://turbine.linkgroup.hu. The ITM-Probe method was used with its emitting model with a damping factor of 0.85. The main script file of the ITM-Probe method was executed separately for every single node in a certain network, and all output was concatenated into a single text file. The "number of visited nodes" measures were extracted from the resulting ITM Probe text file with an awk script.

# Supplementary References